\documentclass[12pt]{iopart}
\usepackage{graphicx}
\usepackage{amsmath}
\usepackage{color}

\begin{document}

\title[Energy dynamics, heat production and heat--work conversion with qubits: towards the development of quantum machines]
{Energy dynamics, heat production and heat--work conversion with qubits: towards the development of quantum machines}

\author{Liliana Arrachea}

\address{Escuela de Ciencia y Tecnolog\'{\i}a and ICIFI, Universidad de San Mart{\'{\i}}n, Av. 25 de Mayo y Francia (1650) Buenos Aires, Argentina}
\ead{larrachea@unsam.edu.ar}

\vspace{10pt}
\begin{indented}
\item[]November 2022
\end{indented}

\begin{abstract}
We present an overview of recent advances in the study of energy dynamics and mechanisms for energy conversion in qubit systems with special focus on realizations in superconducting quantum circuits.
We briefly introduce the relevant theoretical framework to analyze  heat generation, energy transport and energy conversion  in these systems with and without time-dependent driving considering the effect of  equilibrium and non-equilibrium environments.  
We analyze specific problems and mechanisms under current investigation  in the context of qubit systems. These include the problem of energy dissipation and possible routes for its control, energy pumping between driving sources and heat pumping between reservoirs, implementation of thermal machines and mechanisms for energy storage. We highlight  the underlying fundamental phenomena related to geometrical and topological properties, as well as many-body correlations. We also present an overview of recent experimental activity in this field. 
\end{abstract}
\newpage

%
%
%
%
%

\tableofcontents
\thispagestyle{empty}
\clearpage

\newpage

\section{Introduction}

In the middle of the twentieth century an epochal revolution in technology  took place after the development of  semiconductors and  electronic devices, accelerating, in the last decades, towards an impressive and constant miniaturization. Although quantum effects are crucial in the semiconductors, they operate as macroscopic bodies. The scenario has changed considerably with the emergence of  the so called {\it quantum technologies} where the (quantum) devices operate at the level of single atoms, ions and spins, by fully exploiting the quantum-mechanical nature of these systems. 

The building blocks of quantum computers are the qubits and several proposals have been formulated  to realize this fundamental component \cite{platforms}. 
These ongoing technological developments were triggered by several important scientific advances. In solid-state systems, nanofabrication techniques, enable the construction of nano-devices like quantum dots to confine a few electrons \cite{kouwenhoven,semi-qubits}, nanomechanical \cite{nanomec} and optomechanical \cite{optomec} systems, electronic interferometers in topological insulators \cite{interf} as well as superconducting circuits \cite{q-sup,cqed}. The field of atomic and molecular optical (AMO) physics devoted to study trapped atom/ions and  photons has lead to unprecedented level of control from few qubits to complex quantum many-body states \cite{amo1,amo2}.  Nitrogen vacancies (NV) centers in diamond \cite{nv}, has turned into one of the most competitive implementations in many quantum  information processing protocols. An overview of this continuously growing and successful adventure can be found in~\cite{Schleich}. The diversity of operations in quantum devices include, quantum-state manipulation, measurements and the implementation of logic gates. Nowadays, prototypes of quantum computers are already used to implement machine-learning algorithms \cite{qmach-lear} as well as quantum simulations \cite{q-sim}. 

In  quantum devices belonging to the new technological generation, the focus of the performance is naturally the precision and the computational possibilities. In order to benefit from the quantum properties of these devices it is necessary to overcome problems such as miniaturization, error correction, and scalability. 
There are also big expectations on the quantum advantage regarding the energetic optimization. However, it is acknowledged that
this aspect is still unclear \cite{auf}. In this sense, there is an increasing consensus 
 about the need of a special effort at the level of both fundamental and applied research to further understand this important side of the quantum technologies \cite{mani}.
 The understanding and control of the energy dynamics  and entropy generation ubiquitous of all these systems is 
of paramount importance. 
In particular,  the fact that most of their operations require  the challenging  conditions imposed by mK temperatures,  motivates the search for efficient in-chip cooling and the possibility of converting  the generated heat into useful work.

The introduction of thermal machines, like heat engines and refrigerators,  has been at the heart of the industrial revolution that took place between mid-19th century and the beginning of the last century. Similarly, in the last 10 years, the efforts devoted to  investigate heat manipulation/conversion in quantum devices grew enormously \cite{janet,john,benenti,colloquium,landi-pater,deffner}. At the experimental level, one direction of research has been the implementation of the thermodynamic cycles and refrigeration mechanisms in few-level quantum systems, like atoms and ions \cite{ros,mas,chris}, NV centers\cite{klat,peter}. Another direction is the study of thermoelectric effects, energy harvesting and refrigeration in a diversity of solid state devices, like quantum dots \cite{set,kosti,pek1,rafa,thermo1,thermo2,thermo3,thermo4,thermo5, thermo6,thermo7}, superconducting nanostructures \cite{giazotto1,ale,germanese}, nano electro-and optomechanical devices \cite{opto}  and systems in the quantum Hall regime \cite{pierre,banergee,thermohall}. The third direction is the study of the energetics of superconducting qubits in quantum circuits, on which we focus in the present contribution. All these setups belong to the paradigm of {\em open quantum systems}, since they   are based on a configuration where a few-level coherent quantum system is operated out of equilibrium in contact to macroscopic parts which play the role of reservoirs and/or the environment generated by their measurement and manipulation setup.


A qubit is the simplest system to generate a quantum-state superposition. 
So far, qubits realized in superconducting quantum circuits \cite{makhlin,devoret1,devoret2,wendin,blais1} are among the most advanced platforms regarding scalability and degree of concrete implementations. These qubits are akin the atomic realizations in quantum electrodynamic cavities, since the superconducting device is designed to behave as a two-level system while the embedding circuit  behaves akin a photonic cavity. In addition, the quantum dynamics of the qubit-circuit system is formally similar to that induced by the atom-light interaction in the cavity \cite{cavqed}. The coupling between the qubit and the environment is a source of decoherence which is an undesirable but unavoidable effect in quantum information processing. 
In the investigation of many mechanisms related to the energy dynamics the qubit-environment coupling is a key ingredient, which is amenable to being controlled in hybrid circuit quantum electrodynamics (cQED) hosting superconducting qubits \cite{cottet}. This possibility along with the high degree of control of the physical mechanisms (voltage gates and magnetic fluxes) to manipulate quantum states in these systems offer an ideal playground to investigate the fundamental mechanisms of quantum thermodynamics like heat transport, entropy production, energy storage and energy conversion.

Aim of this review is to focus on the body of work that dealt with heat/energy manipulation mechanisms which are proposed to take place or have been illustrated in  qubit systems.
Most of the selected topics are  related to solid-state platforms like cQED. Nevertheless, many of the basic mechanisms and effects 
are ubiquitous in few-level quantum systems embedded in macroscopic or noisy  environments and manipulated by time-dependent processes. 
There are several other very relevant reviews with focus on complementary topics. In particular, on quantum thermodynamics in connection with quantum information~\cite{janet,john}, recent advances in the description of entropy production in non-equilibrium systems \cite{landi-pater} and quantum thermodynamic devices \cite{deffner}. On the specific topic
of the
realization of superconducting qubits there are also several reviews~\cite{makhlin,devoret1,devoret2,wendin,blais1}. Of particular reference for the present work is the Colloquium by Karimi and Pekola~\cite{colloquium}, which focuses on quantum heat transport in condensed matter systems.
We  try to give some perspectives not covered there. In particular, we address the topics of  energy manipulation with time-dependent driving and mechanisms of energy conversion. 

There is a rich variety of problems under the topics covered in this review, which 
are basically defined by the nature of the driving (slow, fast, single or multiple-source) and the characteristics of the environment (thermal bath, noisy, non-equilibrium, with feedback control). The number of physical situations range from 
the expected dissipation of energy to the realization of thermal machines to generate work or refrigerate. 
As in other branches of modern physics geometric and topological properties emerge
in the route through these scenarios, and we devote some space to  analyze them. 













The presentation is organized as follows:

Section 2 is devoted to introduce basic concepts of quantum thermodynamics that will be useful to discuss the main mechanisms addressed in the forthcoming sections. In particular, we start presenting definitions of heat and work 
for quasi-static and finite-time non-equilibrium processes. We briefly introduce the different formalisms to 
describe of time-dependent quantum dynamics in the slow (adiabatic) and fast (Floquet) regimes. We also present the basic tools to describe the heat steady-state transport induced by thermal bias applied at the reservoirs. We finally discuss the description of the measurement processes as a noisy environment for the quantum system. 

Section 3 is devoted to briefly present the simplest models to describe a qubit system coupled to an environment modeled by quantum harmonic oscillators. 

Section 4 is devoted to  the energy dynamics  of a single qubit driven by time-dependent sources and coupled to a single thermal bath. This corresponds to the analysis of the entropy production and energy dissipation introduced by the driving process. A detailed description is possible within the adiabatic regime, where geometrical approaches and control procedures like shortcuts to adiabaticity have been proposed. 

Section 5 is devoted to the mechanism of energy pumping for a qubit coupled to a single thermal bath or isollated. We analyze  the adiabatic as well as the fast-driving regimes and 
discuss the geometrical and topological properties in a common framework.

Section 6 continues with the discussion of  energy pumping, but in configurations with several thermal baths, in which case heat is pumped between the baths as a consequence of driving.

Section 7 introduces the effect of thermal biases between reservoirs along with time-dependent driving. The combination of these effects is the basis for the operation of thermal machines. We consider the case of thermodynamic cycles where the evolution takes place in steps, with the qubit evolving isolated or coupled to a single reservoir, as well as the adiabatic evolution taking place with the qubit coupled to two reservoirs with a thermal bias.

Section 8 is devoted to review setups to storage energy based on qubits.

Section 9 focuses on fundamental problems of many-body physics taking place in qubits coupled to the environment and their impact on the thermal response of this device. This includes, the effects of quantum phase transitions in the spin-boson problem as well as the problem of the thermal drop and non-linear effects leading to rectification mechanisms.
Related experimental works to some of these problems have been recently reviewed in Ref. \cite{colloquium}.

Section 10 is devoted to briefly present the physical properties of superconducting devices hosting Josephson junctions, their coupling to the surrounding circuits and their representations in terms of effective two-level models in contact to harmonic oscillators. Full details have been presented in Refs. \cite{makhlin,devoret1,devoret2,wendin,blais1}

Section 11 is devoted to review experiments on quantum thermodynamics in qubits and quantum circuits. Finally, Section 12 is devoted to summary and conclusions.

\section{Energy dynamics: preliminary concepts}

Goal of this review is to analyze the energy dynamics in systems composed by one or a few interacting qubits coupled to an external environment, in non-equilibrium scenarios.

The system under consideration is generically described by the following Hamiltonian
\begin{equation}\label{gen}
H(t)=\sum_{\alpha} H_{\alpha}+H_{\rm S}[\vec{X}(t)]+H_{\rm cont}.
\end{equation}
The first term  describes the effect of the environment, which can be modeled by one or more Hamiltonian systems (labeled with $\alpha$). The second term describes the  few-level quantum system controlled by $N$ time-dependent parameters. The label $\ell$ will be used to enumerate them and it is convenient for notation purposes to collect them as components of a vector, $\vec{X}(t)=\left(X_1 (t), \ldots,X_{\ell}(t),\ldots, X_N(t) \right)$. The last term describes the contact between the driven quantum system and the environment. Several  mechanisms may take place that are associated to energy/entropy flow, like
dissipation of energy, heat transport, pumping and heat--work conversion. We will address them in the forthcoming sections. The aim of this section is to introduce some formal concepts that will help us to analyze the associated dynamics.

\subsection{Quasi-static vs finite-time regimes}\label{qs-finitet}
The natural question that arises in the discussion of energy dynamics is how to identify heat and work in a time-dependent process. This is rather easy to answer in simple terms when
the evolution is quasi-static with the system contacted to one or more reservoirs at the same temperature $T$. This situation has been discussed in  textbooks \cite{balian} as well as in previous literature \cite{janet,john,landi-pater,deffner}. The useful concept to this end is  the {\em frozen} Hamiltonian $H_t$, which corresponds to taking a snapshot of the parameters $\vec{X}(t)$ at a given time $t$ and define the Hamiltonian of Eq. (\ref{gen}) regarding the parameters as if they were  not time-dependent.
We can define a thermal density operator associated to this Hamiltonian, 
$\rho_{t} = e^{-H_t/(k_B T)}/ \mbox{Tr}\left[e^{-H_t/(k_B T)}\right]$ and use this to calculate the change in the energy stored by the system, $U_t= \mbox{Tr}\left[\rho_t H_{\rm S}\left(\vec{X}(t)\right) \right]$, as the system evolves from $t$ to $t+\delta t$. The result is
\begin{equation}\label{du}
dU= \mbox{Tr}\left[d \rho_t H_{\rm S}\left(\vec{X}(t)\right) \right] + \mbox{Tr}\left[ \rho_t 
\frac{\partial H_{\rm S}\left(\vec{X}(t)\right) }{\partial \vec{X}(t)}\right]\cdot d \vec{X}(t),
\end{equation}
and we identify the first term as heat and the second one as work. 

In time-dependent non-equilibrium processes and also in situations where the reservoirs may have different temperatures, it is more natural to define fluxes. In particular, we can formally define the power developed by the driving source characterized by $X_\ell(t)$ by first defining the force operator 
 \begin{equation}
 {\cal F}_{\ell} =-\frac{\partial H}{\partial X_{\ell}},
 \end{equation}
which results in the following expression for the power developed  on the quantum system,
\begin{eqnarray}\label{power}
P(t)&=& \langle \frac{\partial H[\vec{X}(t)] }{\partial t}\rangle = \sum_\ell P_{\ell}(t), \nonumber \\
P_{\ell}(t)&=& - \langle {\cal F}_{\ell} \rangle \dot{X}_\ell. 
\end{eqnarray}
Similarly we can calculate the rate of change of the energy stored in each piece of the device, $\langle \dot{H}_j \rangle$  with $j\equiv \alpha, {\rm sys}, {\rm cont}$. A quantity  of particular importance is the energy flux into the reservoirs,
\begin{equation}\label{jalpha}
    J_{\alpha}(t)= \langle \dot{H}_{\alpha} \rangle= -\frac{i}{\hbar} \langle \left[{H}_{\alpha}, H \right] \rangle.
\end{equation}
The expectation values in these expressions are taken with respect to non-equilibrium states. In general, Eqs. (\ref{power}) and 
(\ref{jalpha}) cannot be calculated exactly and we must resort to a non-equilibrium quantum many-body approach in order to solve the problem within some degree of approximation. An exception corresponds to the case of bilinear 
Hamiltonians (Hamiltonians that can be expressed  as combinations of products of one creation and one annihilation operator)  for the reservoirs as well as the few-level quantum system and the contact term. Under such conditions, it is possible to exactly calculate these fluxes by recourse to scattering matrix or non-equilibrium Green's functions (Schwinger-Keldysh) formalisms \cite{lan-bu1,lan-bu2,lan-bu3,lan-bu4,bilinear,moskalets-buttiker-floquet,lili-moska}. This is the case of many problems of electron systems. However, this is not the case of few-level systems described by spin models in contact to reservoirs modeled by bosonic excitations. The latter is the relevant picture in many realizacions of qubit systems.

 In a process taking place between the times $t_1$ and $t_2$ we define the work developed by the time-dependent sources  and the heat entering the reservoir $\alpha$ as
 follows,
 \begin{eqnarray}\label{wij-q}
    W_{1\rightarrow 2}&=&\int_{t_1}^{t_2} dt P(t), \nonumber \\
    Q_{\alpha,1 \rightarrow 2 }&=& \int_{t_1}^{t_2} dt J_{\alpha}(t),
 \end{eqnarray}
 with $P(t)=\sum_\ell P_\ell(t)$. While the definition of work is easily connected with the usual definition for a quasi-static process as in Eq.(\ref{du}), the definition of heat in the time-domain has been the subject of several discussions in the context of electron systems strongly coupled to fermionic reservoirs \cite{ludo1,galp,ross,ludo2,nitzan,reac,thomas,stras,riku} and, more recently, also in the context of two-level systems \cite{jacob}. In particular,  there is no full consensus on how to properly account for the energy stored in the contact terms $H_{\rm cont}$ of Eq. (\ref{gen}).
 
We now recall that the expectation values for the power in Eq. (\ref{power}) and of the energy flux in Eq. (\ref{jalpha}) are defined with respect to the non-equilibrium time-dependent state of the full many-body system, reservoirs and couplings. In order to make contact to the conventional thermodynamic description and  Eq. (\ref{du}), it is useful to split these non-equilibrium quantities into two components as follows,
\begin{eqnarray}\label{p-jalpha}
P(t) &= &P^{\rm (cons)}(t) + P^{\rm (non-cons)}(t), \nonumber \\
J_{\alpha}(t) &= & J_{\alpha}^{\rm (qs)}(t) + J_{\alpha}^{\rm (non-eq)}(t).
\end{eqnarray}
The first components correspond to the expectation values for the power and the energy flux  evaluated with the frozen state $\rho_t$ instead of  the full non-equilibrium state.
They define the conservative power  (cons)  and quasi-static (qs) energy exchange between system and reservoirs, associated to  a sequence of instantaneous equilibrium states described by  $\rho_t$,
 \begin{eqnarray}\label{cons-qs}
P^{\rm (cons)}(t)&=& - \sum_{\ell}\mbox{Tr}\left[ \rho_{t} \cal F_{\ell} \right] \dot{X}_{\ell}(t) , \nonumber \\
 J_{\alpha}^{\rm (qs)}(t)&=& - \frac{i}{\hbar} \mbox{Tr}\Big[ \rho_{t} \left[H_{\alpha}, H_t \right] \Big].
\end{eqnarray}
In finite-time processes there is an additional component of the power, $P^{\rm (non-cons)}(t)$,
corresponding to non-conservative processes, including entropy production, which is accompanied by
non-equilibrium energy exchange between the driven system and the reservoirs, $J_{\alpha}^{\rm (non-eq)}(t)$. These components are the most challenging to calculate but they describe the most interesting effects  to be addressed in the present article.

Usually, the coupling between the system and the reservoirs is weak, in which case these quantities can be evaluated by 
means of solving master equations. These have the following general structure,
\begin{equation}\label{masterrho}
\frac{d {\rho}_{\rm S}}{dt}=-\frac{i}{\hbar} \left[H_{\rm S}, {\rho}_{\rm S}\right] + \sum_{\alpha} {\cal L}_{\alpha} 
\left[{\rho}_{\rm S}\right],
\end{equation}
where ${\rho}_{\rm S}$ is the reduced density operator for the quantum system. The first term describes the unitary dynamics of the isolated system, while the "Lindbladian" ${\cal L}_{\alpha}$ depends on superoperators describing the effect of the coupling between system and  the reservoirs \cite{petru}. 
The underlying assumptions in the derivation of this equation are (i) a weak coupling between system and reservoir, justifying 
the treatment of $H_{\rm cont}$ as a perturbation and (ii) reservoir with many degrees of freedom, which can be represented by a continuum density of states with short-memory (Markovian) dynamics. The terms entering ${\cal L}_{\alpha}$ depend on rates, which are functions of the couplings, the density of states and the temperature of the reservoir.
In non-equilibrium situations with several thermal baths, this equation must be used with care. There are also discussions on the appropriate basis to be chosen. The so called global version is based on eigenstates of $H_{\rm S}$, while the local one is based on eigenstates of those operators entering $H_{\rm S}$ which also appear in $H_{\rm cont}$ \cite{chiara,archak,hewgill,hofer,rivas,rivas1,levy}. The equation of motion of the matrix elements of ${\rho}_{\rm S}$ can be derived by means of non-equilibrium Green's function formalism \cite{schon1,schon2,janine-mas,bibek-mas} in which case the natural  basis is the set of eigenstates of $H_{\rm S}$. A similar procedure  can be followed to calculate the currents $J_{\alpha}(t)$.

\subsection{Adiabatic regime: slow dynamics}
\label{sec:adia}

In quantum mechanics the notion of adiabaticity is related to the slow evolution. In the context of closed systems, it  refers to changes in the spectrum of a Hamiltonian as a function of  time-dependent parameters without level crossings and slow evolution in time without transitions between states. This implies a typical time scale for the evolution that is much longer than the internal time scales associated to inter-level transitions.
The scope of the latter definition can be extended to open quantum systems by 
taking into account that the  level life-time related to the coupling between the system and the reservoirs defines an extra internal time scale. 
It is important to stress that in the context of open driven systems adiabatic is not a synonym of quasi-static evolution. Instead, it corresponds to the first non-equilibrium correction to the quasi-static evolution, which is proportional to the rate of change (velocity) of the time-dependent parameters or, equivalently, to the driving period in the case of cyclic protocols \cite{lili-moska,janine-mas,pumping1,moska-adia}. 

In order to provide a more precise meaning of the adiabatic evolution, we  summarize in what follows how to describe this regime in the general framework of the adiabatic linear-response formalism of Refs. \cite{ludo,bibek}, assuming one or more reservoirs at the same temperature.
This procedure is similar to Kubo formalism \cite{bruus}, but implementing the perturbation with respect to the frozen Hamiltonian.
 The adiabatic evolution in time of the expectation values of any observable ${\cal O}$ is expressed as follows
\begin{equation}\label{o-adia}
O(t) = \langle {\cal O} \rangle_t 
+ \sum_{\ell=1}^N \chi_t^{\rm ad}\left[ {\cal O}, {\cal F}_\ell \right] \dot{X}_{\ell}(t),
\end{equation}
where $\langle \cdot \rangle_t$ indicates that the mean value is taken with respect to the thermal distribution $\rho_t$ corresponding to the Hamiltonian frozen at the time $t$. This contribution is similar to the so-called Born-Oppenheimer approximation and it effectively describes a quasi-static evolution where the system is in equilibrium time by time. The other terms are the non-equilibrium adiabatic corrections, which depend on the  adiabatic susceptibilities 
\begin{equation}\label{chiad}
\chi_t^{\rm ad}\left[ {\cal O}, {\cal F}_\ell \right] = -\frac{i}{\hbar} 
\int_{\-\infty}^t dt^{\prime} (t-t^{\prime}) \langle \left[ {\cal O}(t), {\cal F}_\ell(t^{\prime}) \right] \rangle_t.
\end{equation}
 In the context of closed systems, an equivalent scheme was implemented with focus on the evolution of the quantum states \cite{anatoli1,anatoli2}. 

In this framework it is simple to identify the structure of Eqs. (\ref{p-jalpha}) when the energy fluxes in the reservoirs and the forces are evaluated following the previous procedure. The first term of Eq. (\ref{o-adia}) leads to the quasi-static and conservative components
defined in Eq. (\ref{cons-qs}), while the non-equilibrium and non-conservative components read
\begin{eqnarray}\label{jp-non-eq-adia}
P^{\rm (non-cons)}_\ell(t)&=&\dot{X}_\ell(t) \sum_{\ell^{\prime}}\Lambda_{\ell,\ell^{\prime}}(\vec{X})\dot{X}_{\ell^{\prime}}(t),\nonumber \\
J_{\alpha}^{\rm (non-eq)}(t)&=&\sum_{\ell}\Lambda_{\alpha,\ell}(\vec{X})\dot{X}_{\ell}(t),
\end{eqnarray}
being $\Lambda_{\ell,\ell^{\prime}}=-\chi_t^{\rm ad}\left[ {\cal F}_{\ell}, {\cal F}_{\ell^{\prime}} \right]$
and $\Lambda_{\alpha,\ell}= -\chi_t^{\rm ad}\left[ \dot{H}_{\alpha}, {\cal F}_{\ell} \right]$.

\subsection{Adiabatic dynamics of a few-level quantum system weakly coupled to  thermal baths}
\label{sec:adia-weak}
To calculate explicitly quantities like those defined in Eq. (\ref{jp-non-eq-adia}) 
 in the weak-coupling regime and for slow driving we can derive a quantum-master-equation  following
  the procedure of Refs. \cite{janine-mas,pumping4,bibek-mas}. This  is based on expanding the matrix elements of ${\rho}_{\rm S}(t)$ into a frozen and an adiabatic component. 
 Symbolically,  
${\rho}_{\rm S}(t)={\rho}^{(f)}( t)+{\rho}^{(a)}( t)$,
where the upperscript $f$ indicates that we are considering the Hamiltonian frozen at a given time $t_f$, for which 
$\vec{X}(t_f)=\vec{X}$.
The frozen component is the solution of the quantum master equation (\ref{masterrho})
corresponding to the frozen Hamiltonian.

The detailed structure of the Lindbladian  depends on the system under study. In many cases it is proposed on phenomenological grounds.
In Refs. \cite{schon1,schon2,janine-mas,bibek-mas} a systematic derivation was proposed starting from a given Hamiltonian. This treatment  is
formulated in terms of 
Green's function
(Schwinger-Keldysh), by implementing a 
perturbation expansion in the coupling strength. It focuses on stationary situations as well as on  slow driving and applies to  thermal baths at temperatures $T_{\alpha}$. We summarize the outcome because this family of master equations is very useful to analyze the slow dynamics of qubits coupled to several baths \cite{bibek,bibek-mas,pablo}. The starting point is the Hamiltonian 
for the few-level system expressed in the {\em instantaneous basis of eigenstates}, $\left\{|s\rangle, \; j=1, \ldots L\right\}$. The  baths are represented by bosonic excitations described 
 by $H=\sum_{k_{\alpha}} \varepsilon_{k_{\alpha}} a^{\dagger}_{k_{\alpha}} a_{k_{\alpha}} $. The
couplings are expressed as follows 
\begin{eqnarray}
H_{\rm S}(\vec{X})&=&\sum_{j=1}^L \varepsilon_j(\vec{X}) |j(\vec{X}) \rangle \langle j(\vec{X}) |,  \\
H_{\rm cont,\alpha}(\vec{X}) &=&\sum_{j,l=1}^L \sum_{k_{\alpha}} V_{k_{\alpha}} \xi_{\alpha,j,l}(\vec{X}) |j(\vec{X}) \rangle \langle l(\vec{X}) |
\left(  a^{\dagger}_{k_{\alpha}} + a_{k_{\alpha}} \right) ,\nonumber
\end{eqnarray}
being $\xi^{\alpha}_{j,l}(\vec{X})$ elements of the coupling matrix expressed in the instantaneous eigenbasis. The master equation for the frozen 
component, expressed in terms of the matrix elements, reads 
 
\begin{equation}\label{frozen}
\frac{d \rho_{lj}^{(f)}}{dt} = \frac{i \epsilon_{lj}}{h}  \rho_{lj}^{(f)}+ \sum_{m,n,\alpha} \left[\left(M^{jn}_{ml,\alpha} +M^{lm}_{jn,\alpha} \right)\rho^{(f)}_{mn}-
M^{mn}_{jm,\alpha} \rho^{(f)}_{ln} -M^{mn}_{ml,\alpha} \rho^{(f)}_{nj}\right].
\end{equation}
We have introduced the definition $\epsilon_{lj}=\varepsilon_l(\vec{X})-\varepsilon_j(\vec{X})$, and the 
 rate functions 
\begin{equation}\label{rates}
\Gamma_{\alpha}(\varepsilon)=\gamma_{\alpha} n_{\alpha}(\varepsilon),\;\;\;\;\;\; \overline{\Gamma}_{\alpha}(\varepsilon)=\gamma_{\alpha} \left[1+n_{\alpha}(\varepsilon)\right].
\end{equation}
$\gamma_{\alpha}(\varepsilon)= \sum_{k_{\alpha}} V_{k_{\alpha}}/\hbar \delta (\varepsilon- \varepsilon_{k_{\alpha}})$ is the spectral function associated to the coupling to the bath
and $n_{\alpha}(\varepsilon)=1/(e^{\varepsilon/(k_B T_{\alpha})}-1)$ is the Bose-Einstein distribution function.

The adiabatic component can be calculated from
\begin{equation}\label{adia}
    \frac{d \rho_{lj}^{(f)}}{d\vec X} \cdot \dot{\vec{X}}(t) = \frac{i \epsilon_{lj}}{h}  \rho_{lj}^{(a)}+ 
  \sum_{m,n,\alpha} \left[\left(M^{jn}_{ml,\alpha} +M^{lm}_{jn,\alpha} \right)\rho^{(a)}_{mn}-
M^{mn}_{jm,\alpha} \rho^{(a)}_{ln} -M^{mn}_{ml,\alpha} \rho^{(a)}_{nj}\right].  
 \end{equation}
with
\begin{equation}
M^{jn}_{ml,\alpha}= \xi_{\alpha,ml}(\vec{X}) \xi_{\alpha,jn}(\vec{X}) \left\{ 
\Gamma_{\alpha}(\epsilon_{jn})+
\overline{\Gamma}_{\alpha}(\epsilon_{nj}) \right\}.
\end{equation}
It is important to notice that there are two contributions to the derivative of the matrix elements of the frozen density matrix in Eq. (\ref{adia}) with respect to
the parameters. One of the contributions is because of the change of the rates elements $M_{\alpha, j,l}$ as a consequence of the changes of the contacts 
$\xi_{\alpha,j,l}(\vec{X})$ and the energies $\epsilon_{jl}(\vec{X})$. The other 
contribution is because of the instantaneous states $|j(\vec{X})\rangle$. It is also interesting to highlight that the latter remain finite even when 
eliminating the effect of the coupling to the bath. In such case,
Eq. (\ref{adia}) reduces to the adiabatic evolution for a driven closed system as described by  ''adiabatic perturbation theory''
 formulated  in 
Refs. \cite{anatoli1,anatoli2}.

 Similarly, the frozen and adiabatic component of the energy current 
can be calculated from
\begin{equation}\label{j-weak}
    J_{\alpha}^{f/a}(t)= \sum_{m,n,u} \epsilon_{un}(t) \mbox{Re}\left[ M_{mn,\alpha}^{nu} \rho^{(f/a)}(t)\right].
\end{equation}


\subsection{Floquet regime: fast periodic dynamics}\label{sec:floquet}
The opposite limit to the quasi-static and adiabatic regimes, corresponds to very fast driving. In the  regime of strong driving, the notion of a reservoir with a well defined temperature with which the few-level quantum system is contacted is not necessarily useful and the mechanism of thermalization is still under debate \cite{thermalization}. 

A common situation corresponds to periodic driving with one or more commensurate frequencies and the appropriate framework to describe these problems is Floquet theory \cite{floquet} which is the time analog of Bloch theory for spatially periodic systems. This type of driving received significant attention recently for the potential to generate novel collective behavior in quantum systems, which may lead to novel states of the matter \cite{thermalization}. The realization of some of these exotic phases has been recently experimentally realized in  a quantum processor of superconducting qubits \cite{matteoi}. 

In the case of a single frequency $\omega$, the Hamiltonian satisfies $H(t+\tau)= H(t)$, with $\tau=2 \pi/\omega$. This operator can be expanded in Fourier series as
\begin{equation}\label{four}
    H(t)=\sum_k e^{i k \omega t} H_k.
\end{equation}
The Floquet eigenstates  have a structure consistent with this periodicity,
\begin{equation}\label{state}
    |\psi(t)\rangle = e^{i \varepsilon/\hbar t} \sum_{n=-\infty}^{+\infty} e^{-i n \omega t}|\psi_m\rangle.
\end{equation}
 Hence, when substituted in the Schr\"odinger equation and using Eq. (\ref{four}) we find
 \begin{equation}\label{floquet}
     \left(\varepsilon + \hbar \omega n\right) |\psi_n\rangle = \sum_{m=-\infty}^{+\infty}H_{m}|\psi_{n+m}\rangle,
 \end{equation}
 which defines a problem with a tight-binding structure in the synthetic Floquet lattice. The structure of this Eq. also reveals the exchange of "Floquet quanta"   $\hbar \omega$ underlying this dynamics and effectively provides and environment for the driven system.
 
 This formulation can be generalized for the case of $M$ commensurate frequencies, in which case it is convenient to define
 $\vec{\omega}=\left(\omega_1, \ldots, \omega_M\right)$ and also collect the corresponding Floquet indices in a vector $\vec{n}$. Eq. (\ref{floquet}) is generalized to
 \begin{equation}\label{ham-floq}
    \left(\varepsilon + \hbar \vec{\omega}\cdot \vec{ n}\right) |\psi_{\vec{n}}\rangle = \sum_{\vec{m}}H_{\vec{m}}|\psi_{\vec{n}+\vec{m}}\rangle.
 \end{equation}
In this scenario, very interesting ideas have been formulated on the transport or exchange of power between the driving sources,
some of them will be reviewed  in Sec. \ref{topo-floquet}. The relevant quantity to analyze is the power defined in Eq. (\ref{power}). For the case where $\vec{X}(t)=\vec{\omega}t + \vec{\varphi}$, where $\vec{\varphi}$ encloses
$M$ independent phases, we get the following expression for the power developed by the $\ell$-th force,
\begin{equation}
    P_{\ell}(t)=\omega_{\ell} \langle {\cal F}_{\ell} \rangle.
\end{equation}
The expectation value is calculated  with respect to the non-equilibrium state $|\psi(t)\rangle$.

\subsection{Thermal bias: non-equilibrium stationary regime}\label{sec:ther-bias-0}
When the system is not driven ($\vec{X}$ is constant in time) but it is contacted to several reservoirs at different temperatures,  a heat
current is established following the thermal bias through the device. Under these conditions the relevant quantity to analyze is the
steady-state heat flux at each reservoir, $J_{\alpha}$.

The simplest configuration to discuss the mechanism of thermal transport corresponds to a few-level system directly connected to two reservoirs at different temperatures, $T_{\rm h} > T_{\rm c}$. The natural process in this case is a stationary heat flux from the hot to the cold reservoir through the quantum system. On general grounds, we should expect that the energy fluxes into the two reservoirs are amenable to be expressed as power series of  the thermal bias
$\Delta T= T_{\rm h} - T_{\rm c}$,
\begin{equation}\label{jest}
J_{\alpha}= \sum_{n=1}^{\infty} \kappa_{\alpha,n} \left(\Delta T\right)^n, \;\;\;\;\;\;\;\alpha={\rm c, h}.
\end{equation}
The concomitant rate of entropy production at the reservoirs reads
\begin{equation}\label{sdot}
\dot{S}= \sum_{\alpha={\rm c,h}} \frac{J_{\alpha}}{T_{\alpha}},
\end{equation}
and can be also expressed in terms of a Taylor series in $\Delta T$. In the limit of small thermal bias, where only the linear component contributes, energy conservation is assumed. Hence
$J_{\rm c} =-J_{\rm h}=J$, which implies $ \kappa_{\rm h,1}=- \kappa_{\rm c,1}=G_{\rm th}$. The latter parameter is the thermal conductance. When the linear-response contribution is substituted in Eq. (\ref{sdot}) we get $\dot{S}= G_{\rm th} \left( \Delta T\right)^2/T_{\rm c}$. This simple heuristic observation allows us to conclude that  energy dissipation leading to entropy production is a non-linear process, which is consistent with the assumption that energy is conserved if we restrict ourselves to the linear-order contributions. 

The linear regime is properly accounted for the Landauer-B\"uttiker formula, which can be derived exactly  for bilinear 
Hamiltonians \cite{lan-bu1,lan-bu2,lan-bu3,lan-bu4,bilinear}, 
\begin{equation}\label{lan-bu}
J_{\rm c}= \int_{-\infty}^{+\infty} d \varepsilon \; \varepsilon \; {\cal T}(\varepsilon) \left[ n_{\rm h}(\varepsilon)- n_{\rm c}(\varepsilon)\right]= -J_{\rm h}=J, \;\;\;\;\;\;\; {\rm exact\; for \; bilinear~hamiltonians},
\end{equation}
 ${\cal T}(\varepsilon)$  is the so called transmission function which characterizes the transparency of the device to transmit an  amount $\varepsilon$ of energy across it, while
$n_{\alpha}(\varepsilon)= 1/(e^{\beta_{\alpha} \varepsilon}-1)$ is the Bose-Einstein distribution function (assuming that the bath is described as a gas of non-interacting bosonic excitations), depending on the temperature of the reservoir through $\beta_{\alpha}=1/(k_B T_{\alpha})$.
The transmission function depends on the microscopic details of the setup, including the spectral properties of the reservoirs, the central system and the couplings. 
In some cases, the same structure of Eq. (\ref{lan-bu}) is obtained for systems with many-body interactions within linear response, but with a temperature-dependent transmission function ${\cal T}(\varepsilon, T)$. We shall analyze some examples in
Sec. \ref{sec:lin}.

Beyond linear response, and for non-bilinear Hamiltonians, it is necessary to resort to other non-equilibrium many-body techniques to independently calculate the currents at each reservoir.
Notice that many-body interactions are expected to introduce inelastic scattering processes that generate dissipation. That type of effects are expected to generate dissipative components in the fluxes into the reservoirs.
For weak coupling between system and reservoir, quantum master equations is the most popular framework. Following a similar derivation as that
leading to Eq. (\ref{masterrho}), the current at each reservoir can be written as the frozen component of Eq. (\ref{j-weak}).
The time dependence in this case is not introduced by the Hamiltonian, but by the fact that $\hat{\rho}^{f}(t)$ is the solution of the  time-dependent equation (\ref{frozen}). The relevant regime is the long-time solution 
\begin{equation}
    J_{\alpha}=\lim_{t\rightarrow \infty}  J_{\alpha}(t).
\end{equation}
The underlying assumption in this type of calculation is that the  reservoirs have a well defined temperature. For situations where the temperature bias is large, many-body interactions may contribute to build-up an effective temperature profile along the quantum system if it has a spatial distribution.  We shall discuss this aspect in  Sec. \ref{sec:statio}.

\subsection{Slow dynamics in combination with small thermal bias. Thermal geometric tensor and Onsager relations}\label{sec:slow-ther}
The simultaneous effect of a thermal bias and driving is of great interest in the study  of thermal machines operating in contact to thermal baths at different temperatures. In the case of slow dynamics and for a the case of two reservoirs with a small
thermal bias $\Delta T$ -- such that $\Delta T/T$ is a small parameter -- the  adiabatic  formalism introduced in Sec. \ref{sec:adia} can be adapted to include $\Delta T/T$ as an additional entry in an extended vector $\dot{\bf X} =(\dot{\vec{X}},\Delta T/T)$. The extra entry associated to the temperature bias is labeled by $N+1$. This procedure can be naturally implemented in a very general way starting from the Hamiltonian representation of the thermal bias introduced by Luttinger \cite{lut,tatara} and following 
similar steps as in the adiabatic Kubo-like derivation presented in Sec. \ref{sec:adia}. Details can be found  in Ref. \cite{bibek}. Here, we outline the main result, according to which the expressions for the non-conservative power and the non-equilibrium heat fluxes presented in
Eqs. (\ref{jp-non-eq-adia}) are extended to
\begin{eqnarray}\label{jp-non-eq-adia-dt}
P^{\rm (non-cons)}_\ell(t)&=&\dot{X}_\ell(t) \sum_{\ell^{\prime}}\Lambda_{\ell,\ell^{\prime}}(\vec{X})\dot{X}_{\ell^{\prime}}(t)+\dot{X}_\ell(t)\Lambda_{\ell,N+1}(\vec{X})\frac{\Delta T}{T},\nonumber \\
J_{\rm c}^{\rm (non-eq)}(t)&=&\sum_{\ell}\Lambda_{N+1,\ell}(\vec{X})\dot{X}_{\ell}(t)+  \Lambda_{N+1,N+1}(\vec{X})\frac{\Delta T}{T}.
\end{eqnarray}
The notation highlights the fact that the linear-response coefficients define a $(N+1)\times(N+1)$ matrix which is named the {\em thermal geometric tensor}. The geometrical nature
is because of the dependence of all the entries on $\vec{X}$.  The last term of the second equation 
 is proportional to the thermal conductance of the system.

In the case of cycles, and focusing on quantities averaged over the period $\tau$, the same argument of the previous section, regarding energy conservation of power-counting in $\dot{\bf X}$ leads us to conclude that
\begin{equation}
\int_0^{\tau} dt J_{\rm c}^{\rm (non-eq)}(t)=-\int_0^{\tau} dt J_{\rm h}^{\rm (non-eq)}(t)=Q_{\rm c}. \label{qctr}
\end{equation}
Importantly,  the linear response coefficients can  be shown to  satisfy Onsager relations  \cite{ludo,bibek} so that
\begin{equation}\label{onsager}
\Lambda_{N+1,\ell}(\vec{X})=-s_{\ell} \Lambda_{\ell,N+1}(\vec{X}),\;\;\;\;\;\;\;\Lambda_{\ell,\ell^{\prime}}(\vec{X})=s_\ell s_{\ell^{\prime}} \Lambda_{\ell^{\prime},\ell}(\vec{X}).
\end{equation}
The sign $s_{\ell}=\pm$ depends on whether the operators ${\cal F}_{\ell}$ are even/odd under the transformation $t \rightarrow -t$. 

Here, we see that the thermal geometric tensor
has symmetric and antisymmetric components.  The entropy generation is associated to the symmetric component. In fact, the entropy production is associated to the total dissipated work. This contains a component due to the work done by the non-conservative driving forces and another component which accounts for the thermal bias --the latter is the usual contribution taken into account in thermoelectricity (see Ref. \cite{benenti}) -- and the result is
\begin{eqnarray}\label{wdissdt}
W^{\rm (diss)} &= &\int_0^{\tau} dt \left\{\sum_{\ell=1}^N P^{\rm (non-cons)}_\ell(t)+J_{\rm c}^{\rm (non-eq)}(t) \frac{\Delta T}{T} \right\}\nonumber \\
& = & \int_0^{\tau} dt \dot{\vec{X}} \cdot \underline{\Lambda}^{S}(\vec{X})  \cdot \dot{\vec{X}}+
\Lambda_{N+1,N+1} \left(\frac{\Delta T}{T}\right)^2,
\end{eqnarray}
where we have introduced the notation $\underline{\Lambda}^{S}(\vec{X})$ for the matrix containing the symmetric component of $\Lambda_{\ell^{\prime},\ell}(\vec{X})$.

All these results are valid for any system and for any type of coupling to the thermal baths. In the case of weak coupling, the explicit calculation of the coefficients defining the thermal geometric tensor can be accomplished by solving Eqs. (\ref{frozen}), (\ref{adia}) and (\ref{j-weak}) with a small temperature difference $\Delta T$ and performing a linear expansion in this quantity.

\subsection{From Lindblad equation to quantum trajectories}\label{sec:meas}
The description based on quantum master equations, briefly introduced in Sec. \ref{qs-finitet}, resembles the classical Langevin dynamics of the Brownian particle embedded in the bath. Stochastic thermodynamics is the field where this theoretical description is elaborated in classical 
\cite{stoch-thermo,esposito,vdbroek} and quantum-mechanical \cite{parrondo,talkner,strasberg-book} contexts. Recently, there is a surge of interest in analyzing the effect of quantum measurements in the evolution of qubit systems. Measurements are key elements in quantum mechanics in general and are fundamental processes in the operation of quantum computing devices. In principle, a measurement protocol could be represented by a time-dependent term in the Hamiltonian
of Eq. (\ref{gen}). However, if the consequent outcome is fast enough, it is appropriate to represent this effect as a stochastic perturbation. This point of view
was adopted some time ago in the theory of continuous measurements and quantum trajectories \cite{wiseman,brun}. Nowadays, it  
motivates the study of phase transitions induced by the effect of measurements in arrays of qubit systems, which is an active avenue of research \cite{phase-trans1,phase-trans2,phase-trans3}. 

In the context of quantum thermodynamics, the stochastic effect of 
measurements formally plays an analogous role as the thermal bath but of a quantum nature \cite{quantum-meas,elouard1,jacobs,romito,jordan,manzano}. The description of the stochastic dynamics has been combined with Lindblad master equation  for a single qubit weakly coupled to a single reservoir and we summarize below the main ideas. Lindblad equation is formally expressed as Eq. (\ref{masterrho}) with 
\begin{equation}\label{lind-meas}
{\cal L}\left[{\rho}_{\rm S}\right]=\sum_{k=1}^{D^2-1} \Gamma_k \left(L_k \hat{\rho}_{\rm S} L_k^{\dagger} -\frac{1}{2}\left\{{\rho}_{\rm S}, L_k^{\dagger}L_k \right\}\right),
\end{equation}
being $D$ the dimension of the Hilbert space of the system and $\{.,.\}$ denoting anticonmmutation. The quantities $\Gamma_k$ are rates describing the coupling with the reservoir and $L_k$ are "jump" operators, describing the changes 
between the different quantum states of the system.
The effect of the measurement is introduced by considering
  a stochastic unravelling of this equation. This can be implemented by taking the statistical average over a completely positive trace-preserving map that inverts Eq. (\ref{masterrho}). Such a procedure is achieved  by introducing Kraus operators as follows \cite{quantum-meas}
\begin{equation}\label{rhos}
{\rho}_{\rm S}(t+dt) = \sum_{\cal K=1}^{D} M_{\cal K}(t) {\rho}_{\rm S}(t)M^{\dagger}_{\cal K}(t),
\end{equation}
where  ${\cal K}$ denotes the labeling of the eigenbasis defining the measuring apparatus, while
\begin{eqnarray}\label{sol}
    M_{\cal K}(t) &=& \left[1-\frac{i}{\hbar} dt H_{\rm eff}(t) + \sum_{k=1}^{D^2-1} \sqrt{\Gamma_k} dw_k^{\cal K}(t) L_k\right] 
    \times  \sqrt{\prod_k p(dw^{\cal K}_k(t))}, \nonumber\\
    H_{\rm eff}(t) &=& H_{\rm S}(t) -\frac{i}{2}\sum_{k=1}^{D^2-1} \Gamma_k L_k^{\dagger} L_k.
\end{eqnarray}
Here $dw_k^{\cal K}(t)$ is an stochastic increment that satisfies
\begin{equation}
    \langle dw_k^{\cal K}(t)\rangle_{\gamma} =0, \;\;\;\;\;\;\;\; \langle \langle dw_k^{\cal K}(t) dw_l^{\cal K}(t)\rangle_{\gamma} =dt \delta (t-t^{\prime})\delta_{k,l},
\end{equation}
and $p(dw^{\cal K}_k(t))$ is the corresponding probability. 
Each realization of the stochastic increment defines a quantum trajectory. Hence,
the notation $\langle ...\rangle_{\gamma}$ indicates average over quantum trajectories. 

Other   procedures following similar ideas have been recently proposed for the analysis of measurements in non-equilibrium situations \cite{strasberg1,bibek-andrew}. Quantum fluctuations play a relevant role in this context \cite{fluct1,fluct2,fluct3,fluct4,ceri1,ceri2} and several works rely on the formalism based on quantum jumps and the trajectory description of the evolution of the system to analyze this effect \cite{qf1,qf2,qf3,qf4,qf5}.
We shall not address this topic in the present review and we defer the reader to other review articles where it has been covered \cite{janet,john,landi-pater,deffner,talkner,manzano}. 

As we shall discuss in Section \ref{sec:meas}, in quasi-static processes
it is usual to define  heat and work following the same reasoning as in the case  described in Eq. (\ref{du})
upon calculating the traces with the operator ${\rho}_{\rm S}(t)$ after solving Eq. (\ref{rhos}) and taking the average over all the quantum trajectories.

\section{Models for a qubit and the environment of harmonic oscillators}\label{sec:spin-boson}
A qubit is a two-level system amenable to be operated in order to prepare quantum superpositions of the basis states. The paradigmatic
Hamiltonian to describe such a system is,
\begin{equation}\label{hqubit0}
    H_{\rm qubit}= -B_z \sigma^z - B_x \sigma^x,
\end{equation}
where $\sigma^x,\; \sigma^z$ are the Pauli matrices. We are considering the basis  of eigenstates of the first term of Eq. (\ref{hqubit0}), $\{ |\uparrow \rangle, \;  |\downarrow \rangle \}$. The effect of the second term is to generate the mixing of these two states.
This Hamiltonian can be realized in a wide variety of platforms, including atomic systems, semiconductors, NV centers, and superconductors \cite{platforms}. In Sec. \ref{sec:super} we shall briefly explain its realization in superconducting devices. 

We focus on environments which can be represented by sets of quantum harmonic oscillators. This model was introduced by Caldeira and Leggett in Refs. \cite{cal-leg,
caldeira-legget1,caldeira-legget2} and naturally describes a reservoir of photons or phonons. As we shall discuss in Sec. \ref{sec:super}, microwave resonators and transmission lines in cQED are also represented by this type of model. The corresponding Hamiltonian reads
\begin{equation}\label{hres}
    H_{\rm res}= \sum_k \hbar \omega_k a^{\dagger}_k a_k,
\end{equation}
where $a^{\dagger}_k / a_k$ creates/destroys a bosonic mode with frequency $\omega_k$. The number of relevant modes entering this Hamiltonian depends on the problem under study. We shall analyze in the forthcoming sections many situations
where it contains an infinite number of modes, in which case this system behaves as a reservoir or a thermal bath.

A natural and simple coupling between the two systems is 
\begin{equation}\label{qubit-res}
H_{\rm qubit-res}= \sum_k \vec{g}_k \cdot \vec{\sigma} \; \left(a^{\dagger}_k+a_k\right),
\end{equation}
with $\vec{\sigma}=\left( \sigma^x, \sigma^y, \sigma^z \right)$. The coupling to the reservoir depends on the state of the qubit. For instance, in the case of $\vec{g}_k=(g_{k,x},0,0,0)$, the qubit and the bath are coupled when the state of the qubit has a projection along the $x$-direction in the Bloch sphere. 
In Sec. \ref{sec:super} we shall show that this type of coupling is naturally derived in common architectures of superconducting qubits 
coupled to 
transmission lines. 
In the theoretical description it is sometimes convenient  to consider the coupling of Jaynes-Cummings model \cite{jaynes-cummings} of quantum optics,
which reads
\begin{equation}\label{j-c}
H_{\rm J-C}= \sum_k g_k  \; \left(\sigma^+ a_k+ \sigma^- a^{\dagger}_k \right),
\end{equation}
with $\sigma^{\pm}=\sigma^x \pm i \sigma^y$. This coupling is  interpreted as a transition from the ground state to the excited state of the two-level system by absorbing a photon and the opposite process, where a photon is emitted as the state of the system changes from the excited to the ground state. 

The Hamiltonian for the two-level system coupled to a bath of harmonic oscillators defines the celebrated spin-boson model \cite{weiss,caldeira-legget2}.  In the theoretical description it is useful to introduce  an
 hybridization function characterizing the coupling 
of the two-level system with the environment,  
\begin{equation}
    \Gamma(\varepsilon)= 2 \pi \sum_k |g_k|^2 \delta(\varepsilon - \hbar \omega_k).
\end{equation}
For an environment with an infinite number of oscillator modes, this is a continuous function and it is usual to model it by a power-law
$\Gamma(\varepsilon) \propto \varepsilon^{s}$. The case with $s=1$ is named ohmic environment, while the
cases with $s >1, <1$ are named, respectively, super-ohmic and sub-ohmic.
This model has a quantum phase transition at zero temperature depending on the strength of coupling between the system and the reservoir and the spectral properties of the latter (ohmic, subohmic or superohmic) \cite{bm,chack}.  
The phase transition from a state that is a combination of the two qubit states to one where the ground state is localized at one of them
takes place above a critical
coupling. The type of transition strongly depends on the nature of the bath. In the 
sub-ohmic case, the transition is of  second-order  \cite{bulla,vojta1,winter,vojta2,vojta3,chin}, while the ohmic case shows a 
Kosterlitz-Thouless transition \cite{caldeira-legget2,anderson,kosterlitz}. The super-ohmic case does not have a phase transition but exhibits a crossover. Another interesting feature of the ohmic environment is the associated Kondo effect \cite{hewson} at sufficiently low temperatures \cite{guinea1,guinea2}. 
In most of the real  situations, mainly in cQED, the degree of coupling between the qubit and the environment is weak, which prevents  the experimental analysis of this phase transition and justifies its theoretical study by means of Lindblad-type quantum master equations. Quite recently, however, strong-coupling configurations were realized between superconducting circuits and the electromagnetic environment \cite{strong1,strong2,strong3,strong4,strong5}, which enabled the realization of the strongly-coupled spin-boson model in this platform.

In the next sections we shall discuss different mechanisms of energy dynamics with focus on a single qubit. In spite of the simplicity of this system, when the effect of time-dependent driving is included and different settings with one or more reservoirs are considered, a rich variety of phenomena may take place. Our goal is to analyze several of them in  detail. Concretely, we shall consider
the case of the qubit coupled to a single reservoir at finite temperature under time-dependent driving, in which case we identify two interesting mechanism taking place: (i) energy dissipation because of the driving (ii) power exchange between driving sources. Then, we shall turn to analyze configurations where the qubit is coupled to two reservoirs at the same temperature and under time-dependent driving, where the interesting mechanism is the possibility of pumping energy between the two reservoirs as a consequence of the driving. Next, we shall analyze the configurations where the driven qubit is coupled to two reservoirs at different temperatures. This enables the realization of thermal machines: heat-engines and refrigerators. We shall also discuss the realization of batteries in configurations of qubits coupled to reservoirs and isolated. In the case of arrays with two reservoirs at different temperatures, we shall also  analyze the thermal transport in the steady-state regime.

\section{Driving and energy dissipation}\label{sec:en-diss}
\subsection{General considerations}\label{sec:en-diss-gen}
\begin{figure}
    \centering
    \includegraphics[width=0.5\textwidth]{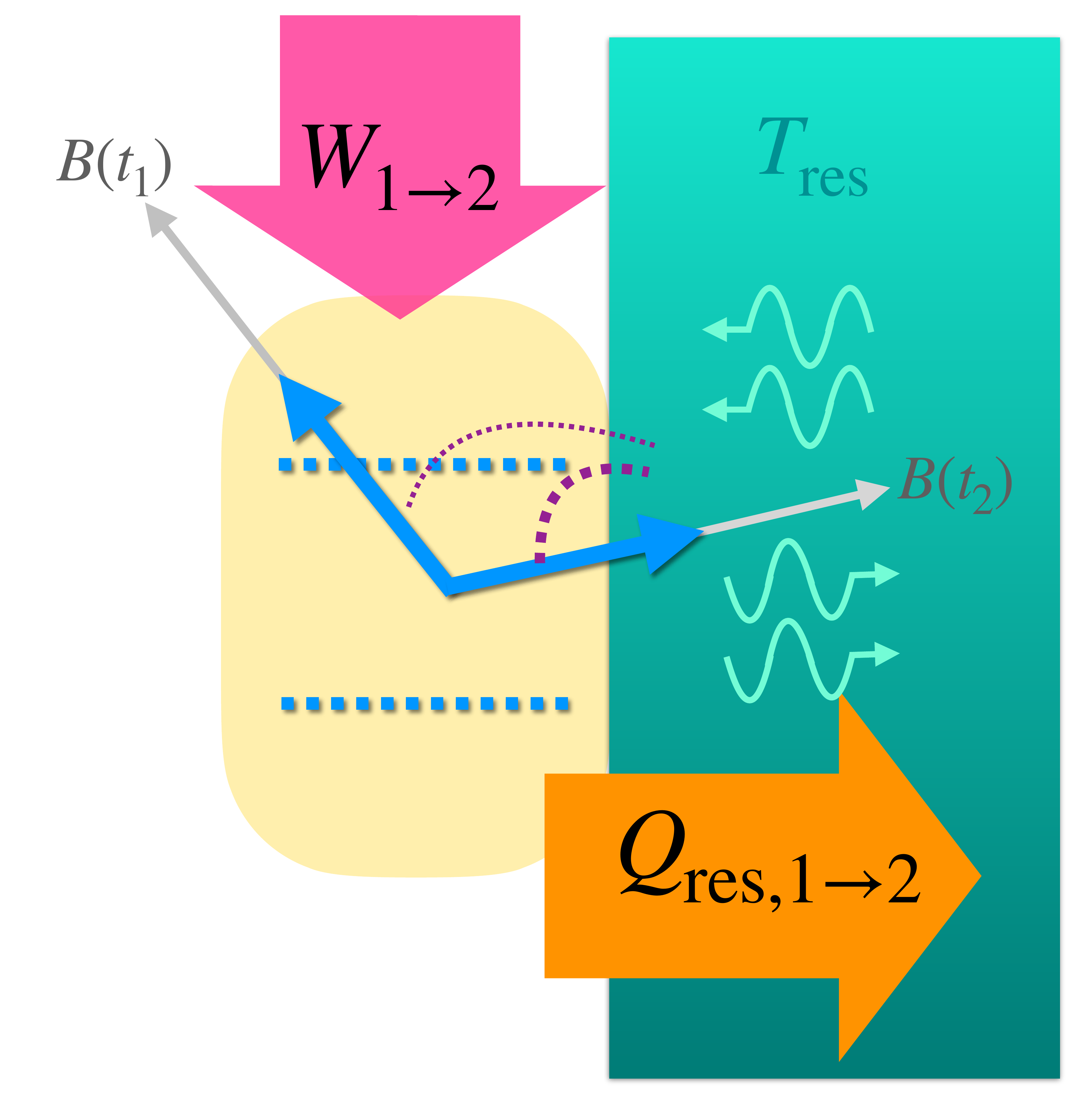}
    \caption{  Illustration of the consequent energy dynamics  of the 
    qubit under a time-dependent operation in contact to a thermal bath. The qubit is represented by two-levels and the vector $\vec{B}$ (grey arrow). The corresponding quantum state in Bloch coordinates is represented by the blue arrow.  The energy dynamics consists in the exchange of work with the
    time-dependent sources  and the exchange of heat between the qubit and the reservoir as described  by Eqs. (\ref{wij-q}) and
    (\ref{p-jalpha}). The net effect in a finite-time process is the transformation of the work invested by the external sources into heat that is dissipated into the reservoir.}
    \label{fig:diss}
\end{figure}
One of the simplest albeit non-trivial configurations is  a single qubit under time-dependent driving coupled to a single bath with a well defined temperature $T$. 
 In terms of the two-level Hamiltonian, these effects can be represented by the following Hamiltonian
\begin{equation}\label{driv-qubit}
H_{\rm qubit}(t) = -\vec{B}\left[\vec{X}(t)\right] \cdot \vec{\sigma},
\end{equation}
with $\vec{B}\left[\vec{X}(t)\right]=\left(B_x\left[\vec{X}(t)\right],B_y\left[\vec{X}(t)\right],B_z\left[\vec{X}(t)\right] \right)$ depending on time through the protocols $\vec{X}(t)$,
while $\vec{\sigma}=\left(\sigma^x,\sigma^y,\sigma^z\right)$ is composed by the three Pauli matrices operating in the qubit Hilbert space. It is 
natural to assume the coupling between the driven qubit and the bosonic reservoir described by Eq. (\ref{qubit-res}).

The full system is described by the Hamiltonian
\begin{equation}\label{full}
H(t)=H_{\rm qubit}\left(t\right)+H_{\rm qubit-res}+H_{\rm res},
\end{equation}
where $H_{\rm res}=\sum_k \hbar \omega_k a^{\dagger}_k a_k$ is the Hamiltonian of the Caldeira-Leggett bath of harmonic-oscillator modes as in  Eq. (\ref{modes}).

The device and  represented by the Hamiltonian of Eq. (\ref{full}) is illustrated in Fig. \ref{fig:diss}. 
As a consequence of the driving represented by $\vec{B}(t)$ 
the qubit state changes in time. In general, this also affects the degree of coupling between the qubit and the reservoir.
In this process, energy is 
 exchanged between the driving sources and the qubit-reservoir system. If it takes place at a finite-time this corresponds to a non-equilibrium evolution 
 of the combined qubit-reservoir system and the net result is the dissipation of 
the supplied energy  in the form of heat into the reservoir. The rest of this section is devoted to analyze this mechanism.

\subsection{Adiabatic regime and thermodynamic length}\label{sec:adia-thermo-length}
We analyze here the mechanism of dissipation introduced by the time-dependent driving in the adiabatic regime. We recall that the main general ideas of this approach were presented in Sec. \ref{sec:adia}. 

The dissipation is accounted for  the non-conservative component of the developed power, Eq. (\ref{jp-non-eq-adia}),  $P^{\rm(diss)}\equiv P^{\rm (non-cons)}$. We  conclude that  the net  entropy production as the system is driven between the times $t_1$ and $t_2$ can be expressed as
 \begin{equation} \label{ts}
 T{\Sigma}=W^{\rm (diss)}=\sum_{\ell,\ell^{\prime}}\int_{t_1}^{t_2} dt  \dot{X}_{\ell}(t) \Lambda^{S}_{\ell,\ell^{\prime}}(\vec{X}) \dot{X}_{\ell^{\prime}}(t),
 \end{equation}
 being $T$ the temperature of the reservoir, $\Sigma$ the total entropy production and $W^{\rm (diss)}$ the net dissipated work.

At this point we can introduce the ideas associated to the
geometric description of the energy dissipation and entropy production of slowly driven systems, which recently became a topic of interest in finite-time thermodynamics.
 A fundamental  concept behind this description is the {\em thermodynamic length} introduced in Refs. \cite{wein,rup} and further elaborated in Refs.  \cite{sal,sal1,nul,schol,diosi,crooks,cam-den-han,pat,pat1,sivak}.

  The key property is the fact that    $\Lambda^{S}_{\ell,\ell^{\prime}}(\vec{X}(t))$ is positive definite because of Eq. (\ref{ts}) and the fact that the second law of the thermodynamics implies $\Sigma \geq 0$. Consequently, this tensor has the necessary properties to be the metric of a Riemannian space. 
 In a Riemannian metric it is possible to define the distance along curves connecting different points. The
length of a curve parameterized by $t$, from $t_1$ to $t_2$ is
\begin{equation}\label{l}
{\cal L} =\int_{t_1}^{t_2} dt \sqrt{ \dot{\vec{X}}(t) \cdot \underline{\Lambda}^{S}(\vec{X}) \cdot \dot{\vec{X}}(t)},
\end{equation}
where we denoted by $\underline{\Lambda}(\vec{X}) $ the matrix with elements
$\Lambda_{\ell,\ell^{\prime}}(\vec{X})$. The curves of  minimal distance are called geodesics. Using Cauchy-Schwartz inequality, 
$\int_{t_0}^{t_1} dt f^2 \int_{t_1}^{t_2} dt g^2 \geq \left[ \int_{t_1}^{t_2} dt fg dt \right]^2$, with $g=1$ and $f$ being the argument of the integral 
of Eq. (\ref{l}) leads to the following relation 
\begin{equation}\label{bound}
T \Sigma \geq \frac{{\cal L}^2}{\delta t},
\end{equation}
with $\delta t=t_2-t_1$.
The dissipated power $T \Sigma$  is a non-geometric quantity because it depends on the way in which the path in the parameter space is circulated.
Nevertheless Eq. (\ref{bound}) tells us that it is lower-bounded by a purely geometric quantity which depends on the path, the {\em thermodynamic length }
 ${\cal L}$. The lower bound of Eq. (\ref{bound}) is saturated when the integrand is constant along the path. This corresponds to
 circulating at a velocity that satisfies a constant dissipation rate at each point of the trajectory. Among all the possible protocols, there exist
  one path --the geodesic--
 for which ${\cal L}$ and therefore the dissipation is minimal
 \cite{sal,sal1,crooks}. This is a remarkable property, which is very useful in the design of optimal finite-time protocols.

In the limit of weak coupling to the reservoir,  this description can be combined with a Lindblad master equation and the adiabatic expansion explained in Sec. \ref{sec:adia}. Here we summarize the main steps.  The starting point is the density matrix expressed in the instantaneous frame of eigenstates of the qubit Hamiltonian. This is achieved by transforming the Hamiltonian of Eq. (\ref{driv-qubit}) with a unitary transformation $U(\vec{X})$ as follows
\begin{equation}\label{hqf}
 H_{\rm qubit}(\vec{X})= |\vec{B}(\vec{X})| U(\vec{X}) \sigma^z U^{-1}(\vec{X}).
 \end{equation}
In the case of a qubit, 
a very convenient representation relies on
introducing vectors with the matrix elements $\rho^{(f/a)}$ in Eqs. (\ref{frozen}) and (\ref{adia}), which are defined from
 the decomposition of the frozen density operator in Pauli matrices as follows
 \begin{equation}\label{rhof-a}
 \rho^{(f)}=\frac{\left(\sigma^0 + \vec{\rho}^{(f)}\cdot \vec{\sigma}\right)}{2},\;\;\;\;\;\;\;\;\;\rho^{(a)}=\frac{ \vec{\rho}^{(a)}\cdot \vec{\sigma}}{2}
 \end{equation}
  with $\vec{\rho}^{(f/a)}=\left(\rho_x^{(f/a)},\rho_y^{(f/a)},\rho_z^{(f/a)}\right)$, and 
  \begin{equation}
  \rho_{x}^{(f/a)}=\rho^{(f/a)}_{12}+ \rho^{(f/a)}_{21}, \;\;\;\;\;\;\;\
 \rho_{y}^{(f/a)}=i(\rho^{(f/a)}_{12}-\rho^{(f/a)}_{21}),\;\;\;\;\;\;\;\
 \rho_z^{(f/a)}=\rho^{(f/a)}_{11}-\rho^{(f/a)}_{22}.
 \end{equation}
 In this notation, the action of  the  Lindbladian operator in the master equation for
 $\rho$
  define a matrix
 ${\cal M}(\vec{X})$ and a vector $\vec{\gamma}$ in terms of which the stationary solution of Eq. (\ref{frozen}) and Eq. (\ref{adia}) can be written as
 follows
 \begin{equation}\label{mgam}
 {\cal M} (\vec{X}) \vec{\rho}^{(f)}(\vec{X})=\vec{\gamma}^{(f)}(\vec{X}), \;\;\;\;\;\;\;\;\;\;\;\;\;\;\;\;\;\;\;\;\;\;  \sum_{\ell} \frac{\partial \vec{\rho}^{(f)}(\vec{X})}{\partial X_{\ell}} \dot{X}_{\ell} =  {\cal M} (\vec{X}) \vec{\rho}^{(a)}(\vec{X}).
 \end{equation}
The solution is 
 \begin{equation}\label{rho}
\vec{\rho}^{(a)}(\vec{X}) =\sum_{\ell}{\cal M}^{-1} (\vec{X}) \frac{d \vec{\rho}^{(f)}(\vec{X})}{d X_{\ell}} \dot{X}_{\ell}.
 \end{equation}
 As mentioned in Sec. \ref{sec:adia}, it is important to consider in $d \vec{\rho}^{(f)}(\vec{X})/d X_{\ell}$, not only the changes because of ${\cal M}(\vec{X})$, but also
 the change of basis introduced by $U(\vec{X})$ in Eq. (\ref{hqf}).
 The dissipated energy because of  the driving mechanism in a time interval between $t_1$ and $t_2$ reads
 \begin{equation}\label{wdiss-qub}
 W^{\rm (diss)}= \sum_{\ell} \int_{t_1}^{t_2} dt \mbox{Tr} \left[\rho^{(a)} \sum_{\ell} \frac{\partial H_{\rm qubit}(\vec{X})}{\partial X_\ell} \right] \dot{X}_{\ell}
 \equiv \sum_{\ell} \int_{t_1}^{t_2} dt \mbox{Tr} \left[\rho^{(a)} \sum_{\ell} {\cal F}_{\ell}(\vec{X}) \right] \dot{X}_{\ell}.
 \end{equation}
 where $H_{\rm qubit}(\vec{X})$ denotes the frozen Hamiltonian of the driven qubit. Introducing the representation
 \begin{equation}\label{forcep}
 {\cal F}_{\ell}(\vec{X}) = \vec{f}_\ell(\vec{X}) \cdot \vec{\sigma},
 \end{equation}
 and substituting Eq. (\ref{rho}) in Eq. (\ref{wdiss-qub}) leads  to the expression of Eq. (\ref{ts}) with
 \begin{equation}\label{lambdas}
 \Lambda^{S}_{\ell^{\prime},\ell} (\vec{X})=  {\cal M}^{-1} (\vec{X}) \frac{d \vec{\rho}^{(f)}(\vec{X})}{d X_{\ell^{\prime}}} \cdot \vec{f}_\ell(\vec{X}).
 \end{equation}
  
 These ideas became recently very useful 
 in the study of the slow evolution of driven qubits.
 They were the guide in the search for optimal protocols to minimize the dissipation in a single driven qubit  weakly contacted to a reservoir modeled by the Hamiltonian of Eq. (\ref{trans-res}) and also for a set of coupled qubits \cite{scan,deffner-bonan}. 
 Notice that while the bound of Eq. (\ref{bound}) is known to exist, the explicit form  of the geodesic  implies solving a non-trivial differential equation. In Ref. \cite{scan} the derivation leading to Eq. (\ref{rho}) was formulated in the language of Lindbladian operators and  the Drazin inverse. Analytical 
 expressions were derived for a  single qubit a bath of harmonic oscillators modeled by 
 Eq. (\ref{full}) with $\vec{X}(t)=\left(B(t), \theta(t), \varphi(t) \right)$ and
 \begin{equation}\label{proto-scandi}
 \vec{B}\left(\vec{X}\right) = B \vec{n}, \;\;\;\;\;\; \vec{n}=\left( \cos\varphi \sin \theta,   \sin\varphi \sin \theta, \cos \theta \right),
 \end{equation}
for which  the unitary transformation entering Eq. (\ref{hqf})  reads
 \begin{equation}
 U(\vec{X})=\left( \begin{array}{cc}
 \cos(\theta/2 ) & -e^{-i\varphi} \sin (\theta/2 ) \\
 e^{i\varphi} \sin (\theta/2) &  \cos (\theta/2 ).
 \end{array}\right)
 \end{equation}
 Details of the calculation are presented in  \ref{appa}. 
 
 The result for the symmetric component of the thermal geometric tensor defining the metric in Eq. (\ref{l}) is \cite{scan}
 \begin{equation}\label{lambda-mar}
 \Lambda^S_{\ell,\ell^{\prime}} = \mbox{Diag}\left(\beta \lambda_B, B \lambda_q  , B \lambda_q   \sin^2\theta \right),
 \end{equation}
 with
  \begin{equation}
     \lambda_B=\frac{ 1}{\Gamma(B)}\frac{\sinh(\beta B)}{\cosh^3(\beta B)} , \;\;\;\;\;\;\;\;\;\; \lambda_q=\frac{1}{\Gamma(B)}  \tanh^2(\beta B).
 \end{equation}
 The function
$ \Gamma(B)= \gamma_0 B^s$ with
  $s=1, >1, <1$ defines the spectral properties of the bosonic bath (Ohmic, super-Ohmic and sub-Ohmic, respectively).

 This behavior is intriguing, since we can see that radial protocols are less dissipative than those evolving tangentially over a solid angle in the low temperature regime where $\beta B>1$. This reveals the different nature  of these two protocols. 
 The radial one affects  an exponentially small fraction of the populations and
 has a small  impact at low enough temperatures. Instead, the tangential one directly affects the off-diagonal elements of the density matrix (coherences). This remarkable different behavior exhibits the relevance of optimizing the protocols in order to reduce the dissipation.

 These ideas were also used to minimize the dissipation in finite-time Otto and Carnot cycles implemented in qubits \cite{paolo,brand}, to analyze the work fluctuations in these systems \cite{miller1,miller2}, and more recently, to minimize the dissipation in cycles implemented in qubits simultaneously coupled to several reservoirs \cite{pablo,brand1}. 
 
 \subsection{Shortcuts to adiabaticity}
The first ideas behind the concept of shortcuts to adiabaticity were proposed by Berry \cite{berrysta} in the context of closed quantum systems
and are formulated in simple terms in the abstract of that paper: \begin{quote}
    For a general quantum system driven by a slowly time-dependent Hamiltonian, transitions between instantaneous eigenstates are exponentially weak. But a nearby Hamiltonian exists for which the transition amplitudes between any eigenstates of the original Hamiltonian are exactly zero for all values of slowness.
\end{quote} 
Such nearby Hamiltonian is constructed by adding "counter-diabatic" terms to the original time-dependent one.  In practice, this implies considering the evolution 
defined by the following effective Hamiltonian
\begin{equation}\label{hef-sta}
    H_{\rm eff}(t) = H(t) + i \hbar \sum_n \left( |\partial_t n \rangle \langle n| - 
    \langle n |\partial_t n \rangle |n \rangle \langle n|  \right), 
\end{equation}
where $\left\{|n\rangle \right\}$ are the instantaneous eigenstates of the original Hamiltonian
$H(t)$.  

For some years now, this is a topic of great interest with many applications and the recent developments are covered in Ref. 
\cite{sta}. Taking into account the analysis of the dissipation in finite-time protocols discussed in Sec. \ref{sec:adia-thermo-length}, 
it is expected that this type of mechanism could be useful to minimize dissipation without requiring slow driving. 
The explicit calculation of the counter-diabatic terms in Eq. (\ref{hef-sta}) is  a non-trivial problem since they depend on the states of the full Hamiltonian. While that calculation  is in principle possible for closed systems, this is a  difficult task
in the case of quantum systems coupled to reservoirs \cite{sels,sa-var}. Furthermore, the additional counter-diabatic terms are expected to generate additional dissipation.

An interesting and different approach amenable to be used in open quantum systems was proposed in Refs. \cite{deffner-bonan,deffner-nonlin}.  The strategy is similar to the adiabatic linear response formalism reviewed in Sec. \ref{sec:adia}. However, instead of considering a slow evolution, an arbitrary fast evolution with a small amplitude is considered. This is basically the  scenario of usual linear response theory and Kubo formalism \cite{bruus}. Concretely, 
small changes in the amplitudes of the parameters  but  arbitrary speed as
the system evolves from $t_0$ to $t_1=t_0+\delta t$ are assumed such that
\begin{equation}
    \vec{X}(t)= \vec{X}(t_0)+ g(t) \vec{\delta X},
\end{equation}
with $g(t_0)=0$ and $g(t_1)=1$. 
In this approach the Hamiltonian of the system is expanded as
\begin{equation}
    H(t)=H(\vec{X}_0)-g(t) \sum_\ell {\cal F}_\ell(\vec{X}_0) \; \delta {X}_{\ell}.
\end{equation}
Expressing the definition of the work given in Eq. (\ref{wij-q}) as
\begin{equation}
    W_{0\rightarrow 1}= -\sum_\ell \int_{t_0}^{t_1} \langle {\cal F}_\ell(\vec{X}) \rangle\; \dot{X}_\ell(t)= W^{\rm (cons)}+W^{\rm (diss)},
\end{equation}
and calculating $\langle {\cal F}_\ell \rangle$ in linear response, it is found
\begin{equation}\label{sca-lr}
    W^{\rm (diss)} = \sum_{\ell,\ell^{\prime}}{\delta X}_\ell \; {\delta X}_{\ell^{\prime}}
    \int_{t_0}^{t_1} dt   \int_{-\infty}^t \;dt^{\prime}  \; \dot{g}(t)
    \Psi_{\ell,\ell^{\prime}}(t-t^{\prime}) \; \dot{g}(t^{\prime})  ,
\end{equation}
with 
\begin{equation}
   -\frac{d \Psi_{\ell,\ell^{\prime}}(t-t^{\prime}) }{d t^{\prime}} =\chi_{\ell,\ell^{\prime}}(t-t^{\prime})=-\frac{i}{\hbar} \Theta(t-t^{\prime}) \langle \left[{\cal F}_\ell(t), 
   {\cal F}_{\ell^{\prime}}(t^{\prime})\right] \rangle_0,
\end{equation}
being $\chi_{\ell,\ell^{\prime}}(t-t^{\prime})$  the retarded susceptibility, $\Theta(t)$ is the Heaviside function and $\langle .\rangle_0$ indicates that the statistical averages are calculated with respect the Hamiltonian $H(\vec{X}_0)$. 
In Ref. \cite{deffner-bonan} it is argued that  there exist  special protocols 
satisfying the small amplitude condition in 
a qubit system where the coupling with reservoir is vanishing small and having
$W^{\rm (diss)}=0$. For finite coupling, we should expect $W^{\rm (diss)} \neq 0$, even in the limit where the coupling is weak.
In such cases, Eq. (\ref{sca-lr}) could be used to minimize $W^{\rm (diss)}$ in fast processes.
The comparison of this approach with the optimization in the  adiabatic regime analyzed 
in Section (\ref{sec:adia-thermo-length}) is an interesting open problem.

\section{Power pumping}\label{sec:pump1}
\subsection{General considerations}
The mechanism of power pumping 
 consists in the exchange of power between driving sources of different nature in quantum systems under  cyclic driving. It  may take place in open as well as in closed quantum systems and has been been studied in qubits in Refs. \cite{gil,gil2,crowley}. 

The net power pumping between  two sources $\ell, \; \ell^{\prime}$ is quantified by the time-average of 
\begin{equation}\label{ppumpel}
P^{\rm (pump)}_{\ell,\ell^{\prime}}(t)= \frac{P_\ell(t)-P_{\ell^{\prime}}(t)}{2},
\end{equation}
where $P_\ell(t)$ is defined in Eq. (\ref{power}).

In the  adiabatic regime discussed in Sec. \ref{sec:adia} substituting the expansion of Eq. (\ref{jp-non-eq-adia}) in
 Eq. (\ref{ppumpel}) leads to 
 \begin{equation}\label{ppump2}
 P^{\rm (pump)}_{\ell,\ell^{\prime}}(t)=\frac{P_\ell(t) - P_{\ell^{\prime}}(t)}{2}=   \dot{X}_{\ell}(t) \; \Lambda^{A}_{\ell,\ell^{\prime}}(\vec{X}) \; \dot{X}_{\ell^{\prime}}(t), \;\;\;\;\mbox{adiabatic},
 \end{equation}
where we have dropped the conservative component  because its average over a cycle vanishes. This equation stresses that
 only the antisymmetric component of $\Lambda_{\ell,\ell^{\prime}}(\vec{X}) $ contributes. Recalling that  in
 Sec. \ref{sec:en-diss} it was shown that only the symmetric component of this tensor contributes to dissipation, we see that power pumping can be viewed as the complementary process to dissipation.   
 In fact, driving forces with an associated antisymmetric adiabatic susceptibility  do not contribute to dissipation but behave similarly to a Lorentz force. This has been  discussed in driven electron systems  Ref. \cite{lu,bode,thomasf} and in models of coupled harmonic oscillators \cite{cam-den-han}. This mechanism is likely to be related to other work-work conversion proposals studied in the literature\cite{cam1,cam2}.




\subsection{Topological power pumping in the adiabatic regime}\label{sec:topo-adia}

One of the most paradigmatic effects associated to the adiabatic evolution  is the generation of a geometric phase that  accumulates in a cyclic protocol. This phase contributes independently to the dynamical time-dependent one and provides signatures of the properties of a quantum system. In quantum mechanics
this concept has been put forward by Berry \cite{Berry1, Berry2, Berry3}, but such a phase is not exclusive of this field and also appears in the slow evolution of classical systems \cite{geo-class}. Importantly, this is a fundamental concept in the characterization of topological phases of matter \cite{Berry-topo,berry-topo1,berry-topo2,berry-topo3,berry-topo4,berry-topo5}. 
In fact, 
Chern insulators, which are one of the best well-known topological states of  matter and include the  quantum Hall state, are characterized by a 
quantized Berry phase normalized by $2\pi$: the Chern number. The  Berry phase can be alternatively  evaluated 
as an area integral of the Berry curvature. 

Remarkably, these geometrical properties can be found in the 
 adiabatic dynamics of single qubit system. 
 Here, the relevant regime corresponds to low enough temperatures where the system evolves close to the instantaneous ground state.
 This was shown in Ref. \cite{gritsev} and it was experimentally verified in Ref. \cite{mike-exp1}, which will be reviewed in Sec. \ref{sec:expe-topo}. Here, we start by briefly reviewing the main ideas of  Ref. \cite{gritsev} expressing them in the same notation of previous sections. 
 In that work, the adiabatic dynamics of the Hamiltonian of Eq. (\ref{driv-qubit}) is analyzed and it is shown that
 some protocols are characterized by a 
 Berry phase.  
In particular, 
spherical coordinates as defined in Eq. (\ref{proto-scandi}) are considered and  a protocol with $B$ constant and constant velocity $\dot{\theta}$ is implemented. The 
adiabatic dynamics of the expectation value of ${\cal F}_{\phi}=-\partial H_{\rm qubit}/\partial \phi$ (see Sec. \ref{sec:adia}) is
\begin{equation}\label{fphi}
\langle {\cal F}_{\phi} \rangle(t) = \langle {\cal F}_{\phi} \rangle_t+ \Lambda_{\phi,\theta} (\theta,\phi) \dot{\theta}(t).
\end{equation}
For the system in the ground state, $|g\rangle$, the calculation of the adiabatic susceptibility leads to
\begin{equation}\label{anti-grit}
 \Lambda_{\phi,\theta}= \Omega_g= i \hbar \left[ \langle \partial_{\phi} g | \partial_{\theta} g \rangle -\langle \partial_{\theta} g | \partial_{\phi} g \rangle\right],
 \end{equation}
 where we see that it is purely antisymmetric and coincides with the definition of the Berry curvature. 
The explicit calculation  gives $\Lambda_{\phi,\theta} =\hbar /2 \sin\theta$,  and the integration over the spherical surface leads to the definition of the Chern number, which is found to be quantized,
\begin{equation}\label{chern}
    C_g=\frac{1}{2\pi}\int_0^{\pi} d \theta \int_0^{2\pi} d\phi \; \Omega_g=1.
\end{equation}
This result indicates that $\langle {\cal F}_{\phi} \rangle$ can be characterized by the Chern number and that it can be quantized for some protocols. As we highlighted in Eq. (\ref{anti-grit}), the Berry curvature is related to the antisymmetric component of $\Lambda_{\ell, \ell^{\prime}}$ and this property has also been noticed in open quantum systems connected to reservoirs at finite temperature \cite{bibek}. 

The fact that  $\Lambda^A_{\ell, \ell^{\prime}}$ defines forces characterized by topological numbers in a qubit
and recalling Eq. (\ref{ppump2}) suggests the possibility of topological power pumping in the adiabatic regime of these systems.
 This is precisely confirmed by analyzing the results reported in Ref. \cite{crowley}. 
In that paper, and also in Refs. \cite{gil,gil2}, the Hamiltonian of Eq. (\ref{driv-qubit}) is considered with 
\begin{equation}\label{ht}
 \vec{B}(\vec{X})= \frac{B_0}{2} \left( \sin X_1, \sin X_2, 2+ \delta - \cos X_1- \cos X_2 \right).
 \end{equation} 
 This model is formally identical to the reciprocal-lattice representation of a model for a Chern insulator \cite{qi,bernevig}. In the present case, there are two driving parameters,  $\vec{X}=\left(X_1(t), X_2(t)\right)$, and the corresponding adiabatic reaction forces are
 \begin{eqnarray}
 {\cal F}_1 &=& -\frac{\partial H}{\partial X_1}=\frac{B_0}{2} \left[\sigma_x \cos X_1+ \sigma_z \sin X_1  \right], 
 \nonumber \\
 {\cal F}_2 & = & - \frac{\partial H}{\partial X_2}=\frac{B_0}{2} \left[\sigma_y \cos X_2+  \sigma_z \sin X_2 \right].
 \end{eqnarray}
 The implemented driving protocol is
 \begin{equation}\label{protox}
 X_{\ell}(t) = \omega_{\ell} t + \varphi_{\ell},
 \end{equation}
  with   frequencies $\omega_{\ell}, \; \omega_{\ell^{\prime}}$ such that they are small enough to justify the adiabatic description. 
 The explicit calculation of the pumped power 
 considering the system in the ground state 
 leads to
\begin{equation}\label{ppump3}
 P^{\rm (pump)}_{1,2}(\vec{X})=\frac{P_1 - P_2}{2}=  \omega_1 \omega_2 \; \Lambda^{A}_{1,2}(\vec{X}),
 \end{equation}
 with $\Lambda^{A}_{1,2}(\vec{X})=\Omega_g$, given by Eq. (\ref{anti-grit}) upon substituting $\phi, \theta$ by $1,2$. Therefore, the pumped power is expressed as a 
 Berry curvature.
  The Hamiltonian of Eq. (\ref{ht}) is periodic in the
  parameters $\vec{X}$. Hence it is natural to focus on the two-dimensional region where $-\pi \leq X_\ell < \pi,\;\ell=1,2$, which behaves as the synthetic Brillouin zone (sBz) of the Hamiltonian of Eq. (\ref{ht}) defined in the synthetic reciprocal space of coordinates $\vec{X}$. Averaging over all possible initial conditions $\varphi_1, \varphi_2$ is regarded as a way to quantify the mean pumped power for irrational frequencies, corresponding to the case where $\omega_1/\omega_2$
  is an irrational number. Such an average can be written as
  \begin{equation}\label{topo-pump}
  \overline{P}^{\rm (pump)}(1,2) = \omega_1 \omega_2 \int_{\rm sBz} \frac{d^2X}{(2\pi)^2} \Omega_g(\vec{X}) \equiv \hbar \omega_1 \omega_2 C_g ,
  \end{equation}
 where $C_g$ is the Chern number. From the formal point of view, this  number is the same as the one characterizing the Chern insulator calculated for the ground state of the
 Hamiltonian of Eq. (\ref{ht}) formulated in the reciprocal space of a square lattice  \cite{qi,bernevig}. The explicit result is
 \begin{equation}
 C_g=
 \begin{cases}
 & 0, \;\;\;\;\;\;\;\;\;\delta > 0,\;\;\;\;\;\;\;\;\;\;\mbox{trivial},\nonumber\\
 & 1/2, \;\;\;\;\;\;\delta = 0,\;\;\;\;\;\;\;\;\;\;\mbox{Dirac},\nonumber\\
 & 1, \;\;\;\;\;\;\;\;\;\delta < 0,\;\;\;\;\;\;\;\;\;\;\mbox{topological},
 \end{cases}
 \end{equation}
 corresponding, respectively, to a trivial phase with a gap between the ground and excited states for all values of $\vec{X}$, a system with a Dirac point for $\vec{X}=0$ and
 a topological gapped phase. Consequently, in the first case the mean pumped power is zero, while in the other cases it is defined by an universal quantity
 proportional to the Chern number.

 \subsection{Topological pumping in the Floquet regime}\label{topo-floquet}
 Topological quantized pumping of energy between two driven sources was originally proposed to take place under 
 non-adiabatic conditions far-from-equilibrium conditions in Refs. \cite{gil,gil2}. 
 These papers focus on the  time-dependent Hamiltonian defined in Eqs. (\ref{driv-qubit}) and (\ref{ht}) with the protocol of Eq. (\ref{protox}) in the Floquet regime. For arbitrary large commensurate frequencies 
 it is possible to solve the driven Hamiltonian by recourse  to the  Floquet theory summarized in Sec. \ref{sec:floquet}.
 
 Introducing the Floquet representation in the Schr\"odinger equation leads to the structure of Eq. (\ref{ham-floq}) with $\vec{n}=(n_1,n_2)$,
 associated to the Floquet modes of $\vec{\omega}=(\omega_1,\omega_2)$.
 In the present problem these coupled equations can be Fourier-transformed to the "momentum" representation $\vec{q}=(q_1,q_2)$ 
 to obtain the effective Hamiltonian
 \begin{equation}
     H=\sum_{\vec{q}} H_{\vec{q}}+\sum_{\vec{n}} H_{\vec{n}},
 \end{equation}
 with $H_{\vec{n}}=-\hbar \vec{n}\cdot  \vec{\omega} $ while $H_{\vec{q}}$ is the Hamiltonian of Eq. (\ref{driv-qubit}) with $\vec{B}$ defined in Eq (\ref{ht}) 
 and  $\vec{X}\equiv \vec{q}$, independent of $t$. In this way, an effective dynamics ruled by a time-independent Hamiltonian defined in a 
 two-dimensional lattice is generated. Using the analogy with the lattice model and relying on semiclasical
 equations of motion for crystals perturbed by slow electromagnetic fields \cite{niu}  an expression for the pumped power is derived. It reads
 \begin{equation}
     P_{1,2}^{\rm(pump)}(t)= \omega_1 \omega_2 |\Omega_{\vec q}|,
 \end{equation}
 where $\Omega_{\vec{q}}$ is the Berry curvature of a given band $\vec{q}$. The structure of this equation is very similar to the
corresponding one in the adiabatic regime expressed in Eq. (\ref{ppump3}). Extending this expression to the case of incommensurate frequencies is interpreted as an average over the Brillouin zone of $\vec{q}$, which leads to the definition of the Chern number as in
Eq. (\ref{topo-pump}). The difference between  the adiabatic  and the Floquet cases is that in the adiabatic one this quantity was calculated over the ground state of the frozen qubit Hamiltonian.
Instead, in the present case it is calculated in the Floquet space. These ideas were further explored and implemented in an electronic spin embedded in an NV center in a diamond \cite{boyers}.
  In Ref. \cite{gil2} this mechanism was analyzed for a qubit coupled to reservoirs. 
 
The usefulness of the topological power exchange to control the spin dynamics in the system coupled to a circuit or cavity QED was analyzed in Ref. \cite{boost} and its realization in platforms containing arrays of qubits was discussed in Refs.\cite{mike1,mike2,mike3}. A unique feature that make
topological phenomena so appealing is their robustness  against perturbations and it is very promising to find such properties in the context of energy exchange in qubit systems.

\section{Heat pumping}\label{sec:heat-pump}
\subsection{General considerations}
Quantum pumping, in particular particle pumping, has been a subject of great interest in the context of electron systems for some time now \cite{pumping1,pumping4,pumping0,pumping2,pumping3,pumping5}. Basically, it consists in inducing a net particle transport between two reservoirs with neither electrical nor thermal bias but by means of a periodic asymmetric modulation the embedded quantum system. 
A widely investigated example is a quantum dot consistent of a double-barrier  structure confining a few quantum levels in contact to electron reservoirs. Without applying any electrical or thermal bias, gate voltages are locally applied at the walls following an asynchronous cyclic protocol. A simple version of such protocols consists in
deforming one of the confining walls enabling electrons to flow from the neighboring reservoir into the quantum dot levels while keeping the other wall
fixed. After the electrons fill-in the quantum-dot levels, the wall is restored to its initial situation while the opposite wall is deformed to facilitate the
flow of the electrons from the quantum dot towards its neighboring reservoir. After the electrons leave the quantum dot, the wall is restored. Hence, the initial configuration is recovered with a net transfer of electrons from one reservoir to the other as a consequence of the deformation of the confining walls. 

This mechanism has a counterpart in the heat \cite{lili-moska,therpump1,therpump2}, phononic \cite{phonon-pump1,phonon-pump2,phonon-pump3,phonon-pump4}, and spin \cite{spin-pump1,spin-pump2,spin-pump3} transport.

\subsection{Adiabatic heat pumping and geometric phase}
We discuss below a concrete and intuitive mechanism for heat pumping operating in a single qubit. Here, we introduce the main ideas to describe this phenomenon in the adiabatic regime. We focus on the slowly driven system
performing a cycle of period $\tau$, so that the  parameters satisfy $\vec{X}(t+\tau)=\vec{X}(t)$ while connected to
 two reservoirs (l, r) at the same temperature $T$. 
It is important to notice that, in order to discern the two reservoirs, there should be some asymmetry in  the couplings. In the case of the qubit, this can be achieved
with  couplings to the reservoirs as in Eq. (\ref{qubit-res}) that are defined by different and non co-linear vectors $\vec{g}_{k,\rm l}$ and $\vec{g}_{k,\rm r}$ for the two reservoirs. 

The starting point is to considering the heat current defined in Eq. (\ref{p-jalpha}). It is possible to introduce the adiabatic expansion given by
 Eq. (\ref{jp-non-eq-adia}) to calculate the heat flowing into each reservoir in a cycle
  \begin{equation}\label{qpumpal}
 Q_{\alpha}= \int_0^{\tau} dt J_{\alpha}(t) =\sum_{\ell} \int_0^{\tau} \Lambda_{\alpha,\ell} (\vec{X}) \dot{X}_{\ell}.\;\;\;\;\;\alpha={\rm l,\;r}
 \end{equation}
 where this term represents the non-equilibrium heat induced by the driving processes. Since the processes associated to
 entropy production are described by Eq. (\ref{ts}) and are second-order in $\dot{\vec{X}}$, we conclude that this first-order component  satisfies
 \begin{equation}\label{heat-pump}
 Q_{\rm l}=- Q_{\rm r}=Q^{\rm (pump)},
 \end{equation}
 which defines the {\em pumped heat} in the slow-driving regime. Here we notice that this definition is similar to the concept of  power pumping previously defined in Eq. (\ref{ppumpel}). In that case the exchange of energy takes place between two driving forces, while in the present case, it takes place as heat exchange between the two baths. In fact, in analogy to Eq.  (\ref{ppumpel}) we can define from Eq. (\ref{heat-pump}),
 $P^{\rm (heat-pump) }= (Q_{\rm l}- Q_{\rm r})/2\tau$. Furthermore, as in the case of power pump, geometrical concepts similar to the Berry phase are useful to describe this mechanism. In fact, for the sake of making the geometrical nature more explicit, it is useful to introduce the vector $\vec{\Lambda}(\vec{X})=\left(\Lambda_{\rm l,1} (\vec{X}), \ldots,\Lambda_{\rm l,N} (\vec{X})\right)$, according to which the pumped heat per cycle reads
  \begin{equation}\label{qpumpal-g}
 Q^{\rm (pump)}=  \oint \vec{\Lambda} (\vec{X}) \cdot \vec{d X}.
 \end{equation}
 The closed integral corresponds to the trajectory in the parameter space defining the cyclic protocol of the driving parameters. This structure is similar to the case of charge pumping, \cite{pumping1,pumping5}. The heat pumping takes place along with energy dissipation. The description of these processes is identical as in 
 Sec. \ref{sec:en-diss}, taking into account the coupling to several thermal baths.
 
 \subsection{Adiabatic heat pumping with a  single qubit}\label{sec:adia-heat-qubit}

We now discuss the mechanism of adiabatic pumping in the context of single-qubit setups. 
We focus on the case studied in Refs. \cite{bibek,pablo}. The model is similar to the one considered in Sec. \ref{sec:en-diss-gen}, in a configuration where the qubit  is coupled to left (l) and right (r) reservoirs  at the same temperature $T$. 
The driven qubit is modeled 
by the Hamiltonian of Eq. (\ref{driv-qubit}) 
with $\vec{B}(t) = \vec{X}(t)=\left(B_x(t),0, B_z(t)\right)$. Importantly, the couplings  to the two reservoirs must be implemented with two non-commuting operators. Otherwise, it would be possible to define  an effective single bath with a  linear combination of modes 
of the different reservoirs. We consider the coupling Hamiltonian of Eq. (\ref{qubit-res}) with $\vec{g}_{\rm l}=(0,0,g^z)$  for the left bath and
$\vec{g}_{\rm r} =(g^x,0,0)$ for the right one. Hence
\begin{equation}\label{contlr}
H_{\rm cont,l}= g^z \sum_{k_l} \left(a^{\dagger}_{k_l}+a_{k_l}\right)\sigma^z, \;\;\;\;\;\;\;\; H_{\rm cont,r}= g^x \sum_{k_r} \left(a^{\dagger}_{k_r}+a_{k_r}\right)\sigma^x.
\end{equation}

 In the weak coupling regime, this problem can be solved with the procedure explained in Sec. \ref{sec:adia-weak}. 
 Details are presented in \ref{appb}.
 With those elements, following Ref. \cite{pablo}
 it is possible to obtain analytical results for the vector $\vec{\Lambda}$ defining the pumped heat in Eq. (\ref{qpumpal-g}),
 \begin{equation}\label{qpum-lamb}
 \vec{\Lambda}(\vec{B})= \frac{\beta B \sin^2 2\theta}{\cosh^2(2\beta B)} \vec{n},
  \end{equation}
 where $\vec{n}$ is the unit vector along the radial direction in the $\left(B_x,B_z \right) \equiv B(\sin \theta, \cos \theta) $ plane.
 In terms of it, the pumped heat reads
 \begin{equation}\label{qpum-ex}
  Q^{\rm (pump)}=\oint \vec{\Lambda} (\vec{B}) \cdot \vec{d B}
 =\int \int \left( \nabla_{\vec{B}} \times \vec{\Lambda} (\vec{B}) \right) \cdot d^2 \vec{B} ,
 \end{equation}
 In the last step we have introduces Stokes' theorem to express the line integral over the protocol in the form of a flux integral through the area enclosed by the protocol.
 This result is interesting because it makes explicit the fact that, in order to have heat pumping  in a cyclic protocol with this device, it is necessary to follow a trajectory
 in the parameter space  where $|\vec{B}|=B$ changes.  The analysis of $\left( \nabla_{\vec{B}} \times \vec{\Lambda} (\vec{B}) \right)_y$ is extremely helpful to visualize the optimal protocols. 
 
\begin{figure}
    \centering
    \includegraphics[width=\textwidth]{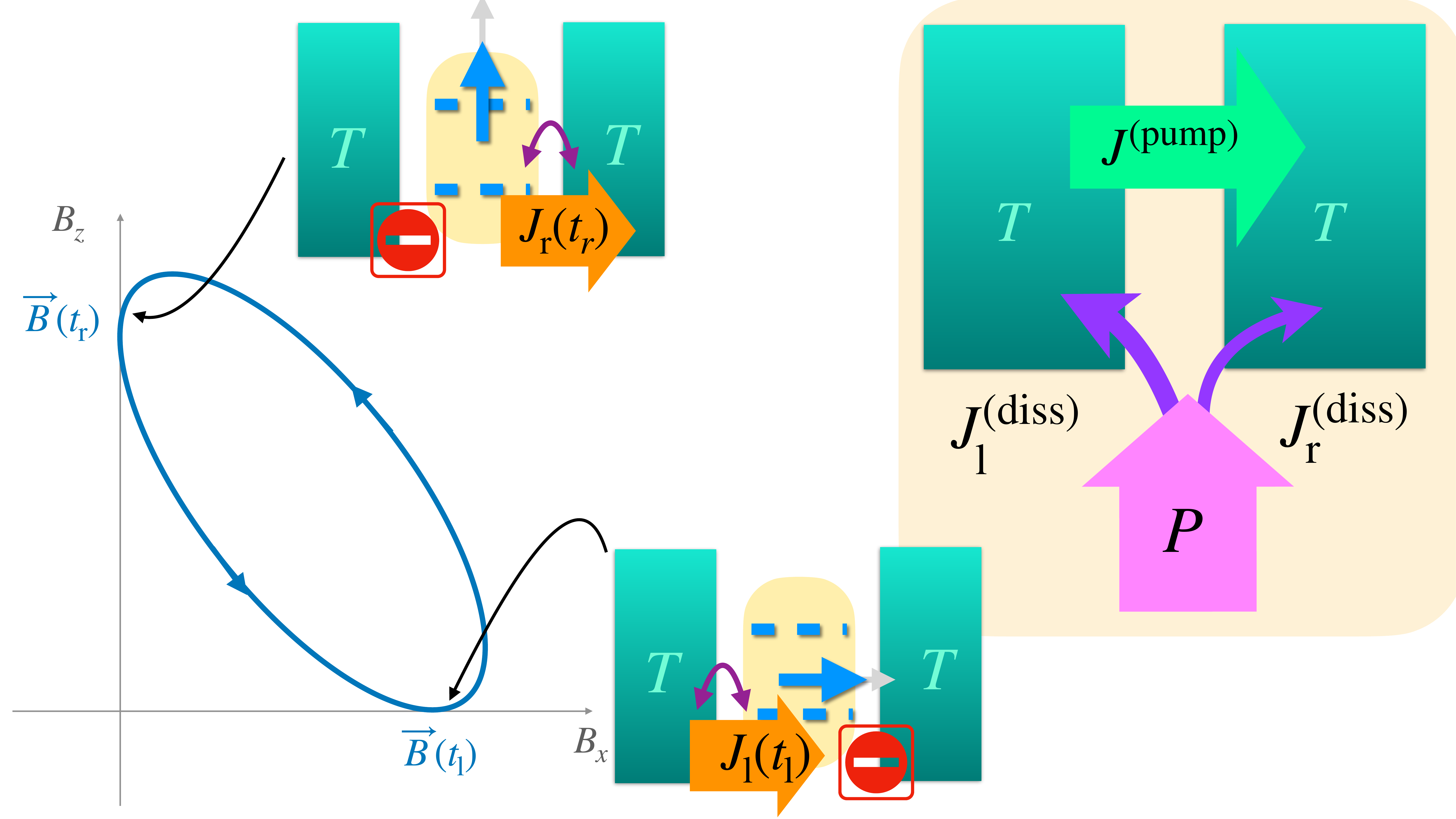}
    \caption{Example of a protocol leading to quantum pumping in the configuration of a driven qubit asymmetrically coupled to reservoirs at the same temperature (see text). The sketches illustrate the instantaneous Hamiltonian and the heat fluxes 
 at the instants $t_{\rm l}$ and $t_{\rm r}$.
    The net result over the cycle is highlighted in the colored box. }
    \label{fig:pump}
\end{figure}

An example of the pumping cycle is sketched  in Fig. \ref{fig:pump}. The two components of $\vec{B}(t)$ evolve cyclically but with a phase lag. At a given instant 
$t_{\rm l}$ the vector $\vec{B}(t_{\rm l})$ points along the $x$ direction and the state is polarized in the $x$ direction of the Bloch sphere. Such a state forbids the energy exchange with the right reservoir.
This can be explicitly seen by noticing that  Eqs. (\ref{ker}) and (\ref{gker}) with $B_z=0$  reduces to the  master equation of the qubit coupled only to the left bath. There is a heat flux $J_{\rm l}(t_{\rm l})=\langle \dot{H}_{\rm l} \rangle$ between the qubit and the left bath. Assume $k_B T > B(t_{\rm l})$, so that  heat  flows  from the left reservoir into the qubit at this time. 
At another instant $t_{\rm r}> t_{\rm l}$, $\vec{B}(t_{\rm r})$ the state is rotated so that it points along $z$ and lets assume that the protocol is such that $B(t_{\rm r}) > B(t_{\rm l})$ with $B(t_{\rm r})> k_BT$.
At this time there is not energy exchange with the left reservoir. In fact, for $B_x=0$, Eqs. (\ref{ker}) and (\ref{gker}) correspond to the qubit coupled only to the right bath.  An  energy
flow  $J_{\rm r}(t_{\rm r})=\langle \dot{H}_{\rm r} \rangle$ takes place between the qubit and the right bath. Since we assumed $B(t_{\rm r})> k_BT$, the heat now flows from the qubit to the right bath.
 If  the cycle is completed by  decreasing $B$ to the value $B(t_{\rm l})$,  the  result is a net transfer of heat from the left to the right as a consequence of the qubit driving. This type of cycles realize the mechanism of heat pumping and the direction of the pumped heat flux can be inverted by reversing  the protocol.
As in the case of the configuration analyzed in Sec. \ref{sec:en-diss} the driving generates heat that is dissipated into the reservoirs, as indicated in the right sketch of  Fig. \ref{fig:pump}. The distribution of this dissipated heat among the two  reservoirs depends on details like their density of states and their coupling with the qubit. 

The total dissipated energy for this problem can be calculated following the steps explained in Sec. \ref{sec:en-diss}. In
Ref. \cite{pablo}, analytical expressions are derived for the present example. The result is
\begin{equation}\label{qdis-ex}
W^{\rm (diss)}= \int_0^{\tau} dt \dot{\vec{B}} \cdot \underline{\Lambda} \cdot \dot{\vec{B}}=\int_0^{\tau} dt \left[ \lambda_B \dot{B}^2(t)
+ \lambda_{\theta} B^2(t) \dot{\theta}^2(t) \right],
\end{equation}
with 
\begin{equation}
\lambda_B =\frac{\beta \sinh(\beta B)}{2 \gamma B \cosh^3(\beta B)},\;\;\;\;\;\;\lambda_\theta=\frac{\gamma }{2  B^2},
\end{equation}
where $\gamma = \gamma_{\rm l}=\gamma_{\rm r}$ is the coupling strength, which is assumed to be the same for the two baths.
We can identify a behavior akin to the example of the driven qubit coupled to a single reservoir discussed in Ref. \cite{scan}
and in Sec. \ref{sec:en-diss}. In fact, radial protocols, weighted with $\lambda_B$, dissipate differently from protocols involving rotations of the
qubit states without changing the spectrum. The latter are weighted with $\lambda_\theta$. The regime where one is dominant with respect to the other
depends on the ratio $B/k_BT$. As already observed in Sec. \ref{sec:en-diss}, radial protocols are associated to changes in the populations and have a small
weight at low temperatures compared to $B/k_B$.

 In Ref. \cite{bibek} the following protocol is implemented in this device,
\begin{eqnarray}\label{proto}
 B_x(t)&=&B_{x,0} + B_{x,1}\cos(\Omega t + \phi), \\
 B_z(t) &=& B_{z,0} + B_{z,1} \cos( \Omega t),
 \end{eqnarray}
 with 
 $B(t)=\sqrt{B_x^2(t)+B_z^2(t)}$. The solution in the weak coupling regime leds to the conclusion
 $Q^{\rm (pump)} \propto \sin(\phi)$. In Ref. \cite{pablo} it was shown that the optimal protocol regarding the
 maximum heat pumping corresponds to a trajectory in the $(B_x,B_z)$ plane that encloses a full quadrant.
 The result is
 \begin{equation}\label{landauer}
 Q^{\rm (pump)}_{\rm Max}=\pm k_B T \log(2).
 \end{equation}
Interestingly, this quantity is known as {\em Landauer limit} and corresponds to the maximum possible entropy change in a two-level system\cite{landauer1,landauer2}. It defines a
fundamental limit for the thermodynamic cost of erasing information. The original proposal constitutes a breakthrough in relating thermodynamics and statistical mechanics with information theory.  It has been  covered in detail in review articles, like
Ref. \cite{janet,john} and it is the motivation of many recent experiments \cite{landauer-exp1,landauer-exp2,landauer-exp3,landauer-exp4,landauer-exp5,landauer-exp6,landauer-exp7,landauer-exp8,landauer-exp9,landauer-exp10,landauer-exp11}. 
In the present context, this limit is achieved in a 
 protocol that consists of an infinite-long trajectory. Such a protocol would take infinite time to be accomplished. Hence, it corresponds to a quasistatic limit
 which defines the bound for any other finite-time protocol. The two signs of this equation correspond to the different ways of circulating the path.
 More discussion will be presented in Sec. \ref{sec:ther-mach} in relation to the role of pumping in the operation of
 thermal machines.

Interestingly, pumping can be also induced in this setup under periodic variations of the temperatures of the two reservoirs, $\delta T_{\rm l}(t)$ and $\delta T_{\rm r}(t)$ with respect to the reference temperature $T$  \cite{segal-nitzan2,therpump3}.
The possibility of realizing topological charge pumping with qubits in cQED was suggested in Ref. \cite{leone} in a quite different setup based on two qubits, and experimental results on charge pumping in  a superconducting circuit 
has been reported in Ref. \cite{sluice2}. 
No experimental studies of  heat pumping in qubits have been reported yet.

\section{Thermal machines}
\subsection{General considerations}
So far we have analyzed effects that are generated purely by time-dependent driving. Thermal machines operate with the cyclically driven working substance -- represented by the qubit in our case of interest -- contacted with reservoirs with a temperature bias. 
In this scenario heat-work conversion is the key mechanism. 
\subsection{Thermodynamic cycles} \label{sec:cycle}
\begin{figure}
    \centering
    \includegraphics[width=\textwidth]{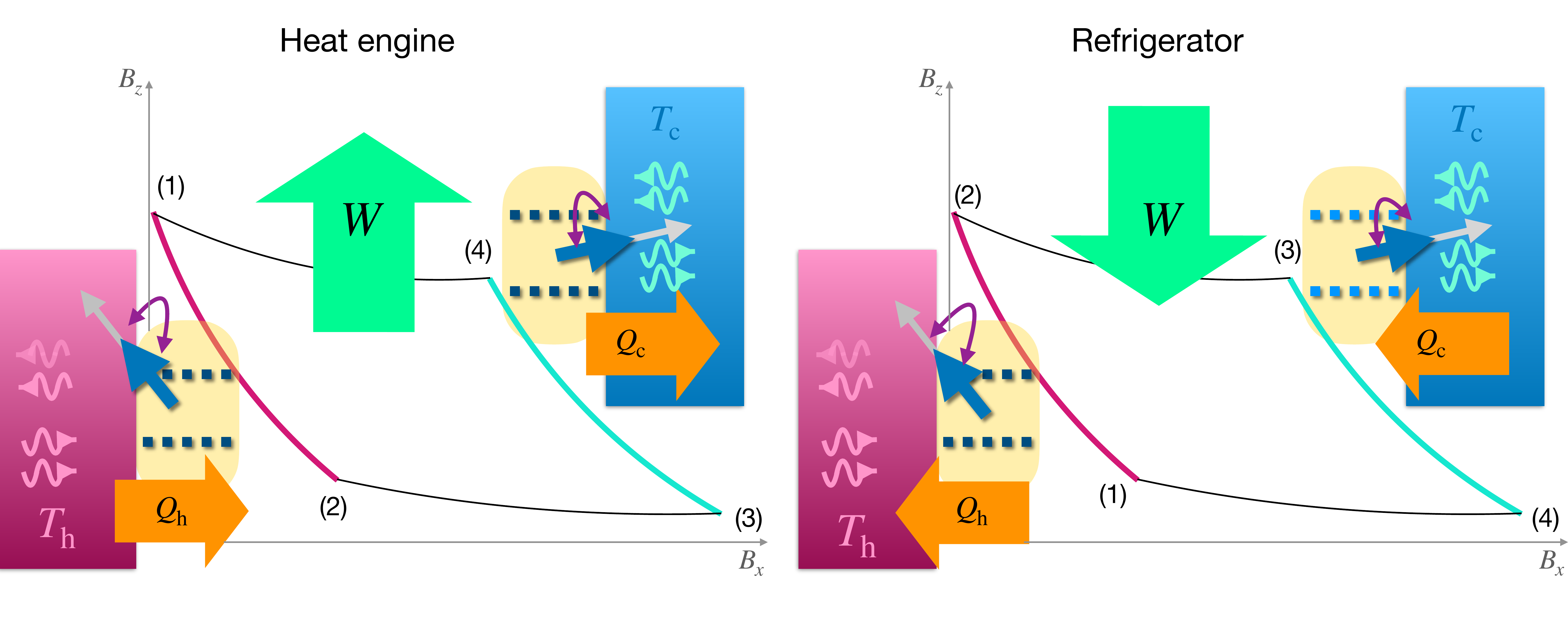}
    \caption{ Ilustration of a Carnot cycle  in a qubit for the heat engine and refrigerator operations. 
    }
    \label{fig:cycle}
\end{figure}

The implementation of quasi-static cycles similar to the classical thermodynamical ones in a single qubit has been considered in the qubit Hamiltonian defined in Eq. (\ref{driv-qubit}). In particular, the well known Carnot, Otto, Stirling and other cycles have been recently studied in Refs. \cite{paolo,brand,carnot1,carnot2,carnot3,cycle1,cycle2,vasco,otto1,otto2,otto3,stir}. Some of the related discussion has been covered in the review
articles of  Refs. \cite{janet,deffner,bhatta,lat}. We briefly summarize some important aspects. 

The relevant mechanisms can be easily understood by expressing the qubit Hamiltonian as follows,
\begin{equation}
H_{\rm qubit}(t)= -B(t) \; \overline{\sigma}^z,
\end{equation}
with 
 $\overline{\sigma}^z$ operating in the instantaneous eigenbasis of Eq. (\ref{driv-qubit}). The {\em Carnot} cycle is sketched in Fig. \ref{fig:cycle} for the qubit operating between a cold and a hot thermal baths at temperatures $T_{\rm c}$ and  $T_{\rm h}$, respectively. 
It consists of 
  four steps: (1) The qubit evolves coupled to the hot reservoir as the parameters  change leading to a change of $B(t)$ from $B_{\rm 1}=B(t_1)$ to $B_{\rm 2}=B(t_2)$. 
  In general, this process takes place in a finite time. In the ideal quasi-static cycle this is assumed to be long enough to justify considering the system in equilibrium with the bath at every instant during the evolution. In this step of the cycle there is exchange of
  heat, $Q_{\rm h}$ and work $W_{1\rightarrow 2}$ with the reservoir. 
  If  the evolution is such that the heat flows from the qubit towards the reservoir, the device operates as a refrigerator, otherwise it operates as a heat engine.   (2) The qubit evolves isolated from the two reservoirs between $B_{2}=B(t_2)$ and $B_3=B(t_3)$. In this process, only work ($W_{2 \rightarrow 3}$)  is exchanged between the system and the driving sources.  (3) The qubit evolves coupled to the cold reservoir from
  $B_3$ to $B_4=B(t_4)$. In this step there is again exchange of both heat $Q_{\rm c}$ and work ($W_{3 \rightarrow 4}$) with the reservoir. (4) The last step is an evolution isolated from the reservoir from $B_4$ to
  $B_5=B(t_5)=B(t_1)$ and takes place without exchange of heat with any of the reservoirs but exchanging work ($W_{4 \rightarrow 1}$) between the qubit and the external driving sources. 
 The  steps (2) and (4) are usually named "adiabatic" as in classical cycles, where this concept applies to processes in which no heat is exchanged. The practical way to implement this type of evolution in classical systems is by means of a fast change of the parameters of the working substance. 
 We recall, however (see  Sec. \ref{sec:adia}), that  in quantum mechanics the concept of "adiabaticity" is instead associated to a very slow evolution and is not necessarily associated to a process where there is no heat exchange in the case of an open quantum system. 
  
  An important concept to qualify the operation of the cycle is the efficiency (in the case of the heat engine) and the coefficient of performance (in the refrigerator). They are, respectively, defined as
  \begin{equation}\label{efic}
  \eta=\frac{W}{Q_{\rm h}}, \;\;\;\;\;\;\;\;\;\;\;\;\;\;\;\;\;\;\;\;\;\;\;\;\;\;\;\;\;\; COP=\frac{-Q_{\rm c}}{W},
  \end{equation}
  where the $W=W_{1\rightarrow 2}+W_{2\rightarrow 3}+W_{3\rightarrow 4}+W_{4\rightarrow 1}$ is the total work done in all the cycle. These definitions follow the convention that $Q_{\alpha}>0$ when it enters 
  the reservoir $\alpha$ and $W>0$ when work is delivered by the  driving sources into the system. In each step $W_{i\rightarrow j}$ may have any sign, depending on the protocol.

  The heat and the work in the different steps of the cycle are defined in  Eqs. (\ref{wij-q}). The conservation of the energy in the cycle implies
   \begin{equation}\label{w-q-cycle}
 W= Q_{\rm h} +Q_{\rm c}.
  \end{equation}

 For a quasi-static evolution we can use the definitions of Eqs. (\ref{cons-qs}) to explicitly verify that this relation holds for the conservative work $W^{\rm (cons)}$, 
 and the quasistatic heat exchanges $Q_{\rm h}^{(\rm qs)}$, $Q_{\rm c}^{(\rm qs)}$.
  In addition, since for the quasi-static processes there is no entropy production, and the total change of the entropy in these reversible processes
  reads
  \begin{equation}\label{ds}
  \sum_{\alpha}\Delta S_{\alpha} = \sum_{\alpha} \frac{Q_{\alpha}^{(\rm qs)}}{T_{\alpha}}=0.
  \end{equation}
  
  Therefore, Eq. (\ref{w-q-cycle}) can be expressed as
  \begin{equation}\label{carnot-qs}
  W^{\rm (cons)}=Q_{\rm h}^{(\rm qs)} \frac{\left( T_{\rm h} - T_{\rm c}\right)}{T_{\rm h}}, \;\;\;\;\;\; {\rm Carnot},
  \end{equation}
  which when substituted in Eqs. (\ref{efic}) leads to the well-known Carnot results for the efficiency and the coefficient of performance,
  \begin{equation}
  \eta_{\rm C} =\frac{T_{\rm h}- T_{\rm c}}{T_{\rm h}}, \;\;\;\;\;\;\;\;\; {\rm COP} =\frac{1}{\eta_{\rm C}},  \;\;\;\;\;\;\;\;\;\;\;\; {\rm Carnot}.
  \end{equation}
  For non-equilibrium finite-time protocols, the components $W^{\rm (non-cons)}$ and $Q_{\alpha}^{\rm (non-eq)}$ contribute \textcolor{red}{as will be  discussed below}. Furthermore, a precise description should also take into account the time-dependent processes of switching on and off the contacts to the reservoirs. The latter  are usually neglected in the literature although, recently, the effect of smooth changes
  in the system-reservoir coupling were found to speed-up the isothermal evolutions of Carnot cycles \cite{marti-speed}. 
 
The Otto cycle is also based on a four-stroke protocol. The main difference with respect to the Carnot one is in the steps (1) and (3) where the evolution in contact with the reservoirs takes place at constant $B$, hence only heat is exchanged in these processes. This cycle is 
very convenient from the  computational point of view, since only heat is exchanged in the strokes (1) and (3) while only work is  exchanged in the strokes (2) and (4), which implies important simplifications in the calculations. 
The implementation of Otto cycles has been the focus of many theoretical and experimental works\cite{janet,deffner,bhatta,lat} with recent focus on many-body effects \cite{otonic} and
speed up protocols \cite{otonic,muk1,muk2,muk3,otop1,otop2,otop3} to enhance the performance. This cycle has been widely analyzed in the literature and we defer the reader to Refs. \cite{janet,deffner,bhatta,lat}.

\subsection{Finite-time Carnot heat engine}\label{finite-carnot}
A good part of the literature on cycles in qubits focuses on a Carnot cycles with a quasi-static evolution in the steps where the system is in contact to reservoirs, and a fast evolution in the steps where it is decoupled \cite{carnot1,carnot2,carnot3,cycle1,cycle2,otto1,otto2,otto3,stir}. Recently,
finite-time effects in the evolution in contact to the reservoirs were addressed  \cite{paolo,brand,vasco}. 
At finite time, besides the efficiency, the other quantities qualifying the performance of the machine are the 
 output power, in the case of the heat engine, and the 
cooling power, in the case of the refrigerator. For a machine operating in a period $\tau$ these quantities read
\begin{equation}\label{power-he}
P^{\rm (he)} =- \frac{W}{\tau}, \;\;\;\;\;\; \;\;\;\;\;\;  \;\;\;\;\;\;  P^{\rm (cool)} =- \frac{Q_{\rm c}}{\tau}.
\end{equation}
The drawback of the finite-time operation is the energy dissipation and entropy production.
The effect of the dissipation in Carnot cycles where the evolution in contact with the reservoirs takes place at finite times was analyzed in the literature in Refs. \cite{espo,seif}.
The main step is to include the contribution of the dissipated energy due to the finite-time evolution during  the strokes  where the system evolves coupled to the  reservoirs.
 The corresponding heat exchanges with the cold and hot baths read
\begin{equation}\label{qal}
Q_{\alpha}=Q_{\alpha}^{(\rm rev)}+W^{\rm diss}_{\alpha} =T_{\alpha} \Delta S_{\alpha} +T_{\alpha}\Sigma_{\alpha}, \;\; \;\; \alpha={\rm c, h},
\end{equation}
where $\Delta S_{\alpha}$ is the reversible change in the entropy, while the dissipative contribution 
associated to the entropy production is denoted by $\Sigma_{\alpha}$. The latter accounts for the finite-time processes and vanishes in the limit of an ideal  cycle. 
According to the sign convention adopted here, the irerversible contributions are always positive, indicating that the dissipated energy flows into the reservoirs.
Instead, the sign of the
reversible component $Q_{\alpha}^{(\rm rev)}$ depends on direction of the heat flux. 

In Ref. \cite{paolo} the conditions to obtain a maximum power in a finite-time Carnot heat engine is studied considering one and several coupled qubits as the working substance and slow  evolutions in the evolutions in contact with the reservoirs. A useful step is to  introduce a 
change  in the notation in order to get an explicit dependence of the entropy production with the duration of the strokes, $\tau_{\alpha}$, $\alpha=$c, h.
This is achieved by  changing of the 
integration variable in Eq. (\ref{ts}) to $s=t/\tau_{\alpha}$, which leads to the following expression for the
entropy production,
\begin{equation}\label{wdis}
W^{\rm diss}_{\alpha}= \frac{1}{\tau_{\alpha}} \sum_{\ell, \ell^{\prime}} \int_0^1 ds \dot{X}_{\ell} \; \Lambda^{( S)}_{\ell,\ell^{\prime}}(\vec{X}) \; \dot{X}_{\ell^{\prime}}
=T_{\alpha}\frac{\overline{\Sigma}_{\alpha}}{\tau_{\alpha}},
\end{equation}
where now  $\dot{X}_{\ell} = dX_{\ell}/ds$. Introducing the same change of variables, the reversible part can be expressed following Eq. (\ref{jp-non-eq-adia}),
\begin{equation}\label{rev}
Q_{\alpha}^{\rm (rev)}= T_{\alpha} \Delta  S_{\alpha} = \sum_{\ell}  \int_{0}^1 ds \; \Lambda_{\alpha, \ell} (\vec{X}) \; \dot{X}_{\ell},
\end{equation}
 with  $\Delta S_{\rm c}=-\Delta S_{\rm h}= \Delta S$.
Eq. (\ref{w-q-cycle}) remains valid, even in the case of non-conservative processes. Hence,
the power of the heat engine defined in Eq. (\ref{power}) as well as the efficiency defined in Eq. (\ref{efic}) read
\begin{equation}\label{phe}
P^{\rm (he)} =- \frac{Q_{\rm c}+Q_{\rm h}}{\tau}, \;\;\;\;\;\;\;\;\;\;\;\;\;\;\;\; \eta^{\rm (he)}= \frac{Q_{\rm c}+Q_{\rm h}}{Q_{\rm h}},
\end{equation}
being $\tau=\tau_{\rm c}+\tau_{\rm h}$. 
The optimization of the durations
 leading to the maximum power of the heat engine  was previously discussed in Ref. \cite{espo,seif}. The result is obtained after substituting Eqs. (\ref{qal}) and (\ref{wdis}) into Eq. (\ref{phe}) 
 \begin{equation}\label{phe1}
 P^{\rm (he)} = -\frac{1}{\tau} \left( -\Delta S \Delta T + T_{\rm c} \frac{\overline{\Sigma}_{\rm c}}{\tau_{\rm c}} + T_{\rm h} \frac{\overline{\Sigma}_{\rm h}}{\tau_{\rm h}}\right).
 \end{equation}
 Maximizing with respect to the durations leads to
 \begin{equation}
\tau_{\rm c}=\tau_{\rm h}\sqrt{\frac{T_{\rm c} \overline{\Sigma}_{\rm c}}{T_{\rm h} \overline{\Sigma}_{\rm h}}}, \;\;\;\;\;\;\;\;\;\;\;\;\;\;\;\;
\tau_{\rm h}=\frac{2 T_{\rm h} \overline{\Sigma}_{\rm h}}{\Delta S \Delta T} \left( 1+ \sqrt{\frac{T_{\rm c} \overline{\Sigma}_{\rm c}}{T_{\rm h} \overline{\Sigma}_{\rm h}}}\right),
\end{equation}
being $\Delta T=T_{\rm h}-T_{\rm c}$.
Substituting these optimal durations in Eq. (\ref{phe}), the expression for the maximum power is obtained,
\begin{equation}
    P^{\rm max} = \frac{(\Delta S)^2}{4 } \frac{\left(T_{\rm h}- T_{\rm c}\right)^2}{\left(\sqrt{T_{\rm h}\overline{\Sigma}_{\rm h}}+\sqrt{T_{\rm c}\overline{\Sigma}_{\rm c}}\right)^2}.
\end{equation}
We see that the  power optimized with respect to the durations of the cycle still depends on on the coefficients $\overline{\Sigma}_{\alpha}$  characterizing the energy dissipation into the bath. These quantities depend on the microscopic details of the driven system, the baths, the couplings to the baths and the evolution protocols. For adiabatic driving, 
the description in terms of the thermodynamic length described in Sec. \ref{sec:adia-thermo-length} can be used to design the optimal protocols. This was precisely the goal of Ref.
 \cite{paolo} with focus on weak coupling between system and reservoirs. 

In these conditions it is shown that the bound for $P^{\rm max}$ is
proportional to the heat capacity of the working substance. Since the heat capacity can scale supraextensively with the number of constituents of the engine, this result is useful in the design of optimal many-body Carnot engines including the effect of many-body interactions and phase transitions \cite{campi-fazio}.

In the case of Otto cycles the finite-time optimization focuses on the steps where the system evolves decoupled from the reservoirs by means of shortcuts-to-adiabaticity. This implies including counter-diabatic terms in the protocols as in Eq. (\ref{hef-sta}). These
 suppress inter-state transitions in the free evolution, which improves the performance in these type of cycles \cite{deffner}.

As mentioned in Sec. \ref{sec:meas}, the identification of the energy exchanges of quantum measurements as heat and work is 
non-trivial. Irrespective of this identification, the measurement mechanism can be regarded as an exchange  of energy  which 
can be used to fuel 
  thermodynamical-like cycles of quantum working substances. There are several implementations in qubit systems \cite{bibek-andrew,jordan,qmes1,qmes2,qmes3,solfa} and similar ideas were also followed in other quantum systems \cite{qmes1ha,qmes2ha,jordan-new}. 
In Sec. \ref{sec:max-dem} we will discuss an experimental implementation.

\subsection{Adiabatic cycles and geometric properties}\label{sec:ther-mach}
\begin{figure}
    \centering
    \includegraphics[width=\textwidth]{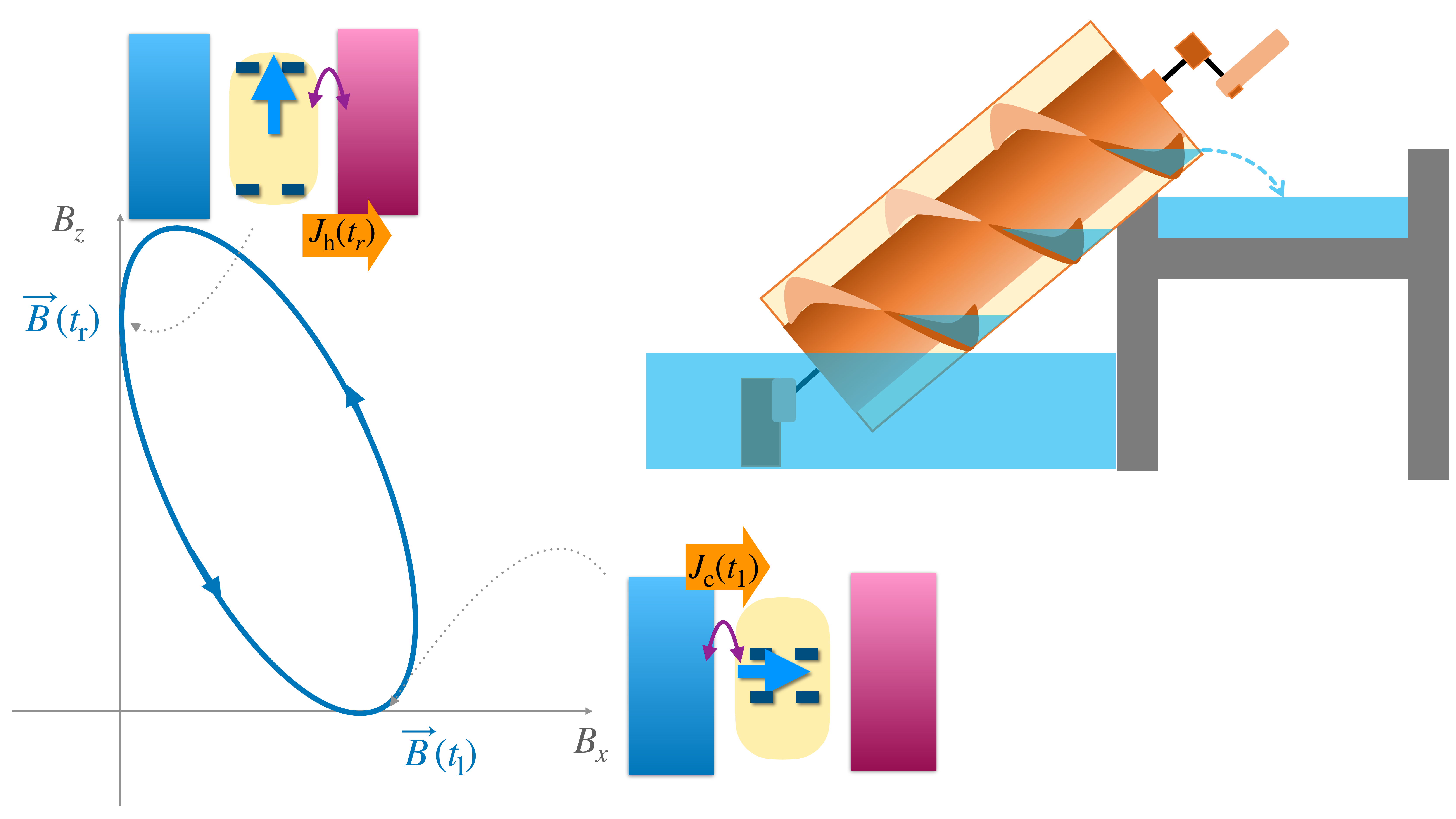}
    \caption{When the protocol of Fig. \ref{fig:pump} is implemented on the qubit between reservoirs at different temperatures, the pumping mechanism can be used against the thermal bias. The device operates as a refrigerator and 
    the mechanism bares resemblance with Archimedes' screw (see sketch). 
    }
    \label{fig:ther-mach}
\end{figure}

 This type of cycles are closely related to the pumping mechanism discussed in Sec. \ref{sec:heat-pump}
 and it can be implemented basically in the same setup 
as the one considered there. The crucial difference is the addition of a thermal bias between the two reservoirs, so that one of them  is hot (say the right one) and the other one is cold, having temperatures $T_{\rm h}= T_{\rm c}+ \Delta T$ and $T_{\rm c}$, respectively. We can now formulate the question of whether it is possible to implement the same pumping mechanism as before to overcome the thermal bias, in order to make a refrigerator out of this device. The answer is yes and the machine can be compared with the Archimedes' screw, which has been used since the time of the ancient Greeks to pump water against gravity by rotating a screw inside a pipe (see sketch of Fig. \ref{fig:ther-mach}). Notice that this also illustrates the pure pumping mechanism discussed in Sec. \ref{sec:heat-pump} if the pipe is placed horizontally. In the case of a thermal bias, the device  relies on the pumping mechanism induced by the driving to overcome the heat leak from the hot to the cold reservoir induced by the temperature bias, in the same way that the Archimedes' pump relies on the rotation of the screw to pump the water upwards. If the pumping mechanism operates in reverse, so that it transfers heat from the right to the left in the sketch of Fig. \ref{fig:ther-mach}, part of the energy transported  because of the thermal bias can be used to drive the machine, which operates as a heat engine in this case. This should be compared with  the Archimedes machine operating in reverse, realizing a generator, in which case the water flowing down through the pipe contributes to rotate the screw. The realization of these machines in the quantum realm has been widely investigated  in the context of electron systems under the operation with slow cyclic driving and  electrical bias and the denomination "motor" has been coined for them \cite{raul1,raul2,lucas1,lucas2,flor1,l-f,misha,janine}. In Ref. \cite{ludo} a unified framework analogous to the theory of thermoelectricity in linear response was presented to describe motors and thermal machines operating in the adiabatic regime  under small electrical and thermal biases. These ideas were further elaborated in Ref. \cite{bibek} with focus on the thermal machines and their geometric properties.

Adiabatic thermal machines usually operate in permanent contact to the reservoirs. The  driving generates  a
heat pumping mechanism in addition to dissipation into the reservoirs.  In order to have pumping, we typically need to break spacial symmetries as in the case of the rotating screw of Archimedes' machine. In the example of the qubit discussed in Sec. \ref{sec:heat-pump} this is achieved by implementing different kind of couplings with  the two reservoirs and  an asymmetric driving  protocol.

As in the case of the thermodynamical cycles, the possible operations are  heat engine and refrigerator. The  relevant energy fluxes are sketched in Fig. \ref{fig:ther-mach-oper}. In Ref. \cite{bibek,pablo} these two operations are analyzed for the machine based on the setup of Fig. \ref{fig:ther-mach} 
with $\vec{X}(t)\equiv \vec{B}(t)$ in the qubit Hamiltonian and asymmetric coupling to the hot and cold reservoirs as in Sec. \ref{sec:heat-pump}.
In what follows we discuss  the  characterization of this machine in the regime where the driving is slow and in addition the temperature difference $\Delta T= T_{\rm h}- T_{\rm c}$ is small to justify 
a linear-response treatment in $\Delta T/T$, being $T$ the reference temperature.
 From the formal point of view this implies combining the linear response description of the thermal bias 
 with the adiabatic description as presented in Sec. \ref{sec:slow-ther}. One of the quantities of interest in the operation of these machines is the heat transported between the reservoirs, given by Eq. (\ref{qctr}) 
 \begin{equation}\label{qcc}
 Q^{\rm (tr)}=\int_0^{\tau} dt \left\{ \sum_{\ell} \Lambda_{N+1,\ell}(\vec{X}) \dot{X}_{\ell} + \Lambda_{N+1,N+1}(\vec{X}) \frac{\Delta T}{T} \right\},
 \end{equation}
 where the label (tr) highlights that this component of the heat is transported between the two reservoirs, in the sense that $ Q^{\rm (tr)}=Q_{\rm c}= - Q_{\rm h}$. We can distinguish the process of pumping in the first term and the response to the thermal bias in the second one. 
 The other quantity of interest is 
 the work performed by the driving forces. Following 
 Eq. (\ref{p-jalpha}) we can identify a conservative and a non-conservative component, $W=W^{\rm (cons)} + W^{\rm (non-cons)}$. The first component corresponds to calculating 
  Eq. (\ref{cons-qs}) over the cycle, which  explicitly leads to $W^{\rm (cons)} =0$. The non-conservative component is obtained from Eq. (\ref{jp-non-eq-adia-dt}) and reads 
  \begin{equation}\label{ww}
 W=  \int_0^{\tau} dt \left\{ \dot{\vec{X}} \cdot \underline{\Lambda}(\vec{X}) \cdot \dot{\vec{X}} +   \sum_{\ell} \dot{X}_{\ell}\Lambda_{\ell,N+1}(\vec{X})\frac{\Delta T}{T}  \right\},
 \end{equation}
 
  For the case of a qubit driven by the protocol sketched in Fig. \ref{fig:ther-mach}, we have $\vec{X}(t)\equiv \vec{B}(t)=(B_x(t), B_z(t))$ and 
  the Onsager relations of Eq. (\ref{onsager}) for this particular protocol can be shown to satify  $\Lambda_{3,\ell}(\vec{B})=- \Lambda_{\ell,3}(\vec{B}),
  $ for $\ell=1,2$. Hence, introducing the notation $\vec{\Lambda}(\vec{B})=\left(\Lambda_{3,1}(\vec{B}), \Lambda_{3,2}(\vec{B})\right)$, 
  Eqs. (\ref{qcc}) and (\ref{ww}) lead to 
\begin{eqnarray}\label{qw-therm}
Q^{\rm (tr)} &=& \oint d \vec{B} \cdot \vec{\Lambda}(\vec{B}) + \kappa \frac{\Delta T}{T}, \nonumber \\
   W &=&  \int dt \; \dot{\vec{B}} \cdot \underline{\Lambda}(\vec{B}) \cdot \dot{ \vec{B}}- \frac{\Delta T}{T} \oint d \vec{B} \cdot \vec{\Lambda}(\vec{B}).
\end{eqnarray}
As already mentioned, the first term of $Q^{\rm (tr)}$ is purely due to pumping 
$Q^{\rm (pump)} = \oint d \vec{B} \cdot \vec{\Lambda}$
 -- compare with Eqs. (\ref{qpumpal-g}) and (\ref{qpum-ex}) -- while the second term is the  thermal leak  induced by the temperature difference ($\kappa \equiv \int_0^{\tau} dt\Lambda_{3,3}$ is proportional the thermal conductance). 
 In the expression of the work
we can identify in the first component the dissipative contribution when  comparing with Eq. (\ref{ts}) and (\ref{qdis-ex}). The second term describes the mechanism of heat-work conversion and it is the fundamental piece for the device to operate as a thermal machine. It reads
\begin{equation}\label{heat-work}
   - W^{\rm (heat-work)}= \frac{\Delta T}{T} Q^{\rm (pump)}.
\end{equation}
This relation is a generalization of 
Eq. (\ref{carnot-qs}) to the case of a finite-time cycle. This term is the dominant contribution to  $W$ in the case of very slow cycles. In fact,
notice that the second term of $W$ in Eq. (\ref{qw-therm}) is first-order in $\dot{\vec{B}}$ while the first one is second-order. This is in accordance with the idea that dissipation decreases as the evolution becomes closer to the quasi-static limit.

The two operational modes depend on the sign of the pumped heat $Q^{\rm (pump)}$. When the pumped heat flows upstream with respect to 
 the temperature bias $Q^{\rm (pump)}<0$ it can compensate the effect of the 
 heat leak represented by the second one, resulting in $Q^{\rm (tr)} <0$ (heat exits the cold reservoir). The machine works as a refrigerator and the pumping generates an extra amount of work through the mechanism of heat-work conversion, $W^{\rm (heat-work)}>0$.  This situation is sketched in the left panel of Fig. \ref{fig:ther-mach-oper}. The reversed operation corresponds to the heat engine, in which case, $Q^{\rm (pump)}>0$ and the pumped heat flows downstream with respect to the temperature bias. In this case, the heat-work conversion offers a mechanism where the second term of $W$ in Eq. (\ref{qw-therm})
 compensates the dissipative one, enabling the possibility of extracting useful work from the machine. This is sketched in the right panel of Fig. \ref{fig:ther-mach-oper}. 
 
We can now proceed as in Ref.  \cite{pablo}
to express Eqs. (\ref{qw-therm}) in terms of geometric quantities. To this end it is convenient to introduce the change of variables
$s=t/\tau$ and to define 
\begin{equation}\label{alkap}
A=\int_0^1 ds \vec{\Lambda} \cdot \dot{\vec{B}}, \;\;\;\; \;\;\;\;  L^2=\int_0^1 ds \dot{\vec{B}}\cdot \underline{\Lambda} \cdot \dot{\vec{B}}, \;\;\;\; \;\;\;\;  \langle \kappa \rangle =\int_0^1 ds \kappa,
\end{equation}
where the names $A$ and $L^2$ are related to the geometrical meaning of these quantities. In fact $L^2$ is related to the thermodynamic length introduced in Sec. \ref{sec:adia-thermo-length}, while 
$A$ can be expressed in terms of an area 
 by recourse to Stokes' theorem, as discussed in Sec. \ref{sec:heat-pump},
\begin{equation}\label{defa}
A=\oint \vec{\Lambda} \cdot d\vec{B} =\int \int  \left(\nabla_{\vec{B}} \times \vec{\Lambda}\right) \cdot d^2 \vec{B},
\end{equation}
where  the surface in the space of parameters is  enclosed by the boundary  defined by the protocol. Hence, this quantity can be interpreted as an area 
in the parameter space
weighted by a Berry-type curvature $\left(\nabla_{\vec{B}} \times \vec{\Lambda}\right)$. Therefore,
 Eqs. (\ref{qw-therm}) reduce to 
\begin{equation}\label{qw-therm-g}
Q^{\rm (tr)} = A + \tau \langle\kappa\rangle \frac{\Delta T}{T}, \;\;\;\;\;\;\;\; \;\;\;\; \;\;\;\;\;\;
   W =  \frac{L^2}{\tau}   - \frac{\Delta T}{T} A.
\end{equation}

 \subsection{Optimal adiabatic heat engine in linear response}
 The  two components of the net work per cycle presented in
 Eq. (\ref{qw-therm-g})
 can be expressed as geometrical properties in the parameter space. In fact, the dissipative one depends on the metric and can be related to the thermodynamic length as discussed in Sec. \ref{sec:adia-thermo-length}, while the heat-work term depends on the  "Berry-curvature" as shown in Eqs. (\ref{defa}). In Ref. \cite{pablo}
 these properties are used to optimize the performance of a qubit following the protocol of Fig. \ref{fig:ther-mach}. We now summarize the procedure for the case of the heat engine operation. The starting point is the definition of the power and the efficiency in terms of the geometric quantities introduced before. This reads
 \begin{eqnarray}\label{peeta}
 P^{\rm (he)}&=&-\frac{W}{\tau} = \frac{\Delta T}{T} \frac{A\left[1- (\tau_D/\tau)\right]}{\tau},\nonumber \\
 \eta&=&-\frac{W}{Q^{\rm tr}} =\eta_C \frac{1-(\tau_D/\tau)}{1+(\tau_D/\tau_{\kappa})}
 \end{eqnarray}
 with the definitions
 \begin{equation}
 \tau_D= \frac{T}{\Delta T} \frac{L^2}{A}, \;\;\;\;\;\; \;\;\;\;\;\; \;\;\;\;\;\;\tau_{\kappa}=\frac{T}{\Delta T} \frac{A}{\langle \kappa \rangle}.
 \end{equation}
 It is easy to calculate the optimal duration of the cycle in order to 
  maximize the power and the efficiency. The results are, respectively, 
 \begin{equation}
 \tau_P=2 \tau_D, \;\;\;\;\;\; \;\;\;\;\;\; \;\;\;\;\;\;\tau_{\eta}= \tau_D+\sqrt{\tau_D(\tau_D+\tau_{\kappa})},
 \end{equation}
 which upon substituting in Eq. (\ref{peeta}) lead to the expressions for the optimal expressions for the power and the efficiency. Particularly interesting is the expression for the maximum power, which reads
 \begin{equation}
     P^{\rm (he)}_{\rm max}=\frac{1}{4} \frac{\Delta T^2}{T^2} \frac{ A^2}{L^2}.
 \end{equation}
 We see that the problem of optimizing the protocol in the parameter space 
 in order to maximize the power developed by the heat engine reduces to the optimization of the ratio between an area and a length in a space with a non-trivial metric.
 In the case of the Euclidean metric the solution is a circle, as is well-known since the time of the ancient Greeks.
 Remarkably, for an arbitrary metric this is still an open problem in geometry, which known as the {\em isoperimetric}  or {\em Cheeger} problem \cite{iso,cheeg1,cheeg2}. 
 
\begin{figure}
    \centering
    \includegraphics[width=0.5\textwidth]{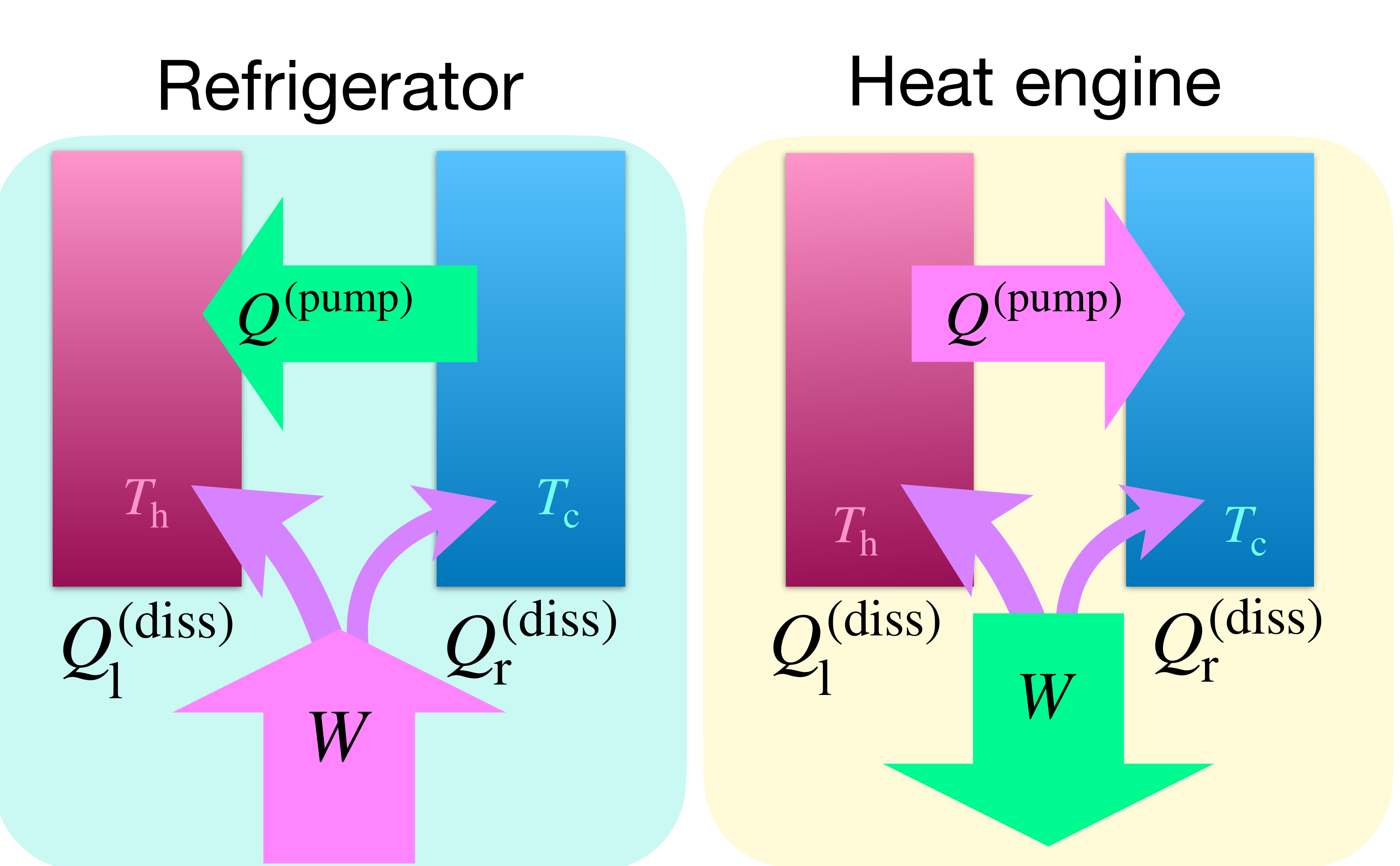}
    \caption{
   Sketch of the operational modes of the thermal machine}
    \label{fig:ther-mach-oper}
\end{figure}

 The geometrical properties introduced before are very useful to the design of optimal protocols in qubit thermal machines. In Ref. \cite{pablo} the problem of the driven qubit asymmetrically coupled to the two thermal baths illustrated in Fig.  \ref{fig:ther-mach-oper} is analyzed in detail. 
 Some results regarding the operation without thermal bias are covered in 
 Sec. \ref{sec:adia-heat-qubit}, including the calculation 
 the pumped heat, $Q^{\rm (pump)}$, and 
 the dissipated energy because of the driving, $W^{\rm (diss)}$. In the linear response regime with respect to $\Delta T$, these quantities define $A$ and $L^2$, respectively, through Eqs. (\ref{alkap}). 
 Recalling  Eq. (\ref{landauer}), we notice that the optimal protocol regarding the maximum $A$ corresponds to a  path evolving from the origin
 along the axis $B_x>0$, followed by a quarter 
 of circumference  of infinite radius and closing  along the axis $B_z$ circulating from infinite to the origin.  Those protocols are, however, not optimal regarding dissipation. The numerical analysis of the optimal solution of this  isoperimetric  problem was found to be elipses in the $(B_x,B_z)$ plane, circulated at
 a $\vec{B}$ dependent speed $\dot{\vec{B}}(\vec{B})$. The latter  is defined in order to get a constant dissipation rate at every point of the trajectory, which
 is the condition leading to a minimum thermodynamic length for a given path, Eq. (\ref{l}), as discussed in Sec. \ref{sec:en-diss}.

\section{Batteries}\label{sec:bat}

Quantum batteries is another subject of active investigation \cite{bat1,bat2,bat3,bat4,bat5,bat6,bat7,bat-mon}. 
These are quantum systems with a discrete-level spectrum manipulated by time-dependent processes in a way that they can store or deliver energy.
The study of quantum batteries was triggered by Ref. \cite{bat1} and has a significant development in the context of qubit systems. The  goal of that paper  is the calculation of the amount of work that can be extracted from a small quantum mechanical system  which
temporarily stores energy: the battery. Previously,  the concept of "ergotropy" had been introduced as the 
 maximum work that can be unitarilly extracted  from a given quantum state given a reference Hamiltonian \cite{alla2004}. Refs. \cite{bat1, bat11} focus on the calculation of the maximum extractable work in the framework of several subsystems and conclude that, in general,  entangled unitary operations extract  more work than independent ones. Refs. \cite{bat2,bat22,batdar} focus on the complementary problem of charging the battery.
 In all these studies the models are purely closed systems, while more complex setups including the baths and dissipation processes are considered in Refs. \cite{bat4,bat6,bat7,bat44,bat47,quach}.
 
A proposal of quantum battery with high charging power in platforms of cQED systems was presented in Ref. \cite{bat3} and further elaborated in Ref. \cite{bat46}. The goal is to extend the comparison of the operation of $N$ qubits in parallel vs the operation of the $N$ entangled qubits as in previous works. In this study each unit is defined by a single qubit coupled to a resonator mode, while the entangled system corresponds to the $N$  qubits coupled to a common resonator mode. The considered model for  the collective configuration is described by Dicke Hamiltonian \cite{dicke}
\begin{equation}
    H_{\lambda}(t) =
    \hbar \omega_c a^{\dagger} a + \omega_0 S_z + 2 \lambda(t)\omega_c S_x \left(a^{\dagger} + a\right),
\end{equation}
with $S_j=\hbar/2 \sum_{\ell=1}^{N} \sigma_\ell^j$, which describes the $N$ qubits coupled to the resonant mode of frequency $\omega_c$ when
$\lambda(t)\neq 0$. The parallel configurations consist of $N$ copies of this Hamiltonian for a single qubit. The battery is operated in a cycle, that is ruled by the time-dependent protocol followed by the coupling term $\lambda(t)$. The latter is defined by charging, storage and discharging times
(respectively $\tau_c, \; \tau_s,\;\tau_d$),
as follows, 
\begin{equation}\label{protob}
\lambda(t)=\begin{cases}
\overline{\lambda}&, \;\;\;\;\; 0<t< \tau_c,\\
0&,\;\;\;\;\; \tau_c\leq t \leq \tau_c + \tau_s ,\\
\overline{\lambda}&, \;\;\;\;\; \tau_c + \tau_s <t< \tau_c + \tau_s +\tau_d.
\end{cases}
\end{equation}
The steps are: (1) The initial state is $|\psi^N(0)\rangle = |N\rangle \bigotimes |g,\ldots,g\rangle$, which corresponds to the $N$ excitations in the resonator and
the ground state of the $N$ decoupled qubits. The coupling to the resonator is switched on during an interval of duration $\tau_c$. (2)  The resonator decouples for a time-interval of duration $\tau_s$. (3) The resonator is coupled again for the discharging step of duration $\tau_d$. The conclusion of these studies is that the charged energy as well as the charging power is enhanced in the correlated array in comparison to the parallel one. The fraction of energy stored in the battery that can be extracted in order to perform thermodynamic work, however is reduced and depends on the details of the preparation of the initial state.

Another type of quantum battery in the context of cQED was proposed in Ref. \cite{bat-mon}. The setup consists in 
a single qubit coupled through a time-dependent protocol similar to  Eq. (\ref{protob}) with a many-modes waveguide. In this setting, a resonant mode in the waveguide acts as a battery while the qubit is the auxiliary system.  A photon beam with a frequency resonant with the qubit ($\omega_0$) is injected into the waveguide with an
 input power  $P_{\rm in}=\hbar \omega_0 \dot{N}_{\rm in}$, being $\dot{N}_{\rm in}$ the incoming flux of photons.
Because of the coupling with the qubit, there is a reflected pulse with power 
$P_{\rm out}=\hbar \omega_0 \dot{N}_{\rm out}$, being $\dot{N}_{\rm out}$ the outgoing flux of photons. The considered Hamiltonian to describe the qubit and its 
coupling to the beam  reads
\begin{equation}\label{hb1}
    H(t)=-\frac{\hbar \omega_0}{2} \sigma_z + i \hbar \sqrt{\gamma(t) \dot{N}(t)} \left(\sigma_- e^{i\omega_0 t} - \sigma_+ e^{-i \omega_0 t}\right),
\end{equation}
where the coupling protocol is determined by $\gamma(t)$.
The output power is calculated by solving the master equation for the qubit coupled to the resonator and the dynamical equation for the reflected outgoing field,
\begin{equation}\label{hb2}
  b_{\rm out}(t)= b_{\rm in}(t)+\sqrt{\gamma}\sigma_-(t), 
\end{equation}
from where it is possible to evaluate $E= \mbox{Tr}[\rho(t) H(t)]$ and 
\begin{equation}\label{hb3}
    P_{\rm out}(t) = P_{\rm in}(t)- \dot{E}(t).
\end{equation}
        The charging process corresponds to take advantage of the spontaneous emission of the qubit ($\dot{E}(t)<0$) injecting extra photons into the waveguide, hence charging the battery. This device was experimentally realized in an implementation of a Maxwell demon \cite{bhuard}, which will be discussed in 
        Sec. \ref{sec:max-dem}.

\section{Steady-state thermal transport}\label{sec:statio}
\begin{figure}
    \centering
    \includegraphics[width=\textwidth]{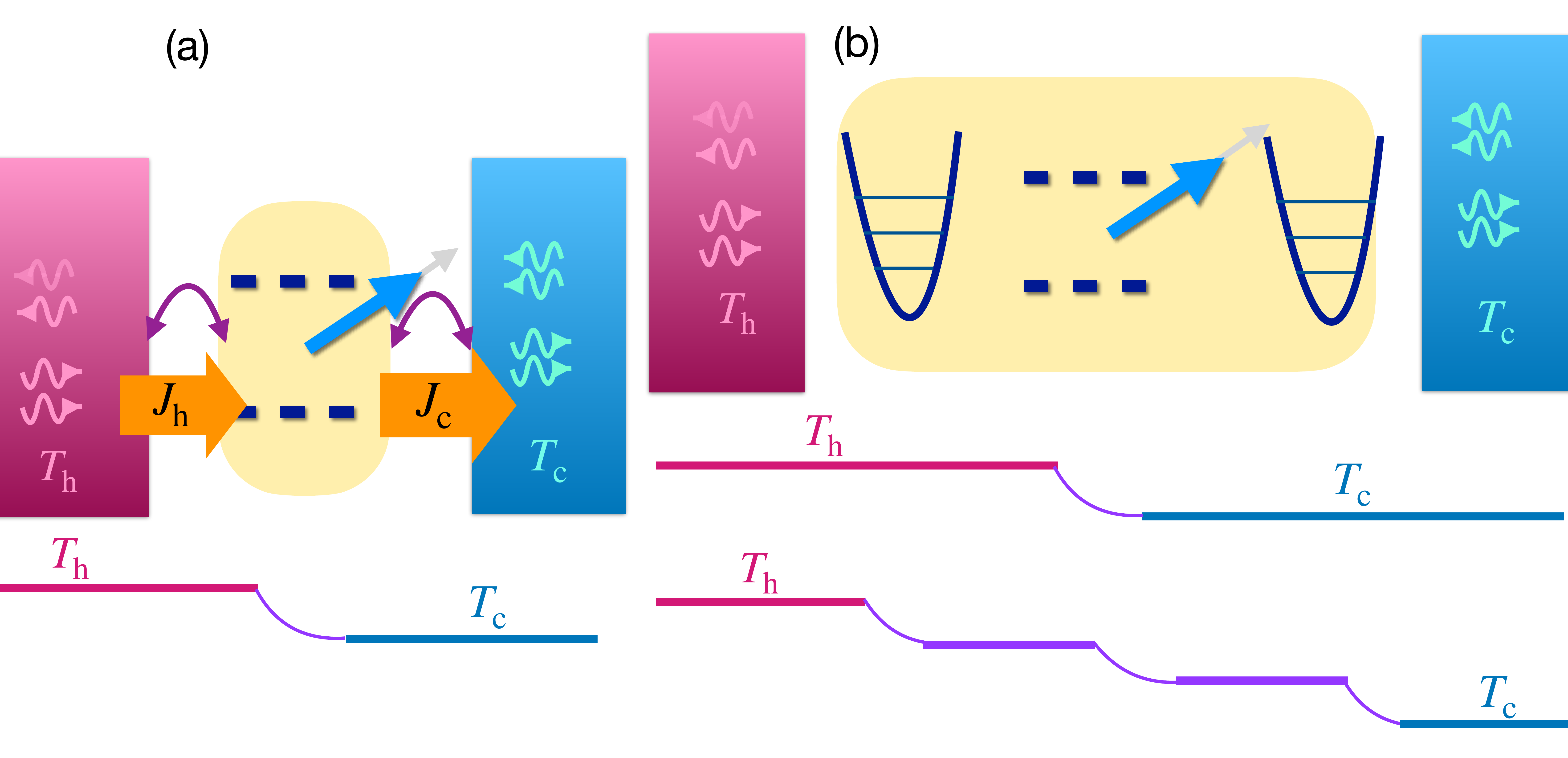}
    \caption{
    Illustration of the thermal transport process when: (a) the qubit is contacted to reservoirs at different temperatures at fixed $\vec{B}$ and (b) the qubit is contacted to reservoirs at different temperatures through resonators. Below each configuration, possible 
    temperature profiles are indicated.}
    \label{fig:transport}
\end{figure}
\subsection{General considerations}
When the qubit is placed in contact to two reservoirs with different temperatures without any driving mechanism,  heat flows through it from the reservoir at high temperature $T_{\rm h}$ to the cold one with temperature $T_{\rm c}$. A sketch is presented in Fig. \ref{fig:transport}.
In this section we review two aspects of the heat transport through this device. The first one is related to the degree of coupling between the qubit and the reservoir. The second one is related to the structure of the reservoir and non-linear effects in the thermal transport.

\subsection{The effect of strong correlations in linear response}\label{sec:lin}
As mentioned before, the Hamiltonian for the two-level system coupled to a bath of harmonic oscillators is equivalent to the spin-boson model \cite{weiss,caldeira-legget2}, whose equilibrium properties  received
a significant attention some time ago. This model has a quantum phase transition at zero temperature depending on the degree of coupling between the system and the reservoir and the spectral properties of the latter (ohmic, subohmic or superohmic) \cite{bm,chack}.  
This quantum phase transition is a consequence of the many-body correlations in this system.


In the context of electron transport, strong correlations are crucial to generate the mechanisms of Coulomb blockade and Kondo, which have a significant impact in the transport properties at low temperatures \cite{coulomb,kondo}. It is, therefore, interesting to investigate similar features in qubit systems.  Many-body effects in the thermal transport of the spin-boson model have been investigated in several works \cite{segal1,ruokola,chen,segal2,yang,wang,taylor}. Refs. \cite{saito1,yamamoto} present a  systematic study of the low-temperature
behavior of the thermal conductance beyond weak coupling between qubit and reservoir, on the basis of several many-body methods and a numerical quantum Monte-Carlo technique. The derived equation for the heat current for the qubit model with $\vec{B}=(B_x,0,0)$ in Eq. (\ref{driv-qubit}) and couplings $\vec{g}_k=(0,0,g_{k}^z)$ from Schwinger-Keldysh non-equilibrium Green's-function technique \cite{meir-wingreen,jauho} reads
\begin{equation} \label{saito-j}
J_{\rm c} = - J_{\rm h} = \frac{ \alpha \gamma}{8}\int d\varepsilon \; \varepsilon \;\mbox{Im} \left[\chi(\varepsilon) \right] I(\varepsilon)
\left[n_{\rm h}(\varepsilon)- n_{\rm c}(\varepsilon) \right].
\end{equation}
It is assumed 
\begin{equation}\label{spec}
\Gamma_{\nu}(\varepsilon)=\alpha_{\nu} I(\varepsilon), \;\;\nu={\rm h,\;c}, \;\;\;\;\;\;\;\;\;\;\;\;\;\;
I(\varepsilon)=2 \frac{\varepsilon}{\hbar} \left(\frac{\varepsilon}{\varepsilon_c}\right)^{s-1}e^{-\varepsilon/\varepsilon_c}, 
\end{equation}
being $\omega_c$
an energy cutoff, $\alpha=\alpha_{\rm c}+\alpha_{\rm h}$, and $\gamma = 4 \alpha_{\rm c} \alpha_{\rm h}/\alpha^2$. The function
\begin{equation}\label{chi}
    \chi(\varepsilon)=-\frac{i}{\hbar}\int_0^t\langle dt \left[ \sigma_z(t), \sigma_z(0)\right]\rangle \;\;
    e^{i \frac{\varepsilon}{\hbar }t},
\end{equation}
is the spin susceptibility. In the linear response regime, the expectation values are calculated with respect to the the many-body equilibrium states with the reservoirs at the mean temperature $T=(T_{\rm c}+T_{\rm h})/2$. Hence, $\chi(\varepsilon)$ is a function
of $T$. 
We see that Eq. (\ref{saito-j}) has the structure of Eq. (\ref{lan-bu}), in spite of the fact that this 
Hamiltonian is non-bilinear but has many-body interactions. It is possible to define a transmission function
${\cal T}(\varepsilon, T)=\alpha \gamma \mbox{Im} \left[\chi(\varepsilon) \right] I(\varepsilon)/8$.
Given this function, it is easy to derive in this regime  the following expression for the thermal conductance, $G_{\rm th} \sim {\cal T}(0,T) G_Q$,
such that $J_{\rm c} = G_{\rm th} \Delta T$ with $G_Q=T_{\rm c}(\pi k_B)^2/(3h) $ being the quantum of thermal conductance, which defines the limit for the quantum thermal transport through a single-channel bosonic or fermionic system \cite{pendry, beck1,beck2}.

Ref. \cite{saito1} focuses on the Kondo signatures at low temperature in this regime, while in Ref. \cite{yamamoto} a more detailed analysis including the comparison of the different methods to evaluate the susceptibility defined in Eq. (\ref{chi}) is presented. A crucial effect of the coupling to the bath is the renormalization of the parameter $B_x$ of the Hamiltonian to an effective value $\Delta_{\rm eff}$ which depends on $\alpha$ and on the bare value of $B_x$. This parameter represents the effective superposition of the two basis states  of the qubit in the presence of the environment and defines the relevant energy scale of the problem.
The approximations considered  in Ref. \cite{yamamoto} are the "sequential tunneling", which corresponds to calculating $\chi(\varepsilon)$ with the reduced density matrix calculated with the Lindblad master equation (\ref{masterrho}). The other  one considered is the so called 
"co-tunneling" approximation, which corresponds to including in the master equation higher order processes in the contact term by recourse to perturbation theory. The third method is the so called "NIBA" (non-interacting-blip approximation), which is described by a rate equation 
derived after integrating out the bosonic bath with the Feynman-Vernon functional technique and dropping some terms \cite{leggett}. The latter is a non-perturbative technique, and the approximation is based on neglecting some family of high-order processes. The results are compared with numerically exact calculations with continuum time quantum Monte Carlo. 
A summary  of the main results is presented below:
\begin{itemize}
    \item {\em Ohmic bath, $s=1$.} The thermal conductance below the critical value $\alpha_c=1$ changes the behavior as a function of $T$. 
    At low temperatures ($k_B T \ll  \Delta_{\rm eff}$), the numerical results agree well with those of the 
    co-tunneling calculation. In this regime, the thermal conductance is always proportional to $T^3$. This universal behavior is typical of the Kondo effect.
    At moderate ($k_B T \sim  \Delta_{\rm eff}$) and high temperatures ($k_B T \gg \Delta _{\rm eff}$), the numerical results deviate from the co-tunneling formula and agree well with NIBA. The thermal conductance  obtained by this method is proportional to $T^{3 - 2\alpha}$ at low temperatures. Sequential tunneling fails to predict the low-temperature behavior and for high temperature it is valid only in the limit
    $\alpha \ll \alpha_c$.
    For $\alpha > \alpha_c$, the renormalized parameter $\Delta_{\rm eff}$ tends to zero and the system is localized in one of the states
    at low temperatures. The behavior of the thermal conductance in this regime is well described by NIBA. 
    \item {\em Sub-Ohmic bath, $s<1$.} In this case, the critical value for the localized-to-delocalized transition, $\alpha_c$, is a function of $s$. Below $\alpha_c$, at moderate and high temperatures, the numerical results agree well with the NIBA.  At low temperatures ($k_B T \ll  \Delta_{\rm eff}$), the numerical results agree well with the co-tunneling formula, showing
$T^{ 2s + 1}$-dependence. The sequential-tunneling formula cannot be applied to the sub-ohmic case.
\item {\em Super-Ohmic bath $s>1$.} At low temperatures ($k_BT \ll \Delta_{\rm eff}$), the numerical results agree with the co-tunneling description and show
$T^{ 2s+1}$-dependence, regardless of the strength of the system-reservoir coupling, while the agreement is better with NIBA at higher temperatures.
\end{itemize}

\subsection{The thermal-bias drop}\label{sec:drop}
One of the fundamental issues in quantum  electron transport induced by a voltage bias applied at the reservoirs is which is the behavior of the voltage drop along the quantum system placed between the two reservoirs at different chemical potentials  and to what an extent such details affect the transport \cite{sdatta}. These aspects are mainly relevant  beyond linear response in the bias and  similar questions can be posed regarding the temperature bias.  Deep inside each of the macroscopic reservoirs the temperature has a well defined value. When the small-size quantum system is composed of  several subsystems which independently intervene  in the contact with the  reservoirs at 
different temperatures it may happen that some of them tend to thermalize with the neighbouring reservoir. This would cause the 
temperature drop to concentrate inside the quantum system. Another possibility is an abrupt drop at the contact (Fig. 
\ref{fig:transport}.b). These questions are fundamental and very timely after recent experimental results \cite{ronzani,recti}, where thermal transport is investigated in a qubit system coupled with resonators which are in turn coupled with reservoirs. The experiments and the interpretations will be discussed in Sec. \ref{sec:exp-ther-trans}. The Hamiltonian for the considered setup reads
\begin{eqnarray}
    H&=&  H_{\rm qubit} + \sigma_z\sum_{\nu={\rm c, h}}\left[ g_{\nu }\left(b_{\nu}^{\dagger} + b_{\nu}\right)+ 
    \hbar \Omega_{\nu} b_{\nu}^{\dagger}  b_{\nu}\right] \nonumber \\
    & & + \sum_{\nu={\rm c, h}, k}\left[ g_{\nu,k}\left(b_{\nu}^{\dagger} + b_{\nu}\right) \left(a_{\nu,k}^{\dagger} + a_{\nu,k}\right)+ 
    \hbar \omega_{\nu,k} a_{\nu,k}^{\dagger}  a_{\nu,k} \right],
\end{eqnarray}
where the first line describes the qubit coupled to the resonators, while the second one describes the resonators coupled to the cold and hot reservoirs.

By analogy with electron transport \cite{sdatta}, we expect the behavior of the thermal drop to be strongly dependent on the details of the system, like the degree of coupling between the different subsystems in comparison with the coupling with the reservoirs and the presence of  many-body interactions, which may generate a mechanism of local or global thermalization.  The transmission properties of a
qubit coupled to resonators which are coupled to reservoirs of many-harmonic oscillators like the experimental configuration of Refs. \cite{ronzani,recti} is very special from the theoretical point of view. This is because it is technically  simple to  include the resonators into the description of the reservoirs by defining a spectral function $\Gamma(\varepsilon)$ for that combined system in order to substitute Eq. (\ref{spec}). This is precisely the procedure
followed in Refs. \cite{grifoni, yama-kato}. The result is Eq. (\ref{spec}) with 
\begin{equation}\label{spec1}
I_{\nu}(\varepsilon)= 2 \alpha_{\nu} \varepsilon \frac{(\hbar \Omega_{\nu})^2}{(\hbar \Omega_{\nu}^2-\varepsilon^2)^2+(2 \gamma_{\nu}\varepsilon)^2},
\end{equation}
where $\alpha_{\nu}$ depends on the couplings $g_{\nu}, g_{\nu_k}$ while $\gamma_{\nu}$ depends on the coupling $g_{\nu,k}$. We see that the functional dependence is a Lorentzian in contrast with the power-law behavior of Eq. (\ref{spec}). In these references, the heat current through the qubit coupled to the effective
reservoirs described by Eq. (\ref{spec1}) is calculated within the NIBA approximation. Notice that this description is equivalent to assume that the temperature drop takes place at the qubit, while the resonators are perfectly thermalized with the neighboring reservoirs. 
In Ref. \cite{yama-kato} the thermal current is analyzed in linear response in the temperature bias. As expected after the discussion of
Sec. \ref{sec:lin} the structure of Eq. (\ref{spec1}) introduces non-trivial and non-universal features in the behavior of the thermal conductance as a function
of the mean temperature. Similar studies in more complex configurations were presented in Refs. \cite{xu,matteo}.
The behavior for  large temperature bias remains an open problem.

In configurations where the couplings of the resonators with  the reservoirs are  weak, this problem may be related to the discussion of the validity of the local master equation vs the global one, which has been the subject of many studies \cite{chiara,hewgill,levy,hofer,rivas,rivas1,vadimov}. The question in most of this literature
 is the appropriateness of  using a local basis to express the reduced density matrix within the framework of master equations of Lindbladian nature, which are derived in perturbation theory with respect to the couplings with the reservoirs. Local master equations are valid when also the coupling between the different subsystems is perturbative, while in the global case, the density matrix is expressed in the eigenbasis of the system 
 composed of several connected parts.

\subsection{Rectification}\label{sec:recti}

For some time now thermal rectification in nano-scale systems is a subject of great interest theoretically and 
experimentally in phononic structures
\cite{therrect1,therrect2,therrect3,therrect4,therrect5,therrect6,therrect7,therrect8,therrect9}, spin chains \cite{spin1,spin2,spin3}, quantum dots
\cite{ruok,aligia,palafox} and superconductors \cite{super1,super2}. 
Thermal rectification or thermal-diode effect means that the thermal transport depends, not only on the magnitude of the temperature bias, but also 
on its direction.  Typically, the origin of such an effect is the existence of non-linearities and asymmetries in the setup and the regime
where it takes place is beyond linear response. 

In the context of qubits coupled to photonic baths thermal rectification was predicted in
Ref. \cite{segal-nitzan1} and this effect was recently  experimentally confirmed in
Ref. \cite{recti} in the configuration where the qubit is coupled to resonators, which are in turn coupled to thermal baths at different temperatures.
This experiment is briefly discussed in Sec. \ref{sec:exp-ther-trans} and already motivated several theoretical works
\cite{diaz, portugal, bibek-rect, giazotto}. This a  difficult theoretical problem which requires the proper treatment of many-body physics along with far from equilibrium conditions, which must be solved approximately. Unlike the case of the thermal conductance, this problem cannot be solved exactly numerically because of its far-from-equilibrium nature. 

In Ref. \cite{bibek-rect} rectification is studied in an anharmonic quantum oscillator coupled to two bosonic thermal baths with Ohmic density of states at different temperatures.
In the limit of very strong anharmonicity the system reduces to a  qubit. The problem is solved with different techniques:   master equation  taking cotunneling into account,  nonequilibrium Green's functions using the Majorana representation for the spin, and exact calculations based on Feynman-Vernon path-integral approach. The calculation based on non-equilibrium Green's functions \cite{meir-wingreen,jauho} similar to the one leading to Eq.
(\ref{saito-j}) and considering the 
hybridization functions of the two reservoirs, defined in Eq. (\ref{spec}),  proportional one another  leads to the following expression for the heat current
in the left reservoir
\begin{equation}\label{non-lin}
J(\Delta T)= 
\int d \varepsilon \; \varepsilon \; {\cal T}(\varepsilon,T, \Delta T) \; \left[n_{\rm l}(\varepsilon)-n_{\rm r}(\varepsilon)\right],
\end{equation}
with $T_{\rm l/r}= T \pm \Delta T/2$. The convention is such that for $\Delta T>,<0$ the heat flow exits/enters this reservoir. 
Importantly, although Eq. (\ref{non-lin}) has a structure which resembles  Eq. (\ref{lan-bu}), the dependence of the transmission function on
$\Delta T$ breaks the symmetry  with respect to the exchange of the hot and cold reservoirs. The explicit expression of this function depends on the regime and the method of calculation.  
The rectification coefficient is defined as
\begin{equation}
    R=\frac{J(\Delta T)+J(-\Delta T)}{J(\Delta T)-J(-\Delta T)},
\end{equation}
where no (perfect) rectification corresponds to $R=0, (1)$, respectively. For weak coupling this parameter is found to be
 upper bounded by $\lambda (T_{\rm h}-T_{\rm c})/(T_{\rm h}+T_{\rm c})$, where  
$\lambda= (\Gamma_{\rm l}-\Gamma_{\rm r})/(\Gamma_{\rm l}+\Gamma_{\rm r})$ is the asymmetry parameter of the bath spectral functions.
For strong coupling and low temperatures there is a strong dependence on the type or coupling to the reservoirs ($\vec{g}_{\rm l}, \vec{g}_{\rm r}$).
Interestingly, all the methods agree in the limit where the energy gap between the two levels is larger than the temperature.
However, this is the regime of lower rectification. For low temperatures, large values of $R$ are found in this setup.

In Ref. \cite{giazotto}  and array of two qubits coupled each of them to two LCR circuits with different temperatures 
was considered. The Hamiltonian describing this system is
\begin{equation}
    H_{\rm qubits}=\sum_{\nu={\rm c,h}}\left[ H_{\nu}+ \sum_{j=1,2} \left( \vec{B}_j \cdot \vec{\sigma}_j + H_{\nu,j}\right)\right]+
    \gamma \sigma_1^z \; \sigma_2^z,
\end{equation}
where $H_{\nu}$ are the Hamiltonians of the baths and  the terms $H_{\nu,j}$ represent the coupling of each qubit to the two baths.
The problem is solved by master equations 
with transition rates that depend on the temperature through the Johnson-Nyquist noise spectral function of the LCR circuits \cite{ojanen}. 
A high rectification ratio  is found in this configuration for asymmetric couplings to the hot and cold reservoirs.

\section{Superconducting qubits and cQED}\label{sec:super}
Superconducting devices with Josephson junctions and capacitances may be designed to realize qubits. This is a very well established platform  which has been proposed some time ago. Several architectures have been implemented and improved, which allow for the control and manipulation of the quantum states  by recourse to magnetic fluxes and gate voltages.  The working principles and details
on the operation of these devices have been covered in at least four review articles \cite{makhlin,devoret1, devoret2, blais1}.  Here, we briefly review the Cooper pair box and the transmon architectures.
Besides variations of these configurations like the charge and flux qubits, there are  recent developments towards the realization of qubits in superconducting Josephson junctions by using the Andreev bound states in the junction \cite{urbina, devoret}. Another intensively explored platform based on superconductors is the topological one, in which case the qubits are realized in the Majorana zero modes localized at the ends of superconducting wires with spin-orbit coupling and with an applied magnetic field \cite{majorana1,majorana2}. 


\subsection{Architectures}
\begin{figure}
    \centering
    \includegraphics[width=0.8\textwidth]{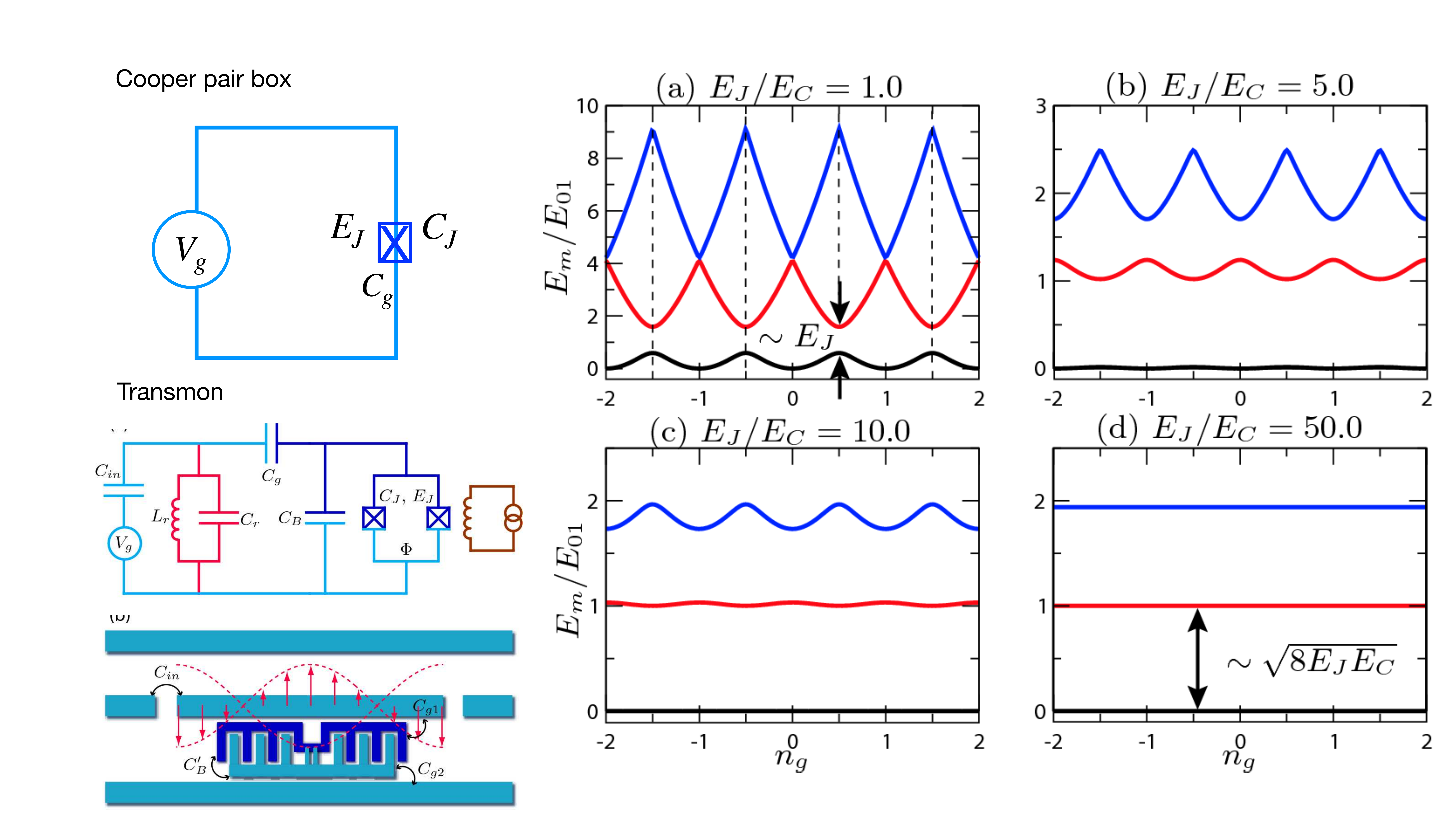}
    \caption{Sketch of the Cooper pair  box (top left) and transmon (bottom left) circuits. The spectrum for different ratios of $E_J/E_C$ is shown in the right panels. (Adapted from Ref.  \cite{koch}).}
    \label{fig:fig1}
\end{figure}

The simplest configuration to realize a superconducting qubit is the {\em Cooper pair box} (CPB) which is sketched in Fig. \ref{fig:fig1} (see left top panel).  It consists in a Josephson junction between two superconductors with a low capacitance $C_{\rm J}$ and a phase bias 
$\varphi$. The knobs to manipulate the device are provided by the magnetic flux and a gate voltage $V_{\rm g}$, contacted to one of the superconductors by means of a capacitance $C_{\rm g}$. The  superconducting gap $\Delta$ is sufficiently large to  prevent the quasiparticle tunneling through the junction and the charge transport takes place  only by the tunneling of Cooper pairs. The Hamiltonian describing this system reads
\begin{equation}\label{qubit}
H_{\rm Junction}= H_{\rm ch} + H_{\rm J}= 4 E_{\rm C} \left(\hat{N}-N_{\rm g}\right)^2- E_{\rm J}\cos\left( \hat{\varphi} \right),
\end{equation}
where the first term describes the capacitive effects while the second one is associated to the Josephson effect.  $\hat{N}$ is the operator describing the difference in the number of Cooper pairs contained by each superconductor, $N=N_1-N_2$ (where $N_l$ is the number of Cooper pairs in the superconductor $l$) and
$N_{\rm g}=C_{\rm g}V_{\rm g}/2e$ is the excess charge controlled by the gate voltage. The corresponding charging energy is
$E_{\rm C}=e^2/2\left(C_{\rm g} + C_{\rm J}\right)$. In this representation, $N$ is the quantum mechanical conjugate of the operator $\hat{\varphi}$, $\hat{N}=-i \hbar \partial/(\partial \varphi)$. 
This Hamiltonian can be written in the basis of eigenstates of $\hat{N}$ as follows
\begin{equation}\label{cpb}
H_{\rm CPB}=\sum_N \left\{ 4 E_{\rm C} \left(N-N_{\rm g} \right)^2|N\rangle \langle N|-\frac{E_{\rm J}}{2} \left(|N\rangle \langle  N+1 |+ |N +1 \rangle \langle  N | \right)\right\},
\end{equation}
where the first term describes the charging energy, while the second one described the tunneling of Cooper pairs through the junction.
The lowest eigenenergies of this Hamiltonian as functions of $N_{\rm g}$ are shown in Fig. \ref{fig:fig1} (a)-(d).  
For $E_{\rm C} \gg E_{\rm J}$, the spectrum is dominated by the first term of the Hamiltonian. The eigenstates have a large component with fixed $N$, except for $N_{\rm g}$ close to half-integer values, corresponding to the degeneracy points of the charging energy. At these points the second term of the Hamiltonian generates an avoided crossing between two adjacent states. The qubit state is realized when focusing, for instance, at gate voltages leading to $N_{\rm g} \simeq 1/2$ which correspond to values close to the degeneracy point of
$N=0,1$ of the charging term. Close to this point,  the previous Hamiltonian can be expressed in the form of Eq. (\ref{hqubit0}) as follows,
\begin{equation}\label{qbit}
H_{\rm qubit}=-B_z \sigma^z-B_x \sigma^x = -B \left[\cos(\theta)  \sigma^z+ \sin(\theta) \sigma^x\right],
\end{equation}
with the definitions 
\begin{equation} \label{bxz}
B_z =  2 E_{\rm C} \left(1-2 N_{\rm g} \right), \;\;  B_x=  E_{\rm J}/2,\;\; B =\sqrt{B_x^2+B_z^2},\;\; \theta= \mbox{tan}^{-1}\left(B_x/B_z \right),
\end{equation}
 while $\sigma^z,\; \sigma^x$ are Pauli matrices expressed in the basis $|\uparrow \rangle \equiv  (1,0)^T$ and
$|\downarrow \rangle \equiv  (0,1)^T$ where $0,1$ are the corresponding values of $N$  and $T$ denotes the transpose operation.

This Hamiltonian has a tunable parameter $B_z$, which can be controlled by the gate voltage that sets $N_{\rm g}$. This qubit configuration can be improved by connecting the two superconductors in a loop.  In this way, the effective Josephson coupling  in Eq. (\ref{qubit}) is given by \cite{makhlin}
\begin{equation}\label{loop}
H_{\rm J}= \left(E_{\rm J_1}+ E_{\rm J_2}\right)\cos(\pi \Phi/\Phi_0)\sqrt{1+d^2\mbox{tan}^2\left(\pi \Phi/\Phi_0 \right)} \cos\left( \hat{\varphi} - \varphi_0 \right),
\end{equation}
where $E_{\rm J_1}=\alpha E_{\rm J_2}$ and $E_{\rm J_2}$ are the tunneling amplitudes for the Cooper pairs at the junctions, $d=(1-\alpha)/(1+\alpha)$ and $\varphi_0=d \;\mbox{tan}\left(\pi \Phi/\Phi_0 \right)$. The effective Hamiltonian for the qubit is given by Eq. (\ref{qubit}) with the Josephson term   modulated by the magnetic flux 
$\Phi$ through the loop in units of the flux quantum $\Phi_0=hc/2e$ as defined in Eq. (\ref{loop}). In this way, not only the gate voltage but also the magnetic flux can be 
used as control parameters to manipulate the effective qubit Hamiltonian of Eq. (\ref{qbit}).

Another popular architecture is {\em the transmon}.
This circuit has been proposed in Ref. \cite{koch} and has the advantage with respect to the CPB of being less sensitive to charge noise. It is sketched in Fig. \ref{fig:fig1} (see bottom left). One of the main differences with respect to the CPB is that this circuit hosts two Josephson junctions shunted by an additional capacitance $C_{\rm B}$. The Hamiltonian for this circuit is also given by Eq. (\ref{qubit}), with
$E_{\rm C}=e^2/2\left(C_{\rm g} + C_{\rm J}+ C_{\rm B} \right)$, while $\hat{\varphi}$ denotes the total phase difference between the two superconductors, which is set by the magnetic flux $\Phi$. The transmon operates in the opposite limit of the CPB, namely $E_{\rm C} \ll E_{\rm J}$, in which case the low-energy spectrum of the Hamiltonian of Eq. (\ref{qubit}) consists in approximately constant energy levels as functions of the offset charge $N_{\rm g}$, as shown in Fig. \ref{fig:fig1} (d).

\subsection{Circuit quantum electrodynamics}
We have previously
presented some architectures for superconducting circuits, and shown that it can effectively behave as  a two-level quantum system. As mentioned before, when we also consider the
surrounding circuit we have
the scenario of a qubit embedded in 
an electrodynamic cavity.
The role  of the cavity is played by the combination of capacitive and inductive circuit elements (resonators). In fact, the dynamics of a transmission line (TL) of length $L$ with a cross section  of dimensions much smaller than $L$ can be effectively described by the following  1D Lagrangian,
\begin{equation}
{\cal L}= \int_{-L/2}^{L/2} dx \left(\frac{l}{2} j ^2- \frac{1}{2c} q^2 \right)=\int_{-L/2}^{L/2} dx \left[\frac{l}{2}\dot{\theta}^2- \frac{1}{2c} (\nabla \theta)^2 \right],
\end{equation}
being $l$ and $c$, respectively,  the inductance and capacitance per unit length, while $j(x,t)$ and $q(x,t)$ are, respectively, the local current and charge densities. In
last identity of the previous equation the following representation was introduced, 
\begin{equation}
\theta(x,t)=\int_{-L/2}^x dx^{\prime} q(x^{\prime},t).
\end{equation}
The equation of motion for this Lagragian is the wave equation with velocity $v=1/\sqrt{lc}$. The solution for open boundary conditions with $\theta(-L/2,t)=\theta(L/2,t)=0$ can be expanded in normal modes as follows
\begin{equation}
\theta(x,t)=\sqrt{\frac{2}{L}} \sum_k  \cos \left(\frac{k \pi x}{L} + \alpha_k \right)\theta_k(t),
\end{equation}
where $\alpha_k = 0, \; (\pi/2)$, for $k$ odd (even) integers, respectively. The quantum-mechanical Hamiltonian for this cavity can be formulated by identifying the coordinates
$\theta_k$ along with the conjugated momenta $\pi_k=l \dot{\theta}_k$ and defining the operators
\begin{equation}\label{modes}
\hat{\theta}_k=\sqrt{\frac{\hbar \omega_k c}{2}} \frac{L}{k\pi} \left(a_k(t)+ a^{\dagger}_k(t)\right),~~~~~~
\hat{\pi}_k= -i \sqrt{\frac{\hbar \omega_k l}{2}} \left(a_k(t) - a^{\dagger}_k(t)\right),
\end{equation}
with $\omega_k=k\pi v/L$ and $\left[ a_k, a^{\dagger}_{k^{\prime}}\right]=\delta_{k,k^{\prime}}$. Therefore, the voltage on the transmission line can be expressed as
\begin{equation}\label{vres}
V(x,t) = \frac{1}{c} \frac{\partial \theta (x,t)}{\partial x}
=  - \sum_{k} \sqrt{\frac{\hbar \omega_k}{Lc}} \sin \left(\frac{k \pi x}{L} + \alpha_k \right)  \left( a_k(t)+ a_k^{\dagger}(t)\right).
\end{equation}

To treat the coupling between a qubit realized in the CPB architecture  and the circuit we must add to the voltage $V_{\rm g}$ in Eq. (\ref{cpb}) the voltage at the connecting point of the TL. Assuming that the qubit is coupled at the coordinate $x=0$ of the TL, 
we  substitute
$N_{\rm g} \rightarrow N_{\rm g}^{\rm dc} + C_{\rm g} V(0,t)/2 e$ in Eq. (\ref{cpb}). Focusing as before on the two lowest-energy states with $N=0,1$ close to  the degeneracy point we get
for the qubit coupled to the LC circuit the following effective Hamiltonian
\begin{equation}
H_{\rm CPB-TL} = - \left[B_z \sigma^z+B_x \sigma^x \right] + \sum_k g_k \left(a^{\dagger}_k+a_k\right) \left( 1- 2 N_{\rm g} - \sigma^z\right).
\end{equation}
We have defined in the previous equation $g_k= -e C_{\rm g} V_k/\left(C_{\rm g}+ C_{\rm J}\right)$ with $V_k=\sqrt{\hbar \omega_k/cL}$ while $B_x, B_z$  are given by Eq. (\ref{bxz}) with
$N_{\rm g}$ given by the dc component $N_{\rm g}^{\rm dc}$. Changing to the basis
that diagonalizes the Hamiltonian for the isolated qubit, it reads
\begin{equation} \label{cpb-res}
H_{\rm CPB-TL} = \frac{\Omega}{2} \overline{\sigma}^z+ \sum_k g_k \left(a^{\dagger}_k+a_k\right) \left[ 1- 2 N_{\rm g} - \cos(\theta) \overline{\sigma}^z+ \sin(\theta) \overline{\sigma}^x\right],
\end{equation}
with $\Omega=\sqrt{E_J^2+\left[4 E_{\rm C}\left(1-2 N_{\rm g}^{\rm dc} \right) \right]^2}$ and $\theta$ defined in Eq. (\ref{bxz}). 

Similarly, for the case of the transmon, the coupling between the qubit and the transmission line reads \cite{koch},
\begin{equation}\label{trans-res}
H_{\rm trans-TL} = \frac{\Omega}{2} \overline{\sigma}^z+ \sum_k \overline{g}_k \left(a^{\dagger}_k+a_k\right) \overline{\sigma}^x,
\end{equation}
with $\overline{g}_k\simeq g_k \left\{E_{\rm J}/(8 E_{\rm C})\right\}^{1/4}$. The qubit coupled to the  electromagnetic environment 
 represented as a set of many quantum-oscillator modes effectively defines a spin system coupled to a Caldeira-Legget type of bath \cite{cal-leg}, as mentioned in Sec. \ref{sec:spin-boson}. Notice that 
Eqs. (\ref{cpb-res}) and (\ref{trans-res}) have the structure of Eq. (\ref{qubit-res}).
  In the literature the rotating wave approximation is sometimes introduced in the calculations, which leads to a coupling of the form
$g_k \sigma^+ a_k+ H.c.$, corresponding to
   the James-Cummings Hamiltonian given by Eq. (\ref{j-c}) \cite{koch}.

 The coupling of a qubit to a single resonator corresponds to considering 
  a single resonant $k$ in Eqs. (\ref{cpb-res}) and (\ref{trans-res}) instead of many modes. Such  circuits are used to
 introduce operations in the qubit, to perform the readout and to implement couplings between several qubits. In particular, couplings, containing $\overline{\sigma}^z$ instead of  $\overline{\sigma}^x$ in Eq. (\ref{trans-res}) 
 have been implemented in designed architectures
of transmons  \cite{didier,touzard,ikonen,olivier}.
 Ref. \cite{blais1} presents a complete overview of the degree of development of this technology.
 The transmon qubit is one the most used one nowadays because of its stability. 
 
 \subsection{Implementing heat transport in quantum electromagnetic circuits}
 Real circuits typically  contain resistive elements (LRC circuits). The resistors produce thermal noise and behave as thermal photon sources. This mechanism is 
 discussed in detail in Ref.  \cite{colloquium,satrya} and is the concrete way to implement a temperature reservoir and to induce thermal transport in these devices.
  The spectral function describing the noise correlations of the voltage defined in Eq. (\ref{vres}), is given by
\begin{equation}\label{noise}
S_V(\omega)= \int_{-\infty}^{+\infty} dt e^{i\omega t} \langle V(0,t) V(0,0 \rangle = 2 R \frac{\hbar \omega}{1- e^{-\beta \hbar \omega}},
\end{equation}
being $R$ the resistance of the circuit and $\beta=1/(k_B T)$. In the limit of $k_B T \gg \hbar \omega$ this expression tends to the classical fluctuation-dissipation result $S_V(\omega)  \rightarrow 2 k_B T R$. In typical devices, the temperature corresponding to the level separation in the superconducting loop is 0.5K, which implies that the  operational temperatures for these devices should be lower than this. 

 Following Ref.  \cite{colloquium,satrya} we summarize the description of the heat transport
 through a qubit or an array of qubits and resonators embedded into   
 two circuits with resistors $R_{\rm h}$ and $R_{\rm c}$ at different temperatures $T_{\rm h}$ (hot)  and $T_{\rm c}$ (cold).  Given a spectral function 
$S_{V_{\rm h}}(\omega)$ defined as in Eq. (\ref{noise}) for the hot resistor, the incident power spectral density 
 at the cold resistor is given by
 \begin{equation}
 S_{P_{\rm c}}(\omega)= \frac{{\cal T}_{\rm c-h}(\omega)}{4 R_{\rm h}}S_{V_{\rm h}}(\omega),
 \end{equation}
 where the (dimensionless) function $ {\cal T}_{\rm c-h}(\omega)$ describes the transmission probability between the two resistive circuits though the array of qubits connecting them. 
 Therefore, the power entering the cold reservoir from the hot one reads
\begin{equation}
P_{\rm c} =\int_{-\infty}^{+\infty} d \omega  S_{P_{\rm c}}(\omega)= \int_{-\infty}^{+\infty} d \omega \hbar \omega {\cal T}_{\rm c-h}(\omega) [n_{\rm h}(\omega)+1/2],
\end{equation}
being $n_{\rm h}(\omega)=1/[1- e^{ \hbar \omega/(k_B T_{\rm h})} ]$ the Bose-Einstein distribution function corresponding to the temperature of the hot bath. 
Following the same reasoning for the power entering the hot reservoir from the cold one and assuming reciprocity such that
$ {\cal T}_{\rm c-h}(\omega)= {\cal T}_{\rm  h-c}(\omega)$, the net heat flux between the two reservoirs is determined from
\begin{equation}\label{plan-but}
P_{\rm net}=P_{\rm c}-P_{\rm h} = \int_{-\infty}^{+\infty} d \omega \hbar \omega {\cal T}_{\rm c-h}(\omega) [n_{\rm h}(\omega)- n_{\rm c}(\omega)],
\end{equation}
which has the same structure as the Landauer-B\"uttiker description of Eq. (\ref{lan-bu}).


\section{Experiments on quantum thermodynamics in cQED}
The possibility of coupling a superconducting qubit to resonators acting as harmonic-oscillators environments motivated the study  of a rich variety of thermodynamic concepts in this setup. The relevant mechanisms have been discussed in the previous sections.  In the present section we review some 
reported experiments in this direction as well as some theoretical proposals of experiments. 

\subsection{Energy dynamics of measurements}

Unlike the ideal picture usually presented in textbooks, the process of measuring a quantum state is not  instantaneous but it takes place in a finite time and it depends on the details of the interaction between the quantum system and the measurement apparatus.

In {\em quantum non-demolition} measurements the goal is the continuous monitoring of the quantum system.
The measurement process implies the coupling of the quantum system to a classical measuring device. In principle, such devices have much more noise than any
 one dictated by  the quantum uncertainty principle. An effective method to reduce these effects is to isolate the  quantum system from the measuring device to perform an {\em indirect measurement}.
This process takes place for the quantum system coupled to a quantum probe and proceeds in two steps: (i) 
 The system undergoes a unitary interaction with the probe, which is initially pre-pared in a known quantum state. (ii) The probe interacts with a classical measurement device.  In experiments, all physical processes occur over some finite timescale. Assuming  the knowledge of the  Hamiltonian describing the coupling between these two systems, this two-step procedure can be used to describe a {\em continuous measurement} if it occurs over an infinitesimally short time step and is repeated continuously. In this case, information is continuously extracted from the probe as it interacts with the measured system. 
  
 The analysis introduced in Ref. \cite{gambe} defines the basis for a set of experiments where the trajectory of a cQED qubit is recorded \cite{nagi,proj-meas1,jordan2,fitchet, proj-meas2}.
  The probe system is a single resonator mode of frequency $f_{\rm res}$ and couples to the qubit as described in Sec. \ref{sec:spin-boson} under the conditions where second-order processes in the coupling  dominates. This defines a dispersive coupling and is modeled by the following Hamiltonian
  \begin{equation}\label{disp-coup}
  H_{\rm disp-coup} = \vec{B} \cdot \vec{\sigma} +  h \left( f_{\rm res} -\frac{ \chi}{2} \sigma^z \right) \left(a^{\dagger} a +\frac{1}{2} \right)
  \end{equation}
  This system has a dispersive coupling in $z$ and can be easily diagonalized for $\vec{B}=(0,0,B_z)$ in the basis  $|\sigma, n\rangle$, where the first entry denotes the state of the qubit and the second one labels the state of the resonator. We see that with this type of dispersive coupling  
  the frequency of the resonator mode is shifted by an amount that depends on the state of the qubit. 
  Hence, by measuring any physical property  that depends on this frequency it is possible to infer the state of the qubit. In practice, this is achieved by sending a signal of a given frequency  to the cavity and measure the
  transmitted or reflected signal. The phase of the transmission signal, in particular, depends on the difference between the  frequency of  the injected wave and the one of the cavity, which provides the information of  the state of the qubit. This measurement procedure is named homodyne detection \cite{bolund} and it basically provides information on $\langle \sigma^z \rangle$. 
  It can be complemented with another optical technique (heterodyne detection) that records the spontaneous fluorescent emission of the qubit, which provides information of the
components $\langle \sigma^x \rangle$ and $\langle \sigma^y \rangle$. The analysis of these measurements enables the reconstruction of the 
  {\em quantum trajectory} and offers the possibility of keeping track on the evolution of the quantum state.   In order to theoretically describe the evolution of the quantum state in contact to the noisy environment including the backaction introduced by the measurement, a useful technique is the stochastic master equation introduced in Sec. \ref{sec:meas}. This is a generalization of the one that describes the dynamics of the quantum state coupled to the reservoirs, with an additional stochastic term to describe the effect of the measurement.

\begin{figure}
    \centering
    \includegraphics[width=\textwidth]{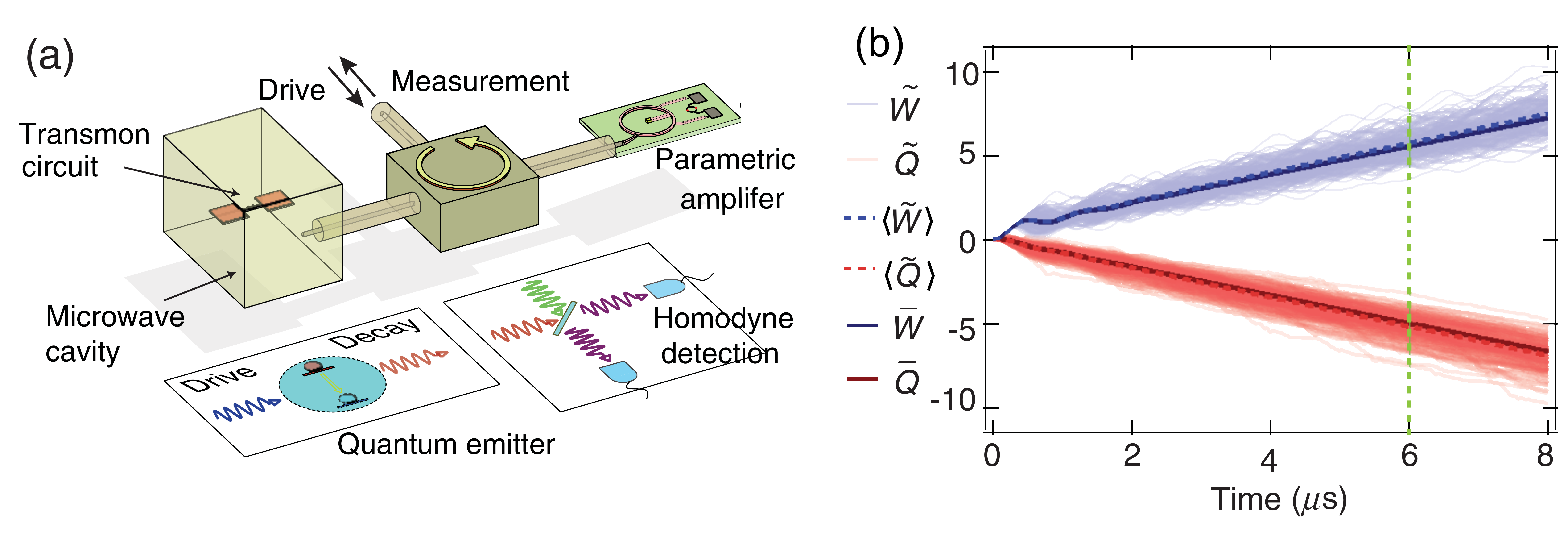}
    \caption{ (a) Experimental setup to drive the qubit and record the quantum trajectories to analyze the work and heat (Adapted from Ref. \cite{proj-meas2}). 
    (b) Recorded quantities. $\tilde{W}$ and $\tilde{Q}$ correspond to work and heat calculated for individual trajectories, while $\langle\tilde{W}\rangle$ and $\langle \tilde{Q} \rangle$ are the corresponding averages over trajectories. The values $\overline{W}$ and  $\overline{Q}$  correspond to the calculations based on the Lindbladian dynamics.}
    \label{fig:meas}
\end{figure}

 Motivated by these experiments, Refs. \cite{proj-meas-rom, proj-measo, proj-meas3} address the description of the concomitant energy dynamics and these ideas were more recently experimentally tested in Ref. \cite{proj-meas2}. 
 In this work it is analyzed   how quantum heat and work can be consistently identified in systems whose environment consists of a continuously coupled quantum detector. 
 
 The Hamiltonian describing the qubit corresponds to
 Eq. (\ref{hqubit0}) with $B_z=0$ and a coupling to the electromagnetic environment is given by the Jaynes-Cummings Hamiltonian Eq. (\ref{j-c}). Hence, the dispersive coupling is along $x,y$ and the information of
 $\langle \sigma_x \rangle$ obtained by homodyne detection is recorded. The sketch of the experimental device is shown in Fig. \ref{fig:meas} (a).
 The evolution of the density matrix is described by an stochastic master equation like the one discussed in Section \ref{sec:meas} with an stochastic term  describing the effect of the measurement. Each trajectory  corresponds to a particular realization of the experiment. As a result,
 for an infinitesimal time interval, the change in the density operator of a quantum trajectory is written as
 $d \tilde{\rho_t} = \delta {\cal W} \left[ \tilde{\rho}_t \right] dt  + \delta {\cal Q} \left[ \tilde{\rho}_t \right] dt$, where
$\delta {\cal W} $ and $\delta {\cal Q}$ are superoperators associated to the respective unitary and non-unitary dynamics. As indicated by the notation, and by analogy to Eqs. (\ref{cons-qs}),
the first term can be used to calculate the conservative component of the work and the second one to calculate the quasi-static heat.
In this way, heat and work for individual  trajectories can be recorded. Results are shown in Fig. \ref{fig:meas}(b), where we can also see that  the averages over quantum trajectories and the calculations based on a pure Lindbladian evolution without any stochastic component show a very good agreement.

\subsection{Maxwell demon}\label{sec:max-dem}
The Maxwell demon is an imaginary character introduced by Maxwell to discuss a {\em gedanken} experiment devoted to illustrate the second law of thermodynamics \cite{max-dem}. The main idea is the implementation of processes  forbidden to  spontaneously occur because of the second law, thanks to the help of an intelligent creature. An example is the split of a mixture of gasses so that those with larger 
kinetic energy are collected in a compartment separated by a wall from those with lower kinetic energy. 
This is accomplished with the help of the demon, who classifies the particles according to their velocities  and separates them into the compartments by conveniently opening or closing a door in the wall. The solution to the seeming paradox is solved by properly accounting for the information processed by the demon. This results in an entropy and energy  balance which is in full agreement with the laws of thermodynamics.

A realization of the Maxwell demon in a transmon qubit coupled to a resonator was reported in Ref.  \cite{bhuard}.  Here we summarize the main idea.
Assume that the qubit  is first coupled to a thermal bath that prepares it in a thermal state and  it is afterwards coupled to a "battery". The latter is of the type proposed in Ref. \cite{bat-mon} and consists of 
a $\pi$-pulse  with a resonant frequency
that switches the populations between the excited state
and the ground state by absorbing or emitting photons, as discussed in Section \ref{sec:bat} [see Eqs. (\ref{hb1}) - (\ref{hb3})]. 
 For a thermal state at a low temperature, the ground state  is more populated than the excited one. 
Therefore, the rate of absorption is larger than that of emission, resulting in a net power transferred  from the battery to the qubit. This process decreases as the temperature of the thermal state increases
and the population of the ground state and the excited one become similar. In the setup studied in Ref. \cite{bhuard} a resonator is coupled to the qubit 
in a way that it plays the role of a demon which favors the process of emission of photons when the system is coupled to the battery by preventing the absorption. This results in a net 
transfer of power from the system to the battery.

\begin{figure}
    \centering
    \includegraphics[width=0.6 \textwidth]{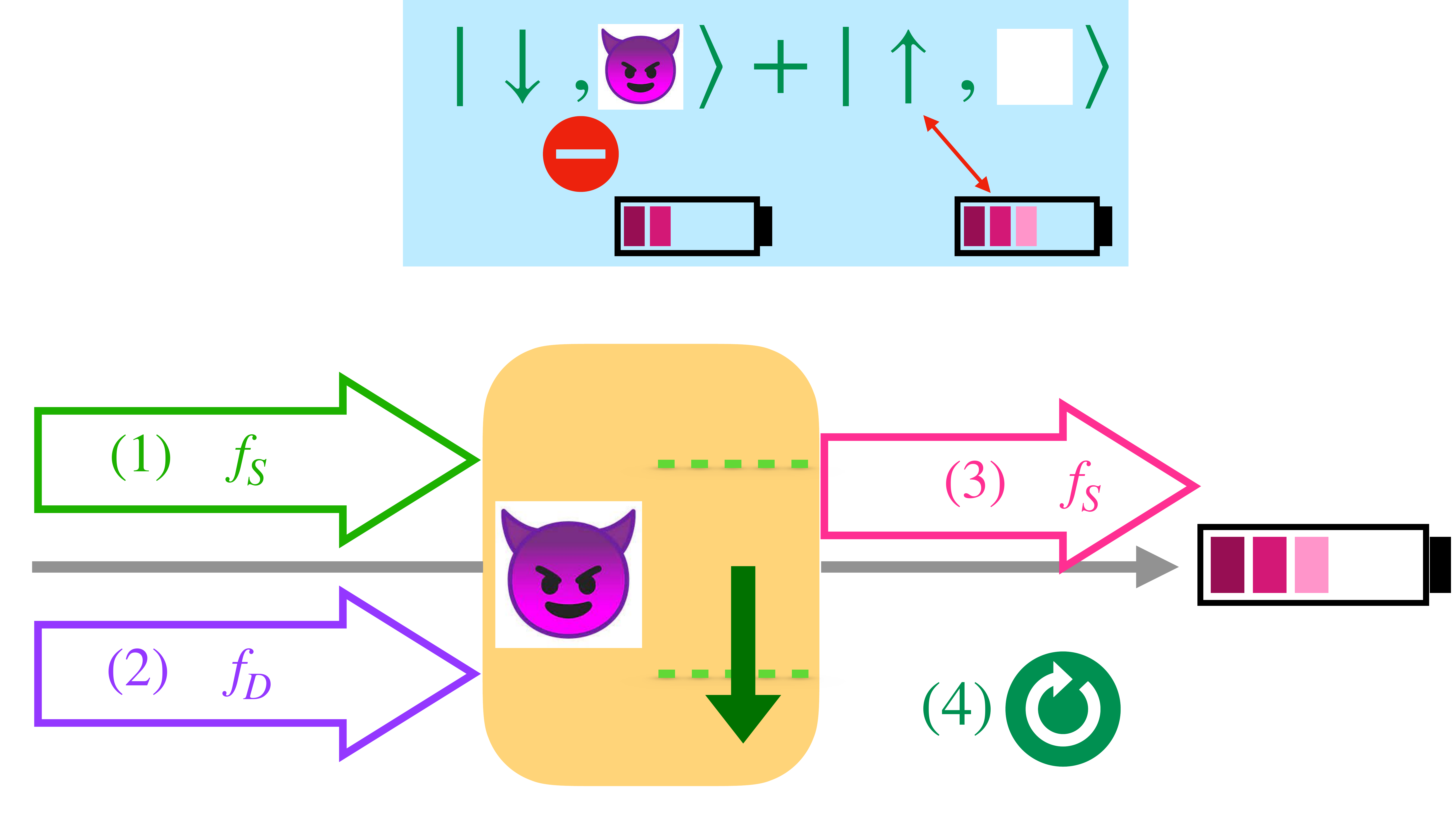}
    \caption{Sketch of the cycle studied in Ref.  \cite{bhuard}). (1) Preparation of the state. (2) The demon is driven and prepared with $n_D$ photons. 
    (3) The battery is connected. Without the demon, it should supply photons to the qubit. Because of the demon, it receives photons from the qubit and it is charged. (4)  Reset step.
    The colored box indicates the superposition state of the qubit and the action of the demon. The latter forbids the absorption of photons from the battery and only allows the 
    emission.}
    \label{fig:demon}
\end{figure}

We now briefly discuss the concrete experiment. The combined qubit-demon  system is described by the Hamiltonian of Eq. (\ref{disp-coup}) with
$\vec{B}\equiv (0,0,h f_S/2)$ and the resonator playing the role of the demon with $f_{\rm res}\equiv f_D$.  
 A cycle is implemented as follows: (1) The state of the qubit is  prepared  in a thermal state at a temperature $T_{\rm h}$ or in a superposition quantum state. (2) The resonator (demon)
is driven with a frequency $f_D$ with a pulse of duration longer than $\chi^{-1}$ and shorter than the coherence times of the qubit and cavity. The process is designed to  create $n_D=\langle a^{\dagger} a \rangle$ photons in the resonator only if the qubit is in the ground state ($|\downarrow\rangle$). The corresponding eigenstates of the Hamiltonian of Eq. (\ref{disp-coup}) are $|\downarrow\rangle \otimes |n_D \rangle$  with energy $E_{\downarrow, n_D}=-hf_S/2+h(f_D+\chi/2 )n_D$  and 
$|\uparrow\rangle \otimes |0\rangle$ with energy $E_{\uparrow}= h f_S/2 $. 
 (3) A $\pi$-pulse, playing the role of a battery, is sent at a frequency $f_S$. 
 The action of such a pulse  with the demon in the 
$n_D=0$ state is the same as without demon and consists in switching the populations of the two qubit states.
 Instead,  any other state of the qubit-demon system with $n_D \neq 0$ 
 is off-resonance with the pulse by an energy $-\chi n_D$ and the absorption of a photon accompanied by a switch of the qubit state is not allowed. The only allowed process for the qubit is the emission of a photon by changing the state from $|\uparrow \rangle $ to $|\downarrow \rangle $. 
 (4) The last step consists in reseting the demon by coupling it to a bath at temperature $T_{\rm c} < T_{\rm h}$. This completes the cycle. See sketch of Fig. \ref{fig:demon}.

 In the experiment of Ref. \cite{bhuard} the power at the battery is recorded in step (3). It is  verified that with the demon with $n_D=0$ the battery releases power to the system, while with 
 $\langle n_D \rangle = 0.9 $  the battery receives power. In a thermal state, the photon emission from the system into the battery  when the demons intervenes ($\langle n_D \rangle \neq 0 $)
 becomes more important as the temperature increases, and this behavior is verified in this experiment. On the other hand, when the state is prepared in a quantum superposition, the outcome is expected to be similar to a high-temperature initial state and this is also observed in the experiment.
 The results  are interpreted  by modeling the coupling of the system to the baths and battery in the framework of Lindblad master equation Eq. (\ref{masterrho}).
 
 A similar setup was theoretically considered in  Ref. \cite{elouard}, where the thermal baths of step (1) is substituted by a  measurement. The idea is very similar, except for the fact that the role of the demon is to perform projective measurements in order to select the excited state of the qubit. These type of  non-demolition measurements can be accomplished by a system that is well described by the Hamiltonian of Eq. (\ref{disp-coup}), where the coupling term commutes with the Hamiltonian of the qubit. The demon resonator is excited with microwave pulses, the response depends on the state of the qubit and it is calibrated as a readout element.

 The qubit-demon device described by the Hamiltonian   of Eq. (\ref{disp-coup}) with the demon used to perform  measurements was also considered in Ref. \cite{masuyama} to experimentally verify a fluctuation theorem. Fluctuation theorems \cite{fluct-theor1,fluct-theor2,fluct-theor3,fluct-theor4,fluct-theor5,fluct-theor6} are regarded as generalizations of the second law of thermodynamics and are valid away from equilibrium. The experiment by Masuyama and coworkers analyzes a fluctuation theorem formulated in Ref. \cite{ueda}, which reads
 \begin{equation}\label{eq:frel}
 \langle e^{-\sigma - I_{\rm Sh}} \rangle_{\rm PM} = 1- \lambda_{\rm fb},
 \end{equation}
 where $\langle \ldots \rangle_{\rm PM}$ denotes the statistical average over the projective-measurement (PM) protocols, $I_{\rm Sh}$ is the stochastic Shannon entropy the demon acquires in the PM and 
 $\sigma=- (W+\Delta F)/k_B T$ is the entropy change  in the qubit, being $W$ the work extracted from the qubit and $\Delta F$  the change in the free energy (in the experiment $\Delta F=0$). The constant $\lambda_{\rm fb}$ in the previous equation is associated to irreversible processes. The implemented protocol  consists in two projective measurements with feedback control. The sequence is: (1) The qubit state is prepared in a thermal state (2) A  measurement by the demon is performed with outcome $x= \uparrow$ or $ \downarrow$. The demon gains stochastic entropy $I_{\rm Sh}=-p(x)$ in this process. 
 (3) A feedback operation is performed. This consists in leaving the state of the qubit unmodified if it is in the ground state or applying a $\pi$-pulse if it is in the excited state. In the latter case
 work is extracted from the qubit-demon, as discussed before. (4) The amount of extracted work is measured by a second projective measurement $z= \uparrow$ or $ \downarrow$.
  The result is $W(x,z)= E(x)-E(z)$, where 
 $E(x), \; E(z)$ are the energies of the states $x$ and $z$ of the qubit. In this way, the quantity 
 $\langle e^{W/k_B T - I_{\rm Sh}} \rangle_{\rm PM}= \sum_{x,z} p(x,z) e^{W(x,z)/k_B T - I_{\rm Sh}(x)}$ is experimentally recorded and compared with the theoretical estimate of the right hand side of Eq. (\ref{eq:frel}). The latter is calculated by recourse to a quantum master that takes into account the qubit relaxation during the pulse sequence. The agreement between experimental and theoretical results is excellent.

A similar device and description was used in Ref. \cite{proj-meas2} to track the quantum trajectories of a driven qubit. The aim of this experiment is to analyze the decomposition of changes in the internal energy of the qubit into heat and work, as discussed in Section \ref{sec:meas}. In the experiment, the results of two-point measurements similar to those in Ref. \cite{masuyama} are compared with the predictions
of quantum master equations based on the Hamiltonian of Eq. (\ref{disp-coup}).
A more complex configuration involving two transmon qubits was considered in Ref. \cite{lebedev}, where a mixed-state target qubit is purified by a pure-state demon qubit connected via an off-resonant transmission line. 
A nice overview and comparison among  these experiments has been presented in Ref. \cite{cottet-huard}.

In Ref.  \cite{mich-md} a different experimental setting to implement a Maxwell demon in cQED was proposed, which bares a closer resemblance to the original idea formulated by Maxwell. 
The gas chambers are substituted by LRC resonators with a tunable frequency at different temperatures and the compartment separating these two reservoirs is realized by a qubit. This system is basically the one discussed in
Sec. \ref{sec:statio} (see Fig. \ref{fig:transport}) and the spontaneous process dictated by the thermodynamics laws is the heat transport from the hot to the cold reservoir through the qubit. The action of the demon is implemented by monitoring the quantum state of the qubit and manipulating the strength of its contact to the reservoirs. This manipulation  can be realized by selecting the frequency of the resonator. When it coincides with the one associated to  the qubit level spacing the contact is switched on  while when   these two frequencies are detuned the contact is switched off. Hence, 
 if the qubit is detected to be in the ground state, it is put in contact with the cold reservoir, in order to favor the absorption of photons from this bath. Instead, when it is detected to be in the excited state it is put in contact with the hot one, in order to favor the emission of photons into this bath. The net effect is a transfer of heat from the cold to the hot reservoir.

 \subsection{Experimental measurement of a topological transition}\label{sec:expe-topo}
 In Sec. \ref{sec:topo-adia} we have briefly reviewed Ref. \cite{gritsev} where the fact that the Berry curvature could be measured in the adiabatic dynamics of a 
 qubit was pointed out. Refs. \cite{mike-exp1,mike-exp2} present the  experimental verification of these ideas. 
 
 \begin{figure}
    \centering
    \includegraphics[width=\textwidth]{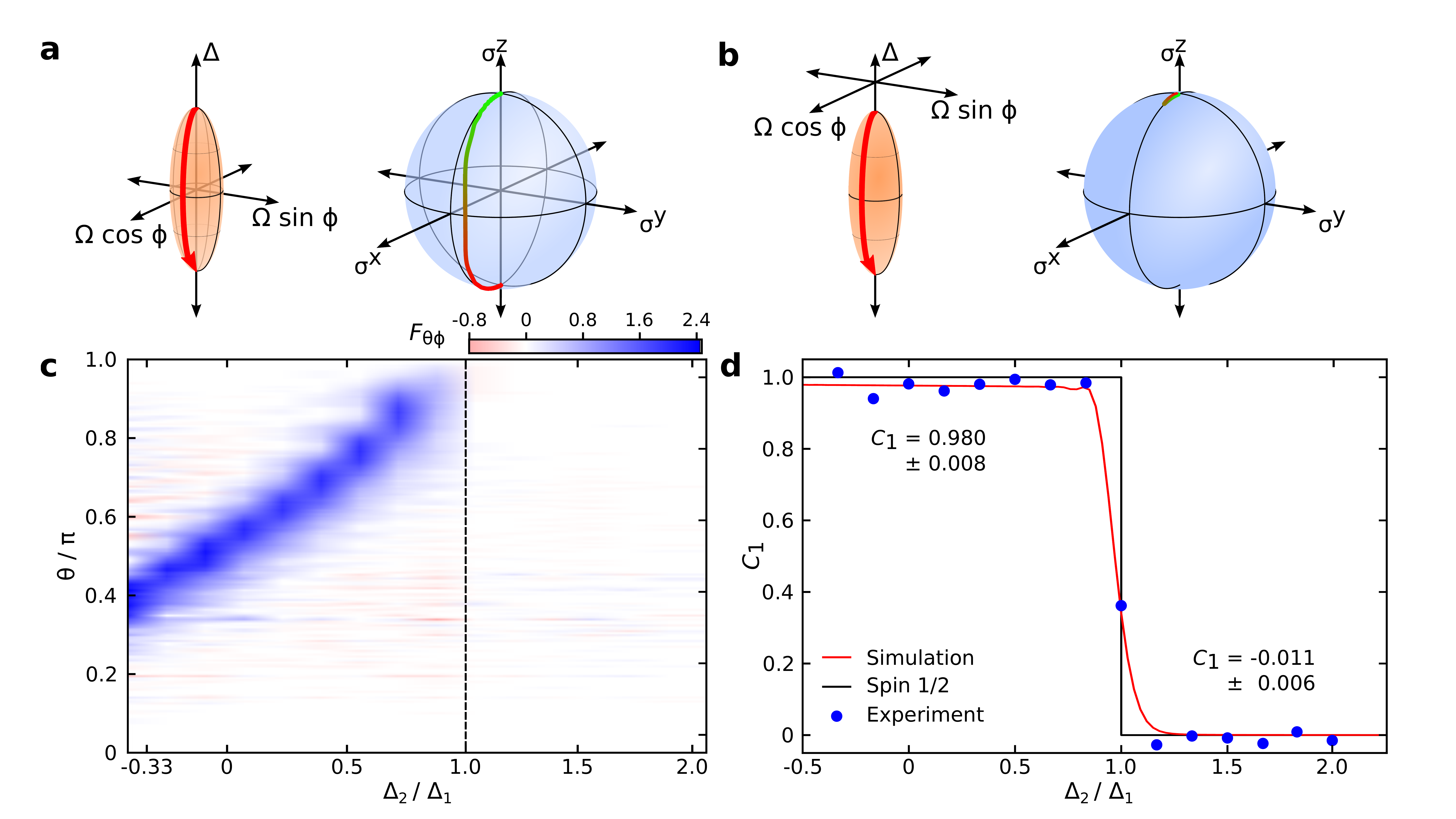}
    \caption{
     Adapted from Ref. \cite{mike1}.
     (a) The parameter space and the implemented protocol for $\Delta_2=0$ in red (left).  The corresponding calculated 
     results for $\langle \vec{\sigma}\rangle$ represented in the Bloch sphere (right). 
      (b) The same as for (a), with $\Delta_2 = 1.5 \Delta_1$ (c) The Berry curvature measured as a function of $\Delta_2/\Delta_1$, which is integrated in (d) to yield the Chern number $C_1$.
     }
    \label{fig:topo}
\end{figure}
 
 In Ref. \cite{mike1} a transmon qubit is evolved in the adiabatic regime following the proposal of Ref. \cite{gritsev}. Concretely, the Hamiltonian of
  Eq. (\ref{driv-qubit}) was considered with 
 \begin{equation}
 \vec{B}(\theta,\phi)=\frac{\hbar}{2}\left(\Delta_1\cos\theta+\Delta_2,\Omega_1 \sin \theta \cos \phi,
 \Omega_1 \sin \theta \sin \phi\right).
 \end{equation}
 This model has a topological transition as a function of $\Delta_2/\Delta_1$,
  where the Chern number jumps from $1$ to $0$, for $\Delta_2=\Delta_1$. 
 The parameters are chosen in order to initialize the qubit in its ground state at $\theta(t=0)=0$ with fixed
 $\phi(t)=0$ and  a ramp with constant $\dot{\theta}$ is implemented. 
 The evolution of the induced force ${\cal F}_{\phi}=-\partial H_{\rm qubit}/\partial\phi$ is monitored and compared with
 the description of  Eq. (\ref{fphi}), with the Berry curvature given by Eq. (\ref{anti-grit}). Experimentally, the latter is
 determined as the linear coefficient of $\langle {\cal F}_{\phi} \rangle$ as a function of $\dot{\theta}$. Given the curvature, the Chern number is calculated from
 Eq. (\ref{chern}). The comparison between theory and experiment is excellent. When  the effect of the environment is considered by simulating the adiabatic evolution with Lindblad master equation  small corrections are found which tend to soften the jump at the transition. 
 
 Results are summarized in Fig. \ref{fig:topo}. The upper panels show the protocols and the calculated
 results for $\langle \vec{\sigma}\rangle$, represented in the Bloch sphere for parameters in the topological phase  (a) and 
 outside it (b). It is clear that in the topological case the state wraps the Bloch sphere and $\langle \sigma_y \rangle \neq 0$.
 Instead, in the non-topological case there is no wrapping of the Bloch sphere and $\langle \sigma_y \rangle = 0$. 
 The measured Berry curvature is shown in panel (c), while panel (d) shows the result for the Chern number. The plots with the label spin $1/2$ correspond to the isolated system while "simulation" labels the results obtained by including the coupling to the environment modeled by a Lindbladian dynamics.

 In Ref. \cite{mike2} a similar protocol and measurement was performed in a cQED qubit with a different architecture and these results were verified. Furthermore,
  by mapping states on the Bloch sphere to wave vectors in the first Brillouin zone of a lattice model,  the topological transition of 
   Haldane model \cite{haldane} is simulated in a qubit. 
   
 \subsection{Thermal machine and Otto cycle}
An experimental setup to implement a non-equilibrium thermal machine  on a superconducting qubit coupled to two resonators with a resistive element (LRC circuits)  was proposed  in Ref. \cite{karimi-pekola-otto} and further analyzed in Ref. \cite{solfa1}. The operation is argued to have a regime that corresponds to an Otto cycle.
 
 \begin{figure}
    \centering
    \includegraphics[width=\textwidth]{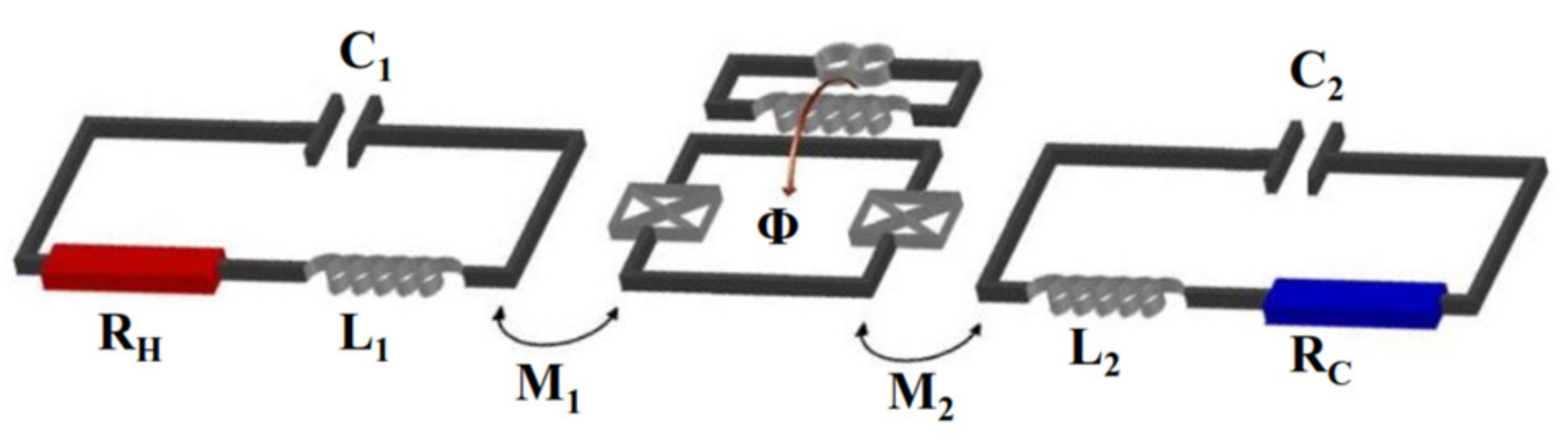}
    \caption{Sketch of the proposed experimental setup (Adapted from Ref. \cite{karimi-pekola-otto}).}
    \label{fig:otto-karimi}
\end{figure}

  The two resonators have 
 different resonant frequencies, $\omega_{LC,1}$ and $\omega_{LC,2}$,
  and different temperatures. 
 The qubit is described by the Hamiltonian of Eq. (\ref{driv-qubit}) with $\vec{B}=E_0\left(\Delta, 0, q(t) \right)$, being $q$ a time-dependent parameter modulated by changing the flux, which follows the protocol,
 \begin{equation}
 q(t)=\frac{1}{4} \left[1+ \cos(2\pi f t) \right].
 \end{equation}
 Hence, the level spacing of the qubit oscillates according to $\hbar \omega(t)= 2 E_0 \sqrt{\Delta^2+q^2(t)}$. 
  Each of the resonators has an associated noise spectral function of the form of Eq. (\ref{noise}) and it is coupled to the qubit operator $\sigma^z$ through an inductive element. The degree of coupling depends on the degree of tuning between $\omega_{LC,j}, \; j=1,2$ and $\omega(t)$.
  
   An ideal Otto cycle would consist in the following four steps: (a) For $q: 0 \rightarrow 1/2$: The qubit decouples from the resonators and evolves isollated from the reservoirs. 
 Its population is determined by the cold resistor. 
 (b) At $q=1/2$ the qubit couples to  the hot resonator and thermalizes with the hot resistor. The energy flows from the qubit to the resonator. (c) $q: 1/2  \rightarrow 0$: The qubit decouples from the resonators and evolves  isolated from the reservoirs. 
 (d) At $q=0$ the qubit couples to the cold resonator and thermalized with the cold resistor. The energy flows from the resonator to the qubit. The full cycle operates as a refrigerator. Although the analogy to the Otto cycle is useful to understand the basics of the operation, the complete decoupling from the reservoirs does never  occur in this setup and
  the  device actually operates as a non-equilibrium thermal machine in permanent contact to the reservoirs, similar to the one discussed in Sec. \ref{sec:ther-mach}. In both Refs. \cite{karimi-pekola-otto} and \cite{solfa1} the problem is solved by means of a non-equilibrium master equation. The description is extended to higher frequencies and the solution is argued to achieve a regime where the performance resembles the one of an Otto cycle. 
Estimates on the cooling power are presented in Ref. \cite{karimi-pekola-otto} and these are consistent with the present state of the art of the experimental detection possibilities. So far,
the actual experiment has not been reported.

\subsection{Thermal transport}\label{sec:exp-ther-trans}
As discussed in Sec. \ref{sec:statio}  when a few-level quantum system  is coupled to two or more baths at different temperatures, there is typically a stationary heat flux from the hot to the cold bath through it. 
\begin{figure}
    \centering
    \includegraphics[width=\textwidth]{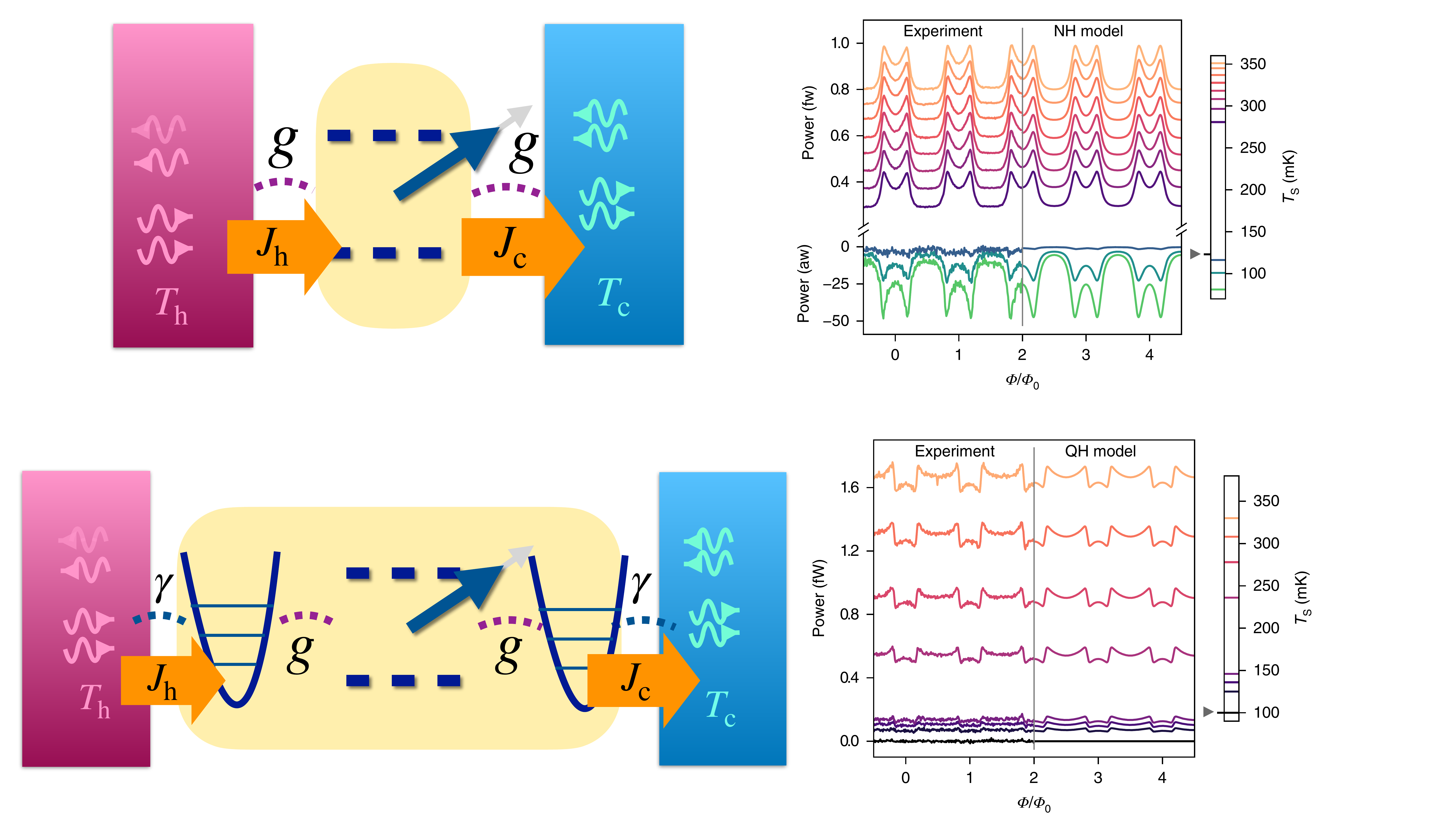}
    \caption{
    Left: Illustration of the thermal transport process when the qubit is contacted to reservoirs at different temperatures through resonators. 
    Right: Experimental results for the total heat power absorbed  by the cold reservoir for these two configurations (Adapted from Ref. \cite{ronzani}). }
    \label{fig:ronzani}
\end{figure}

In Ref. \cite{ronzani}  a setup similar to the one described in that section was considered  to experimentally study the thermal transport through a qubit. The
configuration is similar to the one sketched in Fig. \ref{fig:transport} (b). Instead of coupling the qubit to the reservoirs directly, the coupling is implemented to resonators with a resistive element, which are contacted to metallic reservoirs at different temperatures as sketched in Fig. \ref{fig:ronzani}. Hence, there are two couplings playing a role, which are, respectively, indicated in the Fig. with $g$ (between the qubit and the resonators) and
$\gamma$ (between the resonators and the thermal baths). The proper separation into the quantum system and reservoirs in the description of the thermal transport through the device depends on where the thermal bias effectively drops, as discussed in Sec. \ref{sec:drop} and this depends on the ratio $g/\gamma$. For $g \ll \gamma$, each resonator is thermalized with its reservoir. In this regime, it is appropriate to define  effective cold and hot baths  to describe the combined systems of cold and hot reservoirs in contact to the corresponding resonator. This effective description is sketched in the upper part of Fig. \ref{fig:ronzani}   and implies a proper modeling of the spectral functions for the baths 
to represent the combined systems. In the opposite limit of $g \gg \gamma$, the temperature drop is more likely to take place in the contact between the resonator and the external reservoir. Therefore, in this case it is more appropriate to consider the quantum system consistent of the qubit hybridized with the resonators and this combined system coupled to the reservoirs, as sketched in the lower part of the figure. Both scenarios were successfully analyzed in Ref. \cite{ronzani} where the experimental data was properly modeled by means of master equations.  In both configurations,  the recorded power, corresponding to the heat current through the device through Eq. (\ref{plan-but}), is analyzed as a function of $\Phi$, the magnetic flux through the qubit. Results are shown in the right-hand side of Fig. \ref{fig:ronzani}. An interesting feature there analyzed is the sensitivity of the heat transport to the magnetic flux. This is a signature of quantum coherence in the heat transport, which has been previously observed in the context of superconductors \cite{giazotto}. In the present case, we recall that this flux determines the effective level spacing of the qubit, and this is reflected in the transmission probability for the energy transport through the device.

A related experiment was reported in Ref. \cite{recti}, basically in the same experimental device. Here, the focus was on the heat rectification properties of the system, implying a different amplitude in the heat current when the thermal bias in inverted. Heat rectification is known to be possible only in systems with many-body interactions as discussed in Sec. \ref{sec:recti}.  In this experiment, a 
  thermal rectification of 10 $\%$  was observed. 

 \subsection{Steady-state heat engines and refrigerators}
In Ref. \cite{heat-engine4} a device where two quantum LC resonators is proposed. The two resonators with natural frequencies $\Omega_{\rm c} < \Omega_{\rm h}$
are coupled via a transmission line to baths at temperatures $T_{\rm c} < T_{\rm h}$, respectively. 
They are connected through a Josephson junction, which is biased with a voltage $V= h \left(\Omega_{\rm h}- \Omega_{\rm c} \right)/2e$. 
If the occupation probability of the hot (h) cavity is larger than the cold (c) one, Cooper pairs tunnel against the bias by absorbing a photon in h and creating a photon in c. 
In this way, heat is converted into an electrical supercurrent that tunnels agains the bias. The operation of this setup as a thermometer was analyzed in Ref. \cite{thermo}.
Interestingly, this device has been experimentally realized in Ref. \cite{device}, although no results on the heat transport or the mechanism of heat--work conversion have been reported so far.

A  configuration of two resonators with a temperature bias,  connected by a Josephson coupling was also considered in Ref. 
 \cite{out-ref} to realize an autonomous refrigerator. In this case, in addition to the direct Josephson coupling, the two resonators are coupled through a third one
  with natural frequency $\Omega_{\rm r} $, forming a loop. The loop is threaded by a flux $\phi$, so that the Josephson Hamiltonian
   of the full device reads $E_J \cos\left(2 \hat{\phi}_{\rm c}+ 2 \hat{\phi}_{\rm h} + 2 \hat{\phi}_{\rm r} + \phi \right)$,  where $\hat{\phi}_{\alpha} = \sqrt{\pi e^2 Z_{\alpha}/h}$ is the phase fluctuation
   in each resonator being $Z_{\alpha}$ the corresponding impedance. The non-linearity of this term and imposing the condition $\Omega_{\rm r}= \Omega_{\rm c} + \Omega_{\rm h}$ favors the emission of photons from the cold resonator, with the consequent cooling.

\section{Final remarks}
The problem of the energetics and thermodynamic properties of new quantum devices is a topic of increasing interest, which motivates
basic research while it is also paramount for the development of quantum technologies.

We have presented an overview of the main fundamental problems related to the energy dynamics of qubits, which are currently under active investigation. The rich variety of phenomena ranges from many-body strong correlations to the realization of thermal machines to convert heat into useful work and vice-versa. It also includes the fascinating topological properties which may help to provide a robust framework to control this dynamics.

We have intended to discuss theoretical results and to provide a brief description of the theoretical tools to address them in the different regimes along with their scope and limitations. 

The main focus in connection with concrete realizations of these mechanisms has been put on superconducting devices. For this reason, we devoted some space to briefly explain the operation of these systems. We have also reviewed the main experimental advances in the study of quantum energy dynamics and thermodynamics in this qubit platforms. Nevertheless, 
many of the theoretical discussions are valid or can be simply translated to scenarios based on  other qubit platforms, like quantum dots, NV centers and AMO systems. 

We hope that this contribution may provide a motivating summary to stimulate more experimental research and help in triggering new theoretical ideas in this field.

\section{Acknowledgements}
LA thanks R. Fazio for many discussions and suggestions for this manuscript, as well as  M. Perarnau-Llobet, L. Tosi,  A. Jordan and A. Levy Yeyati for reading the manuscript and for their constructive comments and P. Terren Alonso for the help in checking equations. LA acknowledges support from CONICET-Argentina, FonCyT-Argentina through PICT-2017-2726, PICT-2018-04536, PICT-2020 and the Alexander von Humboldt Fundation (Germany). L.A. thanks KITP for the hospitality in the framework of the activity
Energy and Information Transport in nonequilibrium Quantum Systems
and the support by the National Science Foundation under Grant No. PHY-1748958, as well as the hospitality of ICTP-Trieste and the ICTP-Simons  program. 

\appendix

\section{Details on the calculation of the $\Lambda^S_{\ell,\ell^{\prime}}$ for a driven qubit coupled to a single bath}\label{appa}
We summarize some technical details to calculate Eq. (\ref{lambda-mar}) as reported in Ref. \cite{scandi}. The procedure to calculate the matrix  ${\cal M}(\vec{X})$ is based on the assumption that 
 the master equation for the frozen density matrix is known and  given by 
 \begin{equation}\label{rhof-qubit}
 \frac{d\rho^{(f)}}{dt}= \frac{\Gamma_g}{2} {\cal D}[\sigma^+]\rho^{(f)}+\frac{\Gamma_d}{2} {\cal D}[\sigma^-]\rho^{(f)}.
 \end{equation}
  $\Gamma_g(B)$ and $\Gamma_d$ are gain and damping rates describing the effect of the coupling with the bosonic bath. These read
  $\Gamma_{g}(B)=[1+n_B] \Gamma(B)$  and
 $\Gamma_{d}(B)=n_B \Gamma(B)$, being 
 $n_B=(e^{2B\beta}-1)^{-1}$ the Bose-Einstein distribution function depending on the temperature $T$ of the bath ($\beta=1/k_BT$). The function
$ \Gamma(B)= \gamma_0 B^s$ with
  $s=1, >1, <1$ defines the spectral properties of the bosonic bath (Ohmic, super-Ohmic and sub-Ohmic, respectively). 
  The other quantities in Eq. (\ref{rhof-qubit}) are
 $\sigma^{\pm}= (\sigma^x \pm i\sigma^y)/2$ and the superoperators are 
 ${\cal D}[O]\rho = O \rho O^{\dagger} - \{O^{\dagger} O, \rho \}$. 
   In such a case, the quantities defined in Eq. (\ref{mgam}) read
 \begin{equation}\label{1res}
 {\cal M}=\mbox{Diag}\left(-\Gamma^+(B)/2,-\Gamma^+(B)/2,-\Gamma^+(B)\right), \;\;\;\;\;\;\;\; \vec{\gamma}=\left(0,0,\Gamma^-(B)\right),
 \end{equation}
 being $\Gamma^{\pm}(B)=[\Gamma_d(B) \pm \Gamma_g(B)]/2$. 
 The stationary solution of Eq. (\ref{rhof-qubit}) in the vector notation introduced before
 is $\vec{\rho}^{(f)}=(0,0,-\Gamma^-(B)/\Gamma^+(B)) $, which leads to
 \begin{equation}
 {\cal M}^{-1}\frac{\partial \vec{\rho}^{(f)}}{\partial B}=  \lambda_B
 \vec{z},
 \;\;\;\;\;   {\cal M}^{-1}\frac{\partial \vec{\rho}^{(f)}}{\partial \theta}=\lambda_q \frac{\partial \vec{n}}{\partial \theta},
 \;\;\;\;\;   {\cal M}^{-1}\frac{\partial \vec{\rho}^{(f)}}{\partial \varphi}=\lambda_q \frac{\partial \vec{n}}{\partial \varphi},
 \end{equation}
 with 
  \begin{equation}
     \lambda_B=\frac{ 1}{\Gamma(B)}\frac{\sinh(\beta B)}{\cosh^3(\beta B)} , \;\;\;\;\;\;\;\;\;\; \lambda_q=\frac{1}{\Gamma(B)}  \tanh^2(\beta B).
 \end{equation}
 To compute $\Lambda^{S}_{\ell,\ell^{\prime}}$ from Eq. (\ref{lambdas})  we need to calculate
 the vectors defining the force operator introduced in Eq. (\ref{forcep}). The result is
 \begin{equation}
 \vec{f}_B= \vec{n}, \;\;\;\;\;
 \vec{f}_{\theta}=B \frac{\partial \vec{n}}{\partial \theta},\;\;\;\;\;
 \vec{f}_{\varphi}=B \frac{\partial \vec{n}}{\partial \varphi}.
 \end{equation}

 \section{Master equation for a driven qubit asymmetrically coupled to two reservoirs}\label{appb}
 We present here the  master equation derived as explained in Sec. \ref{sec:adia-weak}
 for a slowly driven qubit weakly coupled to thermal baths with  Eqs. (\ref{contlr}).
 
The instantaneous values of the parameters are  $\vec{B}=(B_x,B_z)$.
 Hence, the matrix elements $\xi_{\alpha,jl}$ of the coupling are obtained by transforming Eqs. (\ref{contlr}) to the instantaneous frame 
 where the Hamiltonian of Eq. (\ref{driv-qubit}) is diagonal for these frozen parameters. The corresponding matrices are
 \begin{equation}
 \xi_{\rm l}=\frac{1}{B}\left(B_z \sigma^z+B_x \sigma^x \right),\;\;\;\;\;\;\;\;\;\;\xi_{\rm r}=- \frac{1}{B}\left(B_x  \sigma^z+ B_z \sigma^x\right).
 \end{equation}
 Assuming an Ohmic bath, the rates of Eq. (\ref{rates}) are 
 \begin{equation}
 \gamma_{\alpha}(\varepsilon)= \varepsilon \gamma_{\alpha,0} e^{-\varepsilon/\varepsilon_C},\;\;\;\;\;\; \varepsilon >0,
 \end{equation}
 being $\varepsilon_C$ and energy cutoff. 
 
Introducing the notation of Eq. (\ref{rhof-a}), the master equations for the
 frozen, Eqs. (\ref{frozen}),  and adiabatic, Eq (\ref{adia}), components read, respectively, for the present case
 \begin{equation}\label{master-q}
\left( {\cal E}+ {\cal M}_{\rm l}+  {\cal M}_{\rm r} \right)\vec{\rho}^{(f)} = \vec{\gamma},\;\;\;\;\;\;\;\;\;\; \sum_{\ell=x,z} \frac{\partial \vec{\rho}^{(f)}}{\partial B_\ell} \dot{B}_{\ell}=\left( {\cal M}_{\rm l}+  {\cal M}_{\rm r} \right)\vec{\rho}^{(a)},
 \end{equation}
 with
 \begin{equation}\label{ker}
 {\cal M}_{\alpha} =   \left( \begin{array}{ccc}
0 & 0 &  \overline{D}_{\alpha}
 \\
0 & D_{\alpha} & 0 \\
0 & 0 & D_{\alpha}   \end{array}
\right),   \;\;\; \;\;\;\;\;\;{\cal E}=   \left( \begin{array}{ccc}
0 & -2B &  0 \\
2 B & 0& 0 \\
0 & 0 & 0 \end{array}
\right),
 \end{equation}
 and 
 \begin{equation}\label{gker}
 \vec{\gamma} = \frac{1}{B^2} \left(B_x B_z (\gamma_{\rm r}-\gamma_{\rm l}) , 0, B_z^2 \gamma_{\rm l} + B_x^2 \gamma_{\rm r} \right),
 \end{equation}
 being $D_{\rm l} = -B_x^2/B^2 \Gamma_{\rm l}, \; \overline{D}_{\rm l} =B_x B_z/B^2 \Gamma_{\rm l} $ and 
 $D_{\rm r} = -B^2_z \Gamma_{\rm r}, \; \overline{D}_{\rm r} =- B_x B_z/B^2 \Gamma_{\rm r} $.
  $\Gamma_{\alpha}=\gamma_{\alpha}  \left(e^{2B \beta_{\alpha}}-1\right)^{-1}$, 
 contains the information of the coupling with the baths with $\beta_{\alpha}=1/(k_B T)$,  and $\gamma_{\alpha}$ being the corresponding coupling strength.

\end{document}